\documentclass[11pt,a4paper,draft]{amsart}
\usepackage{times}
\usepackage{amsfonts,amscd,amsmath,amssymb,amsthm}

\usepackage{curves}

\numberwithin{equation}{section}

\renewcommand{\Im}{{\ensuremath{\mathrm{Im\,}}}}
\renewcommand{\Re}{{\ensuremath{\mathrm{Re\,}}}}

\DeclareSymbolFont{SY}{U}{psy}{m}{n}
\DeclareMathSymbol{\emptyset}{\mathord}{SY}{'306}

\DeclareMathOperator*{\slim}{s-lim} 
 
 \DeclareMathOperator{\tr}{tr}
\DeclareMathOperator{\Ran}{Ran} \DeclareMathOperator{\Ker}{Ker}
\DeclareMathOperator{\Rank}{Rank} 
\DeclareMathOperator{\Dom}{Dom} \DeclareMathOperator{\sign}{sign}

\DeclareMathSymbol{\newtimes}{\mathbin}{SY}{'264}
\DeclareMathOperator*{\Bigtimes}{\newtimes}

\newcommand{\diag}{\mathrm{diag}}

\newcommand{\R}{\mathbb{R}}
\newcommand{\T}{\mathbb{T}}
\newcommand{\C}{\mathbb{C}}
\newcommand{\Z}{\mathbb{Z}}
\newcommand{\N}{\mathbb{N}}

\newcommand{\1}{\mathbb{I}}

\newcommand{\fU}{\mathfrak{U}}
\newcommand{\fN}{\mathfrak{N}}

\newcommand{\fS}{\mathfrak{S}}

\newcommand{\fT}{\mathfrak{T}}

\newcommand{\fM}{\mathfrak{M}}

\newcommand{\cA}{{\mathcal A}}
\newcommand{\cB}{{\mathcal B}}
\newcommand{\cC}{{\mathcal C}}
\newcommand{\cD}{{\mathcal D}}
\newcommand{\cE}{{\mathcal E}}

\newcommand{\cG}{{\mathcal G}}
\newcommand{\cH}{{\mathcal H}}
\newcommand{\cI}{{\mathcal I}}
\newcommand{\cJ}{{\mathcal J}}
\newcommand{\cK}{{\mathcal K}}
\newcommand{\cL}{{\mathcal L}}
\newcommand{\cM}{{\mathcal M}}
\newcommand{\cN}{{\mathcal N}}
\newcommand{\cO}{{\mathcal O}}
\newcommand{\cP}{{\mathcal P}}

\newcommand{\cS}{{\mathcal S}}
\newcommand{\cT}{{\mathcal T}}
\newcommand{\cU}{{\mathcal U}}

\newcommand{\cW}{{\mathcal W}}

\newcommand{\sT}{{\mathsf T}}

\newcommand{\sU}{{\mathsf U}}

\newcommand{\D}{\mathbb{D}}

\newcommand{\sk}{\mathsf{k}}

\newcommand{\bw}{\mathbf{w}}

\newcommand{\bc}{\mathbf{c}}

\newcommand{\ii}{\mathrm{i}}
\newcommand{\e}{\mathrm{e}}

\newtheorem{introtheorem}{Theorem}{\bf}{\it}
{\bf}{\it}
\newtheorem{introcorollary}[introtheorem]{Corollary}{\bf}{\it}
{\bf}{\it}
\newtheorem{introassumption}{Assumption}{\bf}{\it}

\newtheorem{theorem}{Theorem}[section]{\bf}{\it}
\newtheorem{proposition}[theorem]{Proposition}{\bf}{\it}
\newtheorem{corollary}[theorem]{Corollary}{\bf}{\it}
\newtheorem{example}[theorem]{Example}{\it}{\rm}
\newtheorem{lemma}[theorem]{Lemma}{\bf}{\it}
\newtheorem{remark}[theorem]{Remark}{\it}{\rm}
\newtheorem{definition}[theorem]{Definition}{\bf}{\it}
{\bf}{\it}

\title[Inverse Scattering Problem for Metric Graphs]{The Inverse Scattering Problem
for Metric Graphs\\ and the Traveling Salesman Problem}

\author[V. Kostrykin]{Vadim Kostrykin}
\address{Vadim Kostrykin\\ Institut f\"{u}r Mathematik, Technische Universit\"{a}t Clausthal,
Erzstra{\ss}e 1, D-38678 Clausthal-Zellerfeld, Germany}
\email{kostrykin@math.tu-clausthal.de, kostrykin@t-online.de}

\author[R. Schrader]{Robert Schrader}
\address{Robert Schrader\\ Institut f\"{u}r Theoretische Physik\\
Freie Universit\"{a}t Berlin, Arnimallee 14, D-14195 Berlin, Germany}
\email{schrader@physik.fu-berlin.de}

\dedicatory{Dedicated to Arthur Jaffe, Walter Schneider, and Roland Seneor\\
on the occasion of their birthday celebration}

\date{21st February 2006}



\begin{document}

\begin{abstract}
We present a solution to the inverse scattering problem for differential
Laplace operators on metric noncompact graphs. We prove that for almost all
boundary conditions (i) the scattering matrix uniquely determines the graph
and its metric structure, (ii) the boundary conditions are determined
uniquely up to trivial gauge transformations. The main ingredient of our
approach is a combinatorial Fourier expansion of the scattering matrix
which encodes the topology of the graph into analytic properties of the
scattering matrix. Using the technique developed in this work, we also
propose an analytic approach to solving some combinatorial problems on
graphs, in particular, the Traveling Salesman Problem.
\end{abstract}

\maketitle

\tableofcontents

\section{Introduction}\label{sec:1}

The problem of reconstructing geometric objects like obstacles or manifolds
from scattering data is a scattering-theoretic analogue of the famous
question by M.~Kac: \emph{Can one hear the shape of a drum?} \cite{K}. The
question is whether an Euclidean domain (or more generally a compact
Riemannian manifold) is determined by the spectrum of its Laplace-Beltrami
operator. Although there are examples of isospectral (i.e.\ having the same
eigenvalues with the same multiplicities) but non-isometric manifolds
\cite{Gordon:2}, up to now it is unknown whether Kac's question has an
affirmative answer for ``generic'' manifolds. Another open question is how
large the set of manifolds isospectral to a given manifold can be. In
\cite{Osgood:Phillips:Sarnak} Osgood, Phillips, and Sarnak proved that any
isospectral class of two-dimensional domains is at most compact in an
appropriate topology on domains. A number of further ``affirmative'' results
are reviewed in \cite{Zelditch}.

The status of the inverse scattering problem is rather similar. We mention
the following results.

Inverse scattering problem for obstacles: Hassell and Zelditch
\cite{Hassell:Zelditch} proved that the scattering phase (that is, a
complex argument of the determinant of the scattering matrix) for the
Dirichlet Laplace operator of an exterior domain in $\R^2$ with smooth
boundary determines the obstacle up to a compact set of deformations.

Inverse scattering problem on noncompact Riemannian manifolds: An example of
two non-isometric two-dimensional asymptotically Euclidean manifolds with
the same scattering phase has been constructed by Brooks and Perry in
\cite{Brooks:Perry}.

Inverse scattering problem on compact Riemannian manifolds with infinitely
thin horns: Br\"{u}\-ning and Geyler proved in \cite{BG} that the spectrum of
the Laplace operator on a low-dimen\-sional compact Riemannian manifold is
uniquely determined by the scattering matrix of the Laplace operator on this
manifold with an attached semiline $\R_+$. Thus, the inverse scattering
problem is reduced to ``hearing the shape of a drum''.

In the present work we address the inverse scattering problem for Laplace
operators on \emph{noncompact metric graphs} -- one-dimensional noncompact
piecewise linear spaces with singularities at the vertices, that is,
non-smooth deformation retracts of smooth Riemannian manifolds.
Alternatively, a noncompact metric graph is a metric space which can be
written as a union of finitely many intervals, which are either compact or
$[0,\infty)$; any two of these intervals are either disjoint or intersect
only in one or both of their endpoints. The corresponding Laplace operator
arises as a strong limit of Laplace-Beltrami operators on the approximating
manifolds \cite{Exner:Post}, \cite{Kuchment:1}, \cite{Kuchment:2},
\cite{Rubinstein}, \cite{Saito:2}, \cite{Saito:3}. A survey of results on
Laplace operators on metric graphs can be found in \cite{Kuchment:00} and
\cite{Kuchment:05}.

The inverse problem consists of determining the graph, its metric structure
(i.e.\ the lengths of its edges), and the boundary conditions at the
vertices from the scattering matrix. It is known \cite{Kurasov}, \cite{KuSe}
that in general neither the graph nor the boundary conditions can be
determined uniquely from the scattering matrix. A similar situation occurs
in the context of compact metric graphs (that is, unions of finitely many
compact intervals): The articles \cite{Below:1}, \cite{GS} provide examples
of two different metric graphs such that the corresponding Laplace
operators are isospectral.

Below we will prove, that for an arbitrary graph with \emph{generic} metric
structure and \emph{generic} boundary conditions the inverse scattering
problem has a unique solution. This is the main result of our work. Precise
formulations will be presented in the next section.

The main technical ingredient of our approach to the solution of the
inverse scattering problem is a combinatorial Fourier expansion of the
scattering matrix (Theorems \ref{thm:main:harmony}, \ref{verloren}, and
\ref{eindeutigkeit} below). This expansion encodes the topology of the graph
and its metric structure into analytic properties of the scattering matrix.

In \cite{GS} Gutkin and Smilansky have announced a different solution of the
inverse scattering problem for graphs based on a relation between the
scattering phase and lengths of all closed paths on the graph, a kind of
Selberg-Gutzwiller formula \cite{Gutzwiller}. A heuristic derivation of
this relation has been presented by Kottos and Smilansky in \cite{Kottos:2}
and \cite{Kottos:3}. The arguments in \cite{GS} seem to be applicable only
to those boundary conditions which give rise to energy-independent
single-vertex scattering matrices.

A solution of the inverse spectral problem for compact metric graphs has
been recently given by Kurasov and Nowaczyk in \cite{KuNo}. This work gives
a rigorous proof of the solution presented in \cite{GS}. Inverse spectral
problems on finite trees have been studied in \cite{Belishev}.

A different type of the inverse problem for Schr\"{o}dinger operators on metric
graphs (i.e., Laplace operators perturbed by a potential) had been solved
in \cite{G} using results obtained in \cite{GP}. In that work the graph as
well as the boundary conditions were supposed to be known. We mention also
the articles \cite{Ha3}, \cite{Ha4}, \cite{Pivovarchik}, and \cite{Yurko}
devoted to determining the potentials from the spectrum of the Schr\"{o}dinger
operator on a given compact graph.

As an application of concepts and techniques developed in the present work,
we will also propose an analytic approach to solving the symmetric
Traveling Salesman Problem (TSP) as well as some other combinatorial
problems. In particular, we reduce these problems to an analysis of the
scattering matrix of a Laplace operator on the graph. However, it is too
early to decide whether this approach may lead to an effective algorithm.
Expository accounts of TSP can be found, e.g., in \cite{Applegate},
\cite{Jungnickel}, and \cite{LLKS}. The web pages \cite{TSP} provide a
large amount of information on the history of TSP as well as on-going
research. Proofs of NP-completeness of TSP are given in \cite{GJ}, \cite{P}.

For other relations of Laplace operators on metric graphs to combinatorics
we also mention the work \cite{Schanz:Smilansky} by Schanz and Smilansky,
where some results of spectral analysis of Laplace operators on graphs have
been applied to obtain combinatorial identities, as well as our recent work
\cite{KS6} on the calculation of the generating function of walks on graphs.

The article is organized as follows. In Section~\ref{main:results} we
summarize the graph theory terminology used in the present work and
describe the main results -- Theorems~\ref{1:theo:1} and \ref{1:theo:2}. In
Section~\ref{sec:2} we will revisit the theory of (differential) Laplace
operators on metric graphs for general boundary conditions at the vertices.
An important case of local boundary conditions is discussed in detail in
Section~\ref{sec:reconstruction}. Section~\ref{sec:harmony} is devoted to
the study of the scattering matrix as a function of the metric of the
graph. In particular, we prove (Theorem~\ref{thm:main:harmony}) that the
scattering matrix of the Laplace operator on a non-compact graph possesses
an absolutely converging multidimensional Fourier expansion. In
Section~\ref{sec:4} we study walks on a non-compact graph, that is,
sequences of edges with the property that any two successive edges have a
common vertex. The main results of this section are given in Theorem
\ref{verloren}, which expresses the Fourier coefficients as finite sums
over walks with a given combinatorial length, and in Theorem \ref{4:theo:1},
which relates the topology of the graph to analytic properties of the
scattering matrix. Sections \ref{sec:thm1} and \ref{sec:7} are devoted to
the proof of the main results of the present work. The two principal tools
used in the proof of Theorem \ref{1:theo:1} are Theorems
\ref{eindeutigkeit} and \ref{lem:7:7}. The first result is of analytical
nature and the second one is purely combinatorial. The main ingredient of
the proof of Theorem \ref{1:theo:2} is Proposition \ref{rb:propo:1}. In
Section \ref{sec:5} we treat TSP and some other combinatorial problems. The
reader interested in TSP only, may skip Section \ref{sec:7}.

\subsection*{Acknowledgments} The authors would like to thank M.~Gr\"{o}tschel,
M.~Karowski, W.~Klotz, H.~Kurke, M.~Schmidt, F.~Sobieczky, and E.~Vogt for
helpful comments. One of the authors (R.S.) is indebted to the Theory Group
of Microsoft Research for its kind hospitality during his stay in Redmond
in the spring of 2004.

\section{Main Results}\label{main:results}

Before we turn to the description of the main results obtained in the
present work, we summarize the graph theory terminology used below.

A finite noncompact graph is a 4-tuple $\cG=(V,\cI,\cE,\partial)$, where
$V$ is a finite set of \emph{vertices}, $\cI$ is a finite set of
\emph{internal edges}, $\cE$ is a finite set of \emph{external edges}.
Elements in $\cI\cup\cE$ are called \emph{edges}. The map $\partial$ assigns
to each internal edge $i\in\cI$ an ordered pair of (possibly equal) vertices
$\partial(i):=\{v_1,v_2\}$ and to each external edge $e\in\cE$ a single
vertex $v$. The vertices $v_1=:\partial^-(i)$ and $v_2=:\partial^+(i)$ are
called the \emph{initial} and \emph{terminal} vertex of the internal edge
$i$, respectively. The vertex $v=\partial(e)$ is the initial vertex of the
external edge $e$. We write $\partial(i)\bumpeq\{v_1,v_2\}$ if either
$\partial(i)=\{v_1,v_2\}$ or $\partial(i)=\{v_2,v_1\}$. If
$\partial(i)=\{v,v\}$, that is, $\partial^-(i)=\partial^+(i)$ then $i$ is
called a \emph{tadpole}. A graph is called \emph{compact} if
$\cE=\emptyset$, otherwise it is \emph{noncompact}.

Two vertices $v$ and $v^\prime$ are called \emph{adjacent} if there is an
internal edge $i\in\cI$ such that $v\in\partial(i)$ and $v^\prime\in\partial(i)$.
A vertex $v$ and the (internal or external) edge
$j\in\cI\cup\cE$ are \emph{incident} if $v\in\partial(j)$.

We do not require $\partial$ to be injective. In particular, any two
vertices are allowed to be adjacent by more than one internal edge and two
different external edges may be incident with the same vertex. If
$\partial$ is injective and $\partial^-(i)\neq\partial^+(i)$ for all
$i\in\cI$, the graph $\cG$ is called \emph{simple}.\footnote{In a different
terminology (see, e.g., \cite{Diestel}) compact graphs are called digraphs
whereas simple compact graphs are designated as oriented graphs.}

The \emph{degree} $\deg(v)$ of the vertex $v$ is defined as
\begin{equation*}
\deg(v)=|\{e\in\cE\mid\partial(e)=v\}|+|\{i\in\cI\mid\partial^-(i)=v\}|+|\{i\in\cI\mid\partial^+(i)=v\}|,
\end{equation*}
that is, it is the number of (internal or external) edges incident with the
given vertex $v$ by which every tadpole is counted twice. The minimum and
the maximum degree of the graph $\cG$ are defined as
\begin{equation*}
\min_{v\in V} \deg(v)\qquad\text{and}\qquad \max_{v\in V} \deg(v),
\end{equation*}
respectively.

It is easy to extend the First Theorem of Graph Theory (see, e.g.\,
\cite{Diestel}) to the case of noncompact graphs:
\begin{equation*}
\sum_{v\in V} \deg(v) = |\cE| + 2 |\cI|.
\end{equation*}

A vertex is called a \emph{boundary vertex} if it is incident with some
external edge. The set of all boundary vertices will be denoted by $\partial
V$. The vertices not in $\partial V$, that is in $V\setminus\partial V$ are
called \emph{internal vertices}.

The compact graph $\cG_{\mathrm{int}}=(V,\cI,\emptyset,\partial|_{\cI})$
will be called the \emph{interior} of the graph $\cG=(V,\cI,$ $
\cE,\partial)$. It is obtained from $\cG$ by eliminating the external
edges. Similarly, the graph $\cG_{\mathrm{ext}}=(\partial
V,\emptyset,\cE,\partial|_{\cE})$ will be called the \emph{exterior} of the
graph $\cG$. It is obtained from $\cG$ by eliminating all its internal
edges.

Let $\cS(v)\subseteq \cE\cup\cI$ denote the \emph{star graph} of the vertex
$v\in V$, i.e., the set of the edges adjacent to $v$. Also, by $\cS_{-}(v)$
(respectively $\cS_{+}(v)$) we denote the set of the edges for which $v$ is
the initial vertex (respectively terminal vertex). Obviously,
$\cS_{+}(v)\cap \cS_{-}(v)=\emptyset$ if $\cG$ does not contain tadpoles.

We will endow the graph with the following metric structure. Any internal
edge $i\in\cI$ will be associated with an interval $[0,a_i]$ with $a_i>0$
such that the initial vertex of $i$ corresponds to $x=0$ and the terminal
one - to $x=a_i$. Any external edge $e\in\cE$ will be associated with a
semiline $[0,+\infty)$. We call the number $a_i$ the length of the internal
edge $i$. The set of lengths $\{a_i\}_{i\in\cI}$, which will also be treated
as an element of $\R^{|\cI|}$, will be denoted by $\underline{a}$. The map
$\cI\rightarrow\underline{a}$ can be seen as a positive weight on the
interior $\cG_{\mathrm{int}}$ of the graph $\cG$. A compact or noncompact
graph $\cG$ endowed with a metric structure is called a \emph{metric graph}
$(\cG,\underline{a})$. In a different terminology compact metric graphs are
called \emph{positively weighted graphs} or \emph{networks} (see, e.g.,
\cite{Jungnickel}).

To define a (differential) Laplace operator on the metric graph
$(\cG,\underline{a})$ consider the family
$\psi=\{\psi_{j}\}_{j\in\cE\cup\cI}$ of complex valued functions $\psi_{j}$
defined on $[0,\infty)$ if $j\in\cE$ and on $[0,a_{j}]$ if $j\in\cI$. The
Laplace operator is defined as
\begin{equation}\label{diff:expression}
\left(\Delta(A,B,\underline{a})\psi\right)_j (x) = \frac{d^2}{dx^2}
\psi_j(x),\qquad j\in\cI\cup\cE
\end{equation}
with the boundary conditions
\begin{equation}\label{Randbedingungen}
A\underline{\psi} + B\underline{\psi}' = 0
\end{equation}
where
\begin{equation}\label{lin1:add}
\underline{\psi} = \begin{pmatrix} \{\psi_e(0)\}_{e\in\cE} \\
                                   \{\psi_i(0)\}_{i\in\cI} \\
                                   \{\psi_i(a_i)\}_{i\in\cI} \\
                                     \end{pmatrix},\qquad
\underline{\psi}' = \begin{pmatrix} \{\psi_e'(0)\}_{e\in\cE} \\
                                   \{\psi_i'(0)\}_{i\in\cI} \\
                                   \{-\psi_i'(a_i)\}_{i\in\cI} \\
                                     \end{pmatrix},
\end{equation}
$A$ and $B$ are $(|\cE|+2|\cI|)\times(|\cE|+2|\cI|)$ matrices. The operator
$\Delta(A,B,\underline{a})$ is self-adjoint if and only if the matrix
$(A,B)$ has maximal rank and the matrix $A B^\dagger$ is Hermitian
\cite{KS1}. Here and in what follows $(A,B)$ will denote the
$(|\cE|+2|\cI|)\times 2(|\cE|+2|\cI|)$ matrix, where $A$ and $B$ are put
next to each other. Boundary conditions $(A,B)$ and $(A^\prime, B^\prime)$
define the same operator
$\Delta(A,B,\underline{a})=\Delta(A^\prime,B^\prime,\underline{a})$ if and
only if the unitary matrices
\begin{equation*}
\fS(A,B) = - (A+\ii B)^{-1} (A-\ii B)\qquad\text{and}\qquad
\fS(A^\prime,B^\prime) = - (A^\prime+\ii B^\prime)^{-1}
(A^\prime-\ii B^\prime)
\end{equation*}
coincide (see Proposition \ref{unit:neu} below). Conversely, any unitary
matrix $\fS\in\mathsf{U}(|\cE|+2|\cI|)$ defines a self-adjoint
Laplace operator corresponding to boundary conditions
\eqref{Randbedingungen} with $A=\1+\fS$ and
$B=-\ii(\1-\fS)$. Thus, the set of all Laplace operators on a
graph $\cG$ is uniquely parametrized by elements of the unitary group
$\mathsf{U}(|\cE|+2|\cI|)$.

Recall (see \cite{KS1} and \cite{KS5}) that the boundary conditions are
called local if they couple only those boundary values of $\psi$ and of its
derivative $\psi'$ which belong to the same vertex. The precise definition
will be given in Section \ref{sec:reconstruction} below (Definition
\ref{def:local}). The set of all local boundary conditions on a graph $\cG$
is isomorphic to the Lie group
\begin{equation*}
\mathsf{U}_{\cG} = \Bigtimes_{v\in V} \mathsf{U}(\deg(v)),
\end{equation*}
where $\mathsf{U}(n)$ is the group of all unitary transformations of $\C^n$.
The Haar measure on $\mathsf{U}(n)$ induces a Haar measure on
$\mathsf{U}_{\cG}$. The statement that a property holds ``for Haar almost
all local boundary conditions'' means that there is a subset
$\mathcal{U}\subset\mathsf{U}_{\cG}$ of full Haar measure such that this
property holds for all boundary conditions $(A,B)$ with
$\fS(A,B)\in\mathcal{U}$ (see Definition \ref{generic:alg} below).

We start with a discussion of the main results obtained in the present work
with the formulation of the main assumptions made below. We split them into
two groups.

\begin{introassumption}[topological properties]\label{con:graph}
The graph $\cG$ possesses the following properties:
\begin{itemize}
\item[(i)]{$\cG$ is connected, i.e., for any $v,v^\prime\in V$ there is an ordered sequence
$\{v_1=v, v_2,\ldots, v_n=v^\prime\}$ such that any two successive vertices
in this sequence are adjacent. In particular, this implies that any vertex
of the graph $\cG$ has nonzero degree, i.e., for any vertex there is at
least one edge with which it is incident.}
\item[(ii)]{$|\cE|\geq 1$, i.e., the graph has at least one external edge.}
\item[(iii)]{The graph has no tadpoles, i.e., for no edge its initial
    and terminal vertex coincide.}
\end{itemize}
\end{introassumption}

\begin{introassumption}[metric property]\label{2cond}
The lengths $a_{i}\,(i\in\cI)$ of the internal edges of the graph $\cG$ are
rationally independent, i.e., the equation
\begin{equation}\label{indep}
\sum_{i\in\cI}n_{i}\, a_{i} = 0
\end{equation}
with \emph{integer} $n_{i}\in\Z$ has no non-trivial solution.
\end{introassumption}

Let $S(\sk;A,B,\underline{a})$ denote the scattering matrix at the energy
$E=\sk^2>0$ associated with the Laplace operator
$\Delta(A,B,\underline{a})$ on the metric graph $(\cG,\underline{a})$ (see
Subsection \ref{subsec:3.1} below). It is an $|\cE| \times |\cE|$ unitary
matrix indexed by the elements of $\cE$. Its diagonal elements are
reflection amplitudes while the off-diagonal elements are transmission
amplitudes for the incoming plane wave $\e^{-\ii\sk x}$.

\begin{introtheorem}\label{1:theo:1}
Let the metric graph $(\cG,\underline{a})$ satisfy Assumptions
\ref{con:graph} and \ref{2cond}. Then for any external edge $e\in\cE$ the
map
\begin{equation*}
(\cG_{\mathrm{int}},\partial(e),\underline{a}) \mapsto
[S(\cdot;A,B,\underline{a})]_{e,e}
\end{equation*}
is injective for Haar almost all local boundary conditions $(A,B)$.
Furthermore, for any given $e\in\cE$ the maps
\begin{equation*}
(\cG,\underline{a}) \mapsto
\left\{[S(\cdot;A,B,\underline{a})]_{e,e^\prime}\right\}_{e^\prime\in\cE}\qquad\text{and}\qquad
(\cG,\underline{a}) \mapsto
\left\{[S(\cdot;A,B,\underline{a})]_{e^\prime,e}\right\}_{e^\prime\in\cE}
\end{equation*}
are also injective for Haar almost all local boundary conditions $(A,B)$.
\end{introtheorem}

Theorem \ref{1:theo:1} claims that it suffices to know one reflection
amplitude for all $\sk>0$ in order to determine the sets $V$, $\cI$, the
map $\partial|_{\cI}$ and, thus, the interior
$\cG_{\mathrm{int}}=(V,\cI,\emptyset,\partial|_{\cI})$ of the graph $\cG$
as well as the metric $\underline{a}$ and the vertex $\partial(e)$.
Furthermore, if under otherwise the same assumptions one entire row or one
entire column of $S(\sk;A,B,\underline{a})$ is known for all $\sk>0$, then
all boundary vertices are uniquely determined.

Note that ``experimentally'' one may only obtain the transmission and the
reflection probabilities -- the absolute squares of the transmission and
reflection amplitudes -- and not the transmission and reflection amplitudes
themselves. However, as argued in \cite{KS3}, the missing phases may be
obtained by additional ``experiments''.

Although we do not formulate it as a conjecture, based on our discussion
below we expect that Theorem \ref{1:theo:1} remains valid under the much
less stringent condition replacing Assumption \ref{2cond}: \textit{There is
no solution to \eqref{indep} with $|n_{i}|\leq N$ for a suitable integer $N$
depending on the graph $\cG$.}

The following theorem implies that the boundary conditions can also be
determined from the scattering matrix.

\begin{introtheorem}\label{1:theo:2}
Let the metric graph $(\cG,\underline{a})$ satisfy Assumptions
\ref{con:graph} and \ref{2cond}. Then for Haar almost all local boundary
conditions $(A,B)$ on the graph $\cG$ the map
\begin{equation*}
\fS(A,B) \mapsto S(\cdot;A,B,\underline{a})
\end{equation*}
is injective up to trivial gauge transformations.
\end{introtheorem}

The notion of \emph{trivial gauge transformations} is introduced in
Definition \ref{trivial:gauge}. Such transformations correspond to
perturbations of the Laplace operator by vanishing magnetic potentials
\cite{KS5}.

In other words, Theorem \ref{1:theo:2} states that for Haar almost all
boundary conditions $(A,B)$ on the metric graph $\cG$ the equality
$S(\sk;A,B,\underline{a})=S(\sk;A^\prime,B^\prime,\underline{a})$ for all
$\sk>0$ implies that the operators $\Delta(A,B,\underline{a})$ and
$\Delta(A^\prime,B^\prime,\underline{a})$ agree up to trivial gauge
transformations. A partial reconstruction of the boundary conditions from
the spectrum of Laplace operator on compact metric graphs has been
discussed earlier in \cite{Carlson} and \cite{GS}.

The proof of Theorems \ref{1:theo:1} and \ref{1:theo:2} will be given in
Sections \ref{sec:thm1} and \ref{sec:7}, respectively.

By Theorem \ref{thm:cont} below the scattering matrix
$S(\sk;A,B,\underline{a})$ is a meromorphic function in the complex
$\sk$-plane. Therefore, we have the immediate

\begin{introcorollary}\label{cor:1:1}
The function $\R_+\ni\sk\mapsto S(\sk;A;B;\underline{a})$ in Theorems
\ref{1:theo:1} and \ref{1:theo:2} can be replaced by
$\{S(\sk_n;A;B;\underline{a})\}_{n\in\N_0}$, where $\sk_n>0$ is an
arbitrary sequence with at least one finite accumulation point different
from zero.
\end{introcorollary}

\section{Laplace Operators and Scattering Matrices on Metric Graphs}\label{sec:2}

In this section we revisit the theory of Laplace operators on a metric graph
$\cG$ and the resulting scattering theory. A large portion of the material
presented in this section is borrowed from our preceding papers \cite{KS1},
\cite{KS3}, \cite{KS4}.

Given a finite noncompact graph $\cG=(V,\cI,\cE,\partial)$ and a metric
structure $\underline{a}=\{a_i\}_{i\in\cI}$ consider the Hilbert space
\begin{equation}\label{hilbert}
\cH\equiv\cH(\cE,\cI,\underline{a})=\cH_{\cE}\oplus\cH_{\cI},\qquad
\cH_{\cE}=\bigoplus_{e\in\cE}\cH_{e},\qquad
\cH_{\cI}=\bigoplus_{i\in\cI}\cH_{i},
\end{equation}
where $\cH_{e}=L^2([0,\infty))$ for all $e\in\cE$ and
$\cH_{i}=L^2([0,a_{i}])$ for all $i\in\cI$.

By $\cD_j$ with $j\in\cE\cup\cI$ denote the set of all $\psi_j\in\cH_j$
such that $\psi_j(x)$ and its derivative $\psi^\prime_j(x)$ are absolutely
continuous and $\psi^{\prime\prime}_j(x)$ is square integrable. Let
$\cD_j^0$ denote the set of those elements $\psi_j\in\cD_j$ which satisfy
\begin{equation*}
\begin{matrix}
\psi_j(0)=0\\ \psi^\prime_j(0)=0
\end{matrix} \quad \text{for}\quad j\in\cE\qquad\text{and}\qquad
\begin{matrix}
\psi_j(0)=\psi_j(a_j)=0\\
\psi^\prime_j(0)=\psi^\prime_j(a_j)=0
\end{matrix}
\quad\text{for}\quad j\in\cI.
\end{equation*}
Let $\Delta^0$ be the differential operator
\begin{equation}\label{Delta:0}
\left(\Delta^0\psi\right)_j (x) = \frac{d^2}{dx^2} \psi_j(x),\qquad
j\in\cI\cup\cE
\end{equation}
with domain
\begin{equation*}
\cD^0=\bigoplus_{j\in\cE\cup\cI} \cD_j^0 \subset\cH.
\end{equation*}
It is straightforward to verify that $\Delta^0$ is a closed symmetric
operator with deficiency indices equal to $|\cE|+2|\cI|$.

We introduce an auxiliary finite-dimensional Hilbert space
\begin{equation}\label{K:def}
\cK\equiv\cK(\cE,\cI)=\cK_{\cE}\oplus\cK_{\cI}^{(-)}\oplus\cK_{\cI}^{(+)}
\end{equation}
with $\cK_{\cE}\cong\C^{|\cE|}$ and $\cK_{\cI}^{(\pm)}\cong\C^{|\cI|}$. Let
${}^d\cK$ denote the ``double'' of $\cK$, that is, ${}^d\cK=\cK\oplus\cK$.

For any $\displaystyle\psi\in\cD:=\bigoplus_{j\in\cE\cup\cI} \cD_j$ we set
\begin{equation}\label{lin1}
[\psi]:=\underline{\psi}\oplus \underline{\psi}^\prime\in{}^d\cK,
\end{equation}
with $\underline{\psi}$ and $\underline{\psi}^\prime$ defined in \eqref{lin1:add}.

Let $J$ be the canonical symplectic matrix on ${}^d\cK$,
\begin{equation}\label{J:canon}
J=\begin{pmatrix} 0& \1 \\ -\1 & 0
\end{pmatrix}
\end{equation}
with $\1$ the identity operator on $\cK$. Consider the non-degenerate
Hermitian symplectic form
\begin{equation}\label{omega:canon}
\omega([\phi],[\psi]) := \langle[\phi], J[\psi]\rangle,
\end{equation}
where $\langle\cdot,\cdot\rangle$ denotes the scalar product in ${}^d
\cK\cong\C^{2(|\cE|+2|\cI|)}$. Note that the Hermitian symplectic form
$\omega$ differs from the standard (Euclidean) symplectic form
\cite{symplectic}, \cite{McDS}.

Recall that a linear subspace $\cM$ of ${}^d\cK$ is called \emph{isotropic}
if the form $\omega$ vanishes on $\cM$ identically. An isotropic subspace
is called \emph{maximal} if it is not a proper subspace of a larger
isotropic subspace. Every maximal isotropic subspace has complex dimension
equal to $|\cE|+2|\cI|$.

If $\cM$ is a maximal isotropic subspace, then its orthogonal complement
$\cM^\perp$ (with respect to the scalar product $\langle\cdot,\cdot\rangle$)
is also maximal isotropic. Moreover, as in the standard symplectic theory we
have the following result (see \cite[Lemma 2.1]{KS1}):

\begin{lemma}\label{ortho:lemma}
An isotropic subspace $\cM\subset{}^d\cK\cong\C^{2(|\cE|+2|\cI|)}$ is
maximal if and only if $\cM^\perp = J\cM$ such that the orthogonal
decomposition
\begin{equation}
\label{mdecomp4} {}^d\cK =\cM\oplus\,J\cM
\end{equation}
holds.
\end{lemma}

By the symplectic extension theory of symmetric operators on a Hilbert
space (see, e.g., \cite{Akhiezer:Glazman}, \cite{Verdiere}, \cite{Everitt},
\cite{HP}, \cite{Novikov:98}, \cite{Pavlov}) there is a one-to-one
correspondence between all self-adjoint extensions of $\Delta^0$ and
maximal isotropic subspaces of ${}^d\cK$. In explicit terms, any
self-adjoint extension of $\Delta^0$ is the differential operator defined
by \eqref{Delta:0} with domain
\begin{equation}\label{thru}
\Dom(\Delta)=\{\psi\in\cD|\; [\psi]\in\cM\},
\end{equation}
where $\cM$ is a maximal isotropic subspace of ${}^d\cK$. Conversely, any
maximal isotropic subspace $\cM$ of ${}^d\cK$ defines through \eqref{thru}
a self-adjoint operator $\Delta(\cM, \underline{a})$. If $\cI=\emptyset$,
we will simply write $\Delta(\cM)$. In the sequel we will call the operator
$\Delta(\cM, \underline{a})$ a Laplace operator on the metric graph $(\cG,
\underline{a})$.

Let $A$ and $B$ be linear maps of $\cK$ onto itself. By $(A,B)$ we denote
the linear map from ${}^d\cK=\cK\oplus\cK$ to $\cK$ defined by the relation
\begin{equation*}
(A,B)\; (\chi_1\oplus \chi_2) := A\, \chi_1 + B\, \chi_2,
\end{equation*}
where $\chi_1,\chi_2\in\cK$. Set
\begin{equation}\label{M:def}
\cM(A,B) := \Ker\, (A,B).
\end{equation}

\begin{theorem}\label{thm:3.1}
A subspace $\cM\subset{}^d\cK$ is maximal isotropic if and only if there
exist linear maps $A,\,B:\; \cK\rightarrow\cK$ such that $\cM=\cM(A,B)$ and
\begin{equation}\label{abcond}
\begin{split}
\mathrm{(i)}\; & \;\text{the map $(A,B):\;{}^d\cK\rightarrow\cK$ has maximal
rank equal to
$|\cE|+2|\cI|$,}\qquad\\
\mathrm{(ii)}\; &\;\text{$AB^{\dagger}$ is self-adjoint,
    $AB^{\dagger}=BA^{\dagger}$.}
\end{split}
\end{equation}
\end{theorem}

For the proof we refer to \cite{KS1}. We note that Theorem 4 in Section 125
of \cite{Akhiezer:Glazman} presents an analogous result for differential
operators of even order on a compact interval.

By direct calculations one can easily verify the equality
\begin{equation*}
J \cM(A, B) = \cM(B, -A),
\end{equation*}
which by Lemma \ref{ortho:lemma} implies the following result.

\begin{lemma}\label{lem:neu:3.3}
A subspace $\cM(A,B)\subset{}^d\cK$ is maximal isotropic if and only if
\begin{equation}\label{perp}
\cM(A,B)^{\perp} = \cM(B,-A).
\end{equation}
\end{lemma}

We mention also the equalities
\begin{equation}\label{frame}
\begin{split}
\cM(A,B)^\perp & = \bigl[\Ker\,(A,B)\bigr]^\perp  = \Ran\, (A,B)^\dagger,\\
\cM(A,B) & = \Ran(-B,A)^\dagger.
\end{split}
\end{equation}
In the terminology of symplectic geometry (see, e.g., Section 2.3 in
\cite{McDS}) equalities \eqref{frame} have the following interpretation: The
matrix $(A,B)^\dagger$ is a (Lagrangian) \emph{frame} for
$\cM(A,B)^{\perp}$ and the matrix $(-B,A)^\dagger$ is a frame for
$\cM(A,B)$.

\begin{lemma}\label{kerne}
If $(A,B)$ satisfies \eqref{abcond}, then
\begin{itemize}
\item[(i)]{$\Ker A \perp \Ker B$,}
\item[(ii)]{$\Ker A^\dagger \cap \Ker B^\dagger = \{0\}$.}
\end{itemize}
\end{lemma}

\begin{proof}
(i). Choose arbitrary $\chi_1\in\Ker A$ and $\chi_2\in\Ker B$. For any
complex number $c$ we have
\begin{equation*}
\begin{pmatrix} \chi_1 \\ c\chi_2
\end{pmatrix}\in\cM(A,B)\qquad\text{and}\qquad
\begin{pmatrix} c\chi_2 \\ \chi_1 \end{pmatrix}\in\cM(B,-A).
\end{equation*}
{}From \eqref{perp} it follows that
\begin{equation*}
\left\langle\begin{pmatrix} \chi_1 \\ c\chi_2 \end{pmatrix},
\begin{pmatrix} c\chi_2 \\ \chi_1 \end{pmatrix} \right\rangle=0.
\end{equation*}
Hence, $\Re c\langle\chi_1, \chi_2\rangle_{\cK} = 0$ for all $c\in\C$ which
implies that $\langle\chi_1, \chi_2\rangle_{\cK} = 0$.

(ii). Since the linear map $\begin{pmatrix}A^\dagger \\
B^\dagger\end{pmatrix} = (A,B)^\dagger:\, \cK\rightarrow{}^d\cK$ has maximal
rank equal to $|\cE|+2|\cI|$, it follows that $\Ker \begin{pmatrix}A^\dagger \\
B^\dagger\end{pmatrix}=\{0\}$. Noting that  $\Ker \begin{pmatrix}A^\dagger \\
B^\dagger\end{pmatrix}= \Ker A^\dagger \cap \Ker B^\dagger$ completes the
proof.
\end{proof}

\begin{definition}\label{def:equiv}
The boundary conditions $(A,B)$ and $(A',B')$ satisfying \eqref{abcond} are
equivalent if the corresponding maximal isotropic subspaces coincide, that
is, $\cM(A,B)=\cM(A',B')$.
\end{definition}

\begin{proposition}\label{prop:3.3}
The boundary conditions $(A,B)$ and $(A',B')$ satisfying \eqref{abcond} are
equivalent if and only if there is an invertible map $C:\,
\cK\rightarrow\cK$ such that $A'= CA$ and $B'=CB$.
\end{proposition}

\begin{proof}
The ``if" part is obvious. To prove the ``only if" part assume that
$\cM(A,B)=\cM(A',B')$. This equality is equivalent to the condition
\begin{equation*}
\Ker\begin{pmatrix}A, B\end{pmatrix}=\Ker\begin{pmatrix}A', B'\end{pmatrix}.
\end{equation*}
But this condition holds if and only if there is an invertible $C$ such
that $A'= CA$ and $B'=CB$.
\end{proof}

Denote by $\Delta(A,B,\underline{a})$ the Laplace operator corresponding to
the maximal isotropic subspace $\cM(A,B)$, that is,
$\Delta(A,B,\underline{a}):=\Delta(\cM(A,B),\underline{a})$. If
$\cI=\emptyset$, we will simply write $\Delta(A,B)$. {}From the discussion
above it follows immediately that any self-adjoint Laplace operator on
$\cH$ equals $\Delta(A,B,\underline{a})$ for some $A$ and $B$ satisfying
\eqref{abcond}. Moreover,
$\Delta(A,B,\underline{a})=\Delta(A',B',\underline{a})$ if and only if
$\cM(A,B)=\cM(A',B')$. {}From Theorem \ref{thm:3.1} it follows that the
domain of the Laplace operator $\Delta(A,B,\underline{a})$ consists of the
functions $\psi\in\cD$ satisfying the boundary conditions
\begin{equation}
\label{lin2}
 A\underline{\psi}+B\underline{\psi}^{\prime}=0,
\end{equation}
where $\underline{\psi}$ and $\underline{\psi}^{\prime}$ are defined by
\eqref{lin1:add}.

By comparing $\Delta(\cM, \underline{a})$ with the Laplace operators
constructed from $\Delta^0$ by means of the von Neumann extension theory
one immediately concludes that the manifold of all maximal isotropic
subspaces (Lagrangian Grassmannian) is isomorphic to the unitary group
$\mathsf{U}(|\cE|+2|\cI|)$ of all $(|\cE|+2|\cI|)\times(|\cE|+2|\cI|)$
matrices, a result contained in \cite{KS3}.

\begin{proposition}\label{unit:neu}
A subspace $\cM(A,B)\subset {}^d\cK$ is maximal isotropic if and only if
for an arbitrary $\sk\in\R\setminus\{0\}$ the operator $A+\ii\sk B$ is
invertible and
\begin{equation}\label{uuu:def}
\fS(\sk;A,B):=-(A+\ii\sk B)^{-1} (A-\ii\sk B)
\end{equation}
is unitary. Moreover, given any $\sk\in\R\setminus\{0\}$ the correspondence
between maximal isotropic subspaces $\cM\subset {}^d\cK$ and unitary
operators $\fS(\sk;A,B)\in\mathsf{U}(|\cE|+2|\cI|)$ on $\cK$ is
one-to-one.
\end{proposition}

A symplectic theory proof of this fact has been recently given by Arnold
\cite{Arnold} and Harmer \cite{H}. For reader's convenience we present here
an alternative rather elementary proof of this proposition.

\begin{proof}
The ``if'' part. Assume that the operator $A+\ii\sk B$ is invertible and
\eqref{uuu:def} is unitary. Using $\fS^\dagger=\fS^{-1}$
we obtain that $\Im(AB^\dagger)=0$, that is, $AB^\dagger$ is self-adjoint.
By (i) of Lemma \ref{kerne}
\begin{equation*}
\Rank\, (A,B)=\dim\cK-\dim (\Ker\, A\cap\Ker\, B) =\dim\cK
\end{equation*}
is maximal. By Theorem \ref{thm:3.1} this proves that $\cM(A,B)$ is maximal
isotropic.

The ``only if'' part is proven in \cite{KS1}. We recall the arguments.
Assume that $\cM(A,B)\subset{}^d\cK$ is maximal isotropic. Then, by Theorem
\ref{thm:3.1}, $(A,B)$ satisfies \eqref{abcond}. Assume that $\Ker (A +
\ii\sk B)\neq\{0\}$. Then there is a $\chi\in\cK$ such that $(A^\dagger -
\ii\sk B^\dagger)\chi=0$. In particular, we have
\begin{equation*}
\langle(A^\dagger - \ii\sk B^\dagger)\chi, (A^\dagger - \ii\sk
B^\dagger)\chi \rangle = 0.
\end{equation*}
By Theorem \ref{thm:3.1} this implies that
\begin{equation*}
\langle\chi, AA^\dagger \chi\rangle + \sk^2 \langle\chi, BB^\dagger
\chi\rangle = 0.
\end{equation*}
Hence, $\chi\in \Ker A^\dagger \cap \Ker B^\dagger$. Applying (ii) of Lemma
\ref{kerne} we obtain $\chi=0$. Thus, $A + \ii\sk B$ is invertible. To
prove the unitarity of $\fS(\sk;A,B)$ we observe that
\begin{equation*}
(A \pm \ii\sk B)^{-1} = (A^\dagger \mp \ii\sk B^\dagger)(AA^\dagger+\sk^2
BB^\dagger)^{-1}.
\end{equation*}
Combining this with \eqref{uuu:def} we obtain
\begin{equation*}
\fS(\sk;A,B)\fS(\sk;A,B)^\dagger = \fS(\sk;A,B)^\dagger \fS(\sk;A,B) =\1.
\end{equation*}

To prove that for given $\sk>0$ the correspondence between maximal isotropic
subspaces $\cM\subset {}^d\cK$ and unitary matrices
$\fS\in\mathsf{U}(|\cE|+2|\cI|)$ given by \eqref{uuu:def} is
one-to-one, first we show that the map $\cM\mapsto \fS$ is
surjective. Given an arbitrary $\fS\in\mathsf{U}(|\cE|+2|\cI|)$ we
set
\begin{equation}\label{ab:sigma}
A_\fS=-\frac{1}{2}(\fS - \1),\qquad
B_\fS=\frac{1}{2\ii \sk}(\fS+\1)
\end{equation}
such that $-(A_\fS+\ii\sk
B_{\fS})^{-1}(A_\mathfrak{S}-\ii\sk B_{\fS})=\mathfrak{S}$.
Obviously,
\begin{equation}\label{ab:sigma:kreuz}
A_\mathfrak{S} B_\mathfrak{S}^\dagger = \frac{1}{2\sk} \Im\, \mathfrak{S}
\equiv \frac{1}{4\ii\sk}(\mathfrak{S}-\mathfrak{S}^\dagger)
\end{equation}
is self-adjoint. Since $\Ker\, A_\mathfrak{S}\cap\Ker\, B_\mathfrak{S} =
\{0\}$, the map $(A_\mathfrak{S},B_\mathfrak{S})$ has maximal rank. By
Theorem \ref{thm:3.1} this proves that $\cM(A_\mathfrak{S},B_\mathfrak{S})$
is maximal isotropic.

To prove the injectivity by Proposition \ref{prop:3.3} it suffices to show
that given an arbitrary $\mathfrak{S}\in\mathsf{U}(|\cE|+2|\cI|)$ for any
map $(A,B)$ satisfying \eqref{abcond} and such that
\begin{equation}\label{sigma=sigma}
\mathfrak{S}(\sk;A,B)=\mathfrak{S}
\end{equation}
there is an invertible map $C:\, \cK\rightarrow\cK$ such that
\begin{equation*}
A=CA_\mathfrak{S}\qquad\text{and}\qquad B=CB_\mathfrak{S},
\end{equation*}
or, equivalently,
\begin{equation*}
A=-\frac{1}{2} C (\mathfrak{S}-\1)\quad\text{and}\quad B=\frac{1}{2\ii\sk}C
(\mathfrak{S} + \1).
\end{equation*}
Using \eqref{sigma=sigma} these relations can equivalently be written as
\begin{equation*}
A=C (A+\ii\sk B)^{-1} A\quad\text{and}\quad B=C (A + \ii\sk B)^{-1} B.
\end{equation*}
Thus,
\begin{equation*}
C=A+\ii\sk B,
\end{equation*}
which is invertible.
\end{proof}

\begin{remark}
There are several results in the literature related to Proposition
\ref{unit:neu} (see \cite{Kochubej} and references quoted therein).
\end{remark}

Due to Proposition \ref{unit:neu}, given an arbitrary maximal isotropic
subspace $\cM\subset{}^d\cK$, we will write $\mathfrak{S}(\sk;\cM)$ for
$\mathfrak{S}(\sk; A_{\cM}, B_{\cM})$ with $(A_{\cM}, B_{\cM})$ an arbitrary
map satisfying \eqref{abcond} such that $\Ker(A_\cM, B_\cM)=\cM$.

{}From \eqref{perp} and Proposition \ref{unit:neu} it follows that
\begin{equation}\label{sperp}
\mathfrak{S}(\sk;\cM)=-\mathfrak{S}(\sk^{\,-1};\cM^{\perp}).
\end{equation}

{}From \eqref{uuu:def} it follows that
\begin{equation}
\label{suinv} \mathfrak{S}(\sk;AU,BU)=U^\dagger\mathfrak{S}(\sk;A,B)U.
\end{equation}
for arbitrary $U\in\sU(|\cE|+2|\cI|)$. We cast this relation into the
following form. Let $^{d}U:\ {}^d\cK \rightarrow {}^d\cK$ denote the
``double'' of $U$,
\begin{equation}\label{muinv}
{}^{d}U=\begin{pmatrix}U&0\\0&U\end{pmatrix}.
\end{equation}
Since $^{d}U^{\dagger}J\;^{d}U=J$, the transformation $^{d}U$ leaves the
Hermitian symplectic structure invariant. Therefore, if $\cM$ is maximal
isotropic, then so is ${}^{d}U\cM$. Furthermore, by \eqref{M:def} we have
\begin{equation*}
{}^{d}U\cM(A,B) = \cM(AU^{\dagger},BU^{\dagger}).
\end{equation*}
Combining this equality with \eqref{suinv} gives that
\begin{equation}\label{usinv}
\mathfrak{S}(\sk\,;\;^{d}U\cM) = U\mathfrak{S}(\sk\,;\cM)U^{\dagger}
\end{equation}
holds for any maximal isotropic space $\cM$ and any unitary $U$.

{}From \eqref{uuu:def} we obtain that the matrix $\mathfrak{S}(\sk;A,B)$ for
an arbitrary $\sk\in\R\setminus\{0\}$ can be obtained from
$\mathfrak{S}(\sk_0;A,B)$ with $\sk_0\in\R\setminus\{0\}$ via
\begin{equation}
\label{sinv}
\mathfrak{S}(\sk;A,B)=\bigl((\sk-\sk_{0})\mathfrak{S}(\sk_{0};A,B)+(\sk+\sk_{0})\bigr)^{-1}
\bigl((\sk+\sk_{0})\mathfrak{S}(\sk_{0};A,B)+(\sk-\sk_{0})\bigr).
\end{equation}
Note that for all $\sk\in\R\setminus\{0,\sk_0\}$ the inequality
$|(\sk+\sk_{0})/(\sk-\sk_{0})| \neq 1$ holds such that
$((\sk-\sk_{0})\mathfrak{S}(\sk_{0};A,B)+(\sk+\sk_{0}))$ is invertible.

The relation \eqref{sinv} implies the somewhat surprising fact that the
operators $\mathfrak{S}(\sk;A,B)$ for different $\sk$ form a commuting
family, i.e.,
\begin{equation}
\label{scomm}
\left[\mathfrak{S}(\sk_1;A,B),\mathfrak{S}(\sk_{2};A,B)\right]=0\quad\text{for
all}\quad \sk_1,\sk_2\in\R\setminus\{0\},
\end{equation}
where $[\cdot,\cdot]$ denotes the commutator.

\begin{remark}\label{k-independent}
The matrix $\mathfrak{S}(\sk;A,B)$ is $\sk$-independent if and only if
$\mathfrak{S}(\sk;A,B)$ is self-adjoint for at least one $\sk>0$ and, thus,
for all $\sk>0$. Indeed, assume that $\mathfrak{S}(\sk;A,B)$ is
$\sk$-independent. Then for any eigenvector $\chi\in\cK$ with eigenvalue
$\lambda$
\begin{equation*}
(\lambda+1)A\chi + \ii\sk (\lambda-1) B\chi =0
\end{equation*}
holds for all $\sk>0$. By Lemma \ref{kerne} this implies $\lambda\in\{-1,1\}$. Thus,
$\mathfrak{S}(\sk;A,B)$ is self-adjoint.

Conversely, assume that $\mathfrak{S}(\sk;A,B)$ is self-adjoint for some
$\sk>0$. By \eqref{sinv} it is self-adjoint for all $\sk>0$. Hence, any of
its eigenvalues is either $+1$ or $-1$. By \eqref{scomm} the eigenspaces do
not depend on $\sk$. The same argument as above shows that the eigenvalues
of $\mathfrak{S}(\sk;A,B)$ are $\sk$-independent. Thus,
$\mathfrak{S}(\sk;A,B)$ does not depend on $\sk>0$.

To sum up, the matrix $\mathfrak{S}(\sk;A,B)$ is $\sk$-independent if and
only if it is of the form $\mathfrak{S}(\sk;A,B)=\1-2P$, where $P$ denotes
an orthogonal projection.

Alternatively, the matrix $\mathfrak{S}(\sk;A,B)$ is $\sk$-independent if
and only if $A B^\dagger =0$. To prove this fact we note that
$\mathfrak{S}(\sk;A,B)$ is self-adjoint if and only if $A_{\mathfrak{S}}
B_{\mathfrak{S}}^\dagger =0$ with $A_{\mathfrak{S}}$ and $B_{\mathfrak{S}}$
defined in \eqref{ab:sigma}. Since the boundary conditions $(A,B)$ and
$(A_{\mathfrak{S}}, B_{\mathfrak{S}})$ are equivalent, from
\eqref{ab:sigma:kreuz} we obtain the claim.
\end{remark}

\subsection{Scattering Theory}\label{subsec:3.1}

On the exterior $\cG_{\mathrm{ext}}=(\partial
V,\emptyset,\cE,\partial|_{\cE})$ of the graph $\cG=(V,\cI,\cE,\partial)$ we
consider the Laplace operator $-\Delta(A_{\cE}=0, B_{\cE}=\1)$
corresponding to Neumann boundary conditions. Let $\cP_{\cE}:\;
\cH\rightarrow \cH_{\cE}$ be the orthogonal projection in $\cH$ onto
$\cH_{\cE}$. By the Kato-Rosenblum theorem (see, e.g., \cite[Theorem 6.2.3
and Corollary 6.2.4]{Yafaev}) the two-space wave operators
\begin{equation*}
\Omega^\pm(-\Delta(A,B,\underline{a}), -\Delta(A_{\cE}=0,
B_{\cE}=\1);\cP_{\cE})=\slim_{t\rightarrow\mp\infty}
\e^{-\ii\Delta(A,B,\underline{a}) t}\; \cP_{\cE}\; \e^{\ii \Delta(A_{\cE}=0,
B_{\cE}=\1) t}
\end{equation*}
exist and are $\cP_{\cE}$-complete. Thus, the scattering operator
\begin{equation}\label{S.matrix:2spaces}
S(-\Delta(A,B,\underline{a}),-\Delta(A_{\cE}=0,B_{\cE}=\1);\cP_{\cE})=
(\Omega^-)^\dagger\Omega^+:\cH_{\cE}\rightarrow\cH_{\cE}
\end{equation}
is unitary and its layers $S(\sk;A,B,\underline{a}):\,
\cK_{\cE}\rightarrow\cK_{\cE}$ (in the direct integral representation with
respect to $-\Delta(A_{\cE}=0,B_{\cE}=\1)$) are also unitary for almost all
$\sk>0$.

The resulting scattering matrix $S(\sk):=S(\sk;A,B,\underline{a})$
associated to $\Delta(A,B,\underline{a})$ has the following interpretation
in terms of the solutions to the Schr\"{o}dinger equation (see \cite{KS1} and
\cite{KS4}). For given $k\in\cE$ consider the solutions $\psi^{k}(\sk)$ of
the stationary Schr\"{o}dinger equation at energy $\sk^2>0$,
\begin{equation*}
-\Delta(A,B,\underline{a})\psi^{k}(\sk)=\sk^2\psi^{k}(\sk).
\end{equation*}
This solution is of the form
\begin{equation}
\label{10} \psi^{k}_{j}(x;\sk)=\begin{cases} [S(\sk)]_{jk} \e^{\ii\sk x}
&\text{for}
                                          \;j\in\cE, j\neq k,\\
  \e^{-\ii\sk x} + [S(\sk)]_{kk} \e^{\ii\sk x} & \text{for}\;j\in\cE, j=k,\\
                                  [\alpha(\sk)]_{jk} \e^{\ii\sk x}+
        [\beta(\sk)]_{jk} \e^{-\ii\sk x} & \text{for}\; j\in\cI. \end{cases}
\end{equation}
Thus, the number $[S(\sk)]_{jk}$ for $j\neq k$ is the transmission amplitude
from channel $k\in\cE$ to channel $j\in\cE$ and $[S(\sk)]_{kk}$ is the
reflection amplitude in channel $k\in\cE$. Their absolute squares may be
interpreted as transmission and reflection probabilities, respectively. The
``interior'' amplitudes $[\alpha(\sk)]_{jk}$ and $[\beta(\sk)]_{jk}$ are
also of interest, since they describe how an incoming wave moves through a
graph before it is scattered into an outgoing channel.

The condition for the $\psi^{k}(\sk)\;(k\in \cE)$ to satisfy the boundary
conditions \eqref{lin2} immediately leads to the following solution for the
scattering matrix $S(\sk):\; \cK_{\cE}\rightarrow\cK_{\cE}$ and the linear
maps $\alpha(\sk)$ and $\beta(\sk)$ from $\cK_{\cE}$ to
$\cK_{\cI}:=\cK_{\cI}^{(-)}\oplus\cK_{\cI}^{(+)}$. Indeed, by combining
these operators into a map from $\cK_{\cE}$ to
$\cK=\cK_{\cE}\oplus\cK_{\cI}\oplus\cK_{\cI}$ we obtain the linear equation
\begin{equation}\label{11}
Z(\sk;A,B,\underline{a})\begin{pmatrix} S(\sk)\\
                      \alpha(\sk)\\
                    \beta(\sk)\end{pmatrix}
                      =-(A-\ii\sk B)
              \begin{pmatrix} \1\\
                               0\\
                               0 \end{pmatrix}
\end{equation}
with
\begin{equation}\label{Z:def}
Z(\sk;A,B,\underline{a})= A X(\sk;\underline{a})+\ii\sk B
Y(\sk;\underline{a}),
\end{equation}
where
\begin{equation}
\label{zet}
X(\sk;\underline{a})=\begin{pmatrix}\1&0&0\\
                                  0&\1&\1\\
               0&\e^{\ii\sk\underline{a}}&\e^{-\ii\sk\underline{a}}
               \end{pmatrix}\qquad\text{and}\qquad
Y(\sk;\underline{a})=\begin{pmatrix}\1&0&0\\
                                  0&\1&-\1\\
               0&-\e^{\ii\sk\underline{a}}&\e^{-\ii\sk\underline{a}}
               \end{pmatrix}.
\end{equation}
The diagonal $|\cI|\times |\cI|$ matrices $\e^{\pm \ii\sk\underline{a}}$
are given by
\begin{equation}
\label{diag} [\e^{\pm \ii\sk\underline{a}}]_{jk}=\delta_{jk}\e^{\pm \ii\sk
a_{j}}\quad
                       \text{for}\quad j,k\in\;\cI.
\end{equation}

{}From \eqref{11} it follows that the scattering matrix is a meromorphic
function in the complex $\sk$-plane. In particular, $S(\sk)$ is real
analytic in $\sk\in\R$ except for a discrete set of values with no finite
accumulation points. Indeed, the determinant $\det Z(\sk)$ of
$Z(\sk):=Z(\sk;A,B,\underline{a})$ is an entire function in $\sk$ and,
therefore, $Z(\sk)^{-1}$ is meromorphic. As proven in \cite[Theorem
3.1]{KS1} the set of real zeros of $\det Z(\sk)$ is discrete. This implies,
in particular, that $S(\sk)$ is known if it is known for a countable set of
values $\sk\in\R$ with at least one finite accumulation point. So Corollary
\ref{cor:1:1} is a direct consequence of Theorem \ref{1:theo:2}.

Observe that
\begin{equation}\label{verloren:3:neu}
\begin{split}
A X(\sk;\underline{a})&+\ii\sk B Y(\sk;\underline{a}) = (A+\ii\sk
B)R_+(\sk;\underline{a}) + (A-\ii\sk B)R_-(\sk;\underline{a})\\
&=(A+\ii\sk B)\big[\1+(A+\ii\sk B)^{-1}(A-\ii\sk
B)T(\sk;\underline{a})\big]R_+(\sk;\underline{a})\\
&=(A+\ii\sk B)\big[\1-\mathfrak{S}(\sk;A,B)
T(\sk;\underline{a})\big]R_+(\sk;\underline{a}),
\end{split}
\end{equation}
where
\begin{equation}\label{U:def:neu}
R_+(\sk;\underline{a}):=\frac{1}{2}[X(\sk;\underline{a})+Y(\sk;\underline{a})]
=
\begin{pmatrix} \1 & 0 & 0 \\ 0 & \1 & 0 \\
0 & 0 & \e^{-\ii\sk\underline{a}}
\end{pmatrix},
\end{equation}
\begin{equation*}
R_-(\sk;\underline{a}) :=
\frac{1}{2}[X(\sk;\underline{a})-Y(\sk;\underline{a})] =
\begin{pmatrix} 0 & 0 & 0 \\ 0 & 0 & \1 \\ 0 & \e^{\ii\sk\underline{a}} & 0
\end{pmatrix},
\end{equation*}
and
\begin{equation}\label{T:def:neu}
T(\sk;\underline{a})=\begin{pmatrix} 0 & 0 & 0 \\ 0 & 0 &
\e^{\ii\sk\underline{a}} \\ 0 & \e^{\ii\sk\underline{a}} & 0
\end{pmatrix}
\end{equation}
with respect to the orthogonal decomposition \eqref{K:def}. Thus, from
\eqref{11} we get
\begin{equation}\label{12}
 K(\sk;A,B,\underline{a}) \begin{pmatrix} S(\sk)\\
                      \alpha(\sk)\\
                    \e^{-\ii\sk\underline{a}}\beta(\sk)\end{pmatrix}
                      =\mathfrak{S}(\sk;A,B)
              \begin{pmatrix} \1\\
                               0\\
                               0 \end{pmatrix},
\end{equation}
where
\begin{equation}\label{K:def:neu}
K(\sk;A,B,\underline{a}) := \1-\mathfrak{S}(\sk;A,B)T(\sk;\underline{a})
\end{equation}
and $\begin{pmatrix} \1\\
                               0\\
                               0 \end{pmatrix}:\; \cK_{\cE}\rightarrow \cK$
                               such that $\begin{pmatrix} \1\\
                               0\\
                               0 \end{pmatrix}\chi = \chi\oplus 0\oplus 0$
                               with respect to orthogonal decomposition
                               \eqref{K:def}.

\begin{theorem}[= Theorem 3.2 in \cite{KS1}]\label{3.2inKS1}
Let the boundary conditions $(A,B)$ satisfy \eqref{abcond}. For any
$\sk\in\R\setminus\{0\}$
\begin{equation}\label{inklusion}
\Ran\; \mathfrak{S}(\sk;A,B) \begin{pmatrix} \1 \\ 0 \\ 0
\end{pmatrix}\subset \Ran K(\sk;A,B,\underline{a}).
\end{equation}
Thus, equation \eqref{12} has a solution even if $\det
K(\sk;A,B,\underline{a})=0$ for some $\sk\in\R\setminus\{0\}$. This
solution defines the scattering matrix uniquely. Moreover,
\begin{equation}\label{S-matrix}
S(\sk)=-\begin{pmatrix} \1 & 0 & 0 \end{pmatrix}
\left(K(\sk;A,B,\underline{a})|_{\Ran
K(\sk;A,B,\underline{a})^\dagger}\right)^{-1} P_{\Ran
K(\sk;A,B,\underline{a})} \mathfrak{S}(\sk;A,B) \begin{pmatrix} \1 \\
0 \\ 0
\end{pmatrix}
\end{equation}
is unitary for all $\sk\in\R\setminus\{0\}$.
\end{theorem}

Here $P_{\Ran K(\sk;A,B,\underline{a})}$ denotes the orthogonal projection
in $\cK$ onto $\Ran K(\sk;A,B,\underline{a})$, the range of
$K(\sk;A,B,\underline{a})$ as a map on $\cK$.

We present an elementary proof of Theorem \ref{3.2inKS1}.

\begin{proof}
For any $\chi\in\Ker K(\sk;A,B,\underline{a})$ we have
\begin{equation*}
T(\sk;\underline{a}) \chi = \mathfrak{S}(\sk;A,B)^\dagger \chi.
\end{equation*}
Multiplying this equality by $\begin{pmatrix} \1 & 0 & 0 \\
0 & 0 & 0 \\ 0 & 0 & 0
\end{pmatrix}$ from the left we obtain $\begin{pmatrix} \1 & 0 & 0 \\
0 & 0 & 0 \\ 0 & 0 & 0
\end{pmatrix} \mathfrak{S}(\sk;A,B)^\dagger \chi = 0$.
Thus,
\begin{equation*}
\Ker K(\sk;A,B,\underline{a}) \subset \Ker\; \begin{pmatrix} \1 & 0 & 0 \\
0 & 0 & 0 \\ 0 & 0 & 0
\end{pmatrix} \mathfrak{S}(\sk;A,B)^\dagger,
\end{equation*}
which implies the inclusion
\begin{equation}\label{spaet:ref:1}
\Ran K(\sk;A,B,\underline{a})^\dagger \supset \Ran\; \mathfrak{S}(\sk;A,B)
\begin{pmatrix} \1 & 0 & 0 \\
0 & 0 & 0 \\ 0 & 0 & 0
\end{pmatrix}.
\end{equation}

Observe that for any $\sk>0$ the operator
$\mathfrak{S}(\sk;A,B)T(\sk;\underline{a})$ is a partial isometry with
initial subspace $\cK_{\cI} := \cK_{\cI}^{(-)}\oplus \cK_{\cI}^{(+)}$ and
final subspace $\mathfrak{S}(\sk;A,B)\cK_{\cI}$. In particular, it is a
contraction. Thus (see, e.g.\ \cite[Theorem I.3.2]{Foias} or \cite[Theorem
1.6.6]{Horn}), $\Ran K(\sk;A,B,\underline{a})$ and $\Ran
K(\sk;A,B,\underline{a})^\dagger$ are equal. Hence, noting that
\begin{equation*}
\Ran\;
\mathfrak{S}(\sk;A,B) \begin{pmatrix} \1 & 0 & 0 \\
0 & 0 & 0 \\ 0 & 0 & 0
\end{pmatrix} = \Ran\; \mathfrak{S}(\sk;A,B) \begin{pmatrix} \1 \\ 0 \\ 0
\end{pmatrix},
\end{equation*}
inclusion \eqref{spaet:ref:1} proves \eqref{inklusion}. Combined with
\eqref{12} this also establishes \eqref{S-matrix}.

To prove the unitarity we note that equation \eqref{12} implies that
\begin{equation*}
\begin{pmatrix}S(\sk) \\ \alpha(\sk) \\ \e^{-\ii \sk\underline{a}}\beta(\sk) \end{pmatrix} =
\mathfrak{S}(\sk;A,B,\underline{a}) \begin{pmatrix} \1 \\ \beta(\sk) \\
\e^{\ii \sk\underline{a}} \alpha(\sk)
\end{pmatrix}.
\end{equation*}
Since $\mathfrak{S}(\sk;A,B,\underline{a})$ is unitary we obtain
$S(\sk)^\dagger S(\sk)=\1$.
\end{proof}

We turn to the study of the analytic properties of the scattering matrix as
a function of $\sk\in\C$.

\begin{proposition}\label{proposition:2.3}
Let the boundary conditions $(A,B)$ satisfy \eqref{abcond}. If $\det(A +
\ii\sk B)=0$ for some $\sk\in\C$, then $\sk=\ii\kappa$ with $\kappa\in\R$.
For any sufficiently large $\rho>0$ there is a constant $C_\rho>0$ such that
\begin{equation}\label{absch}
\|(A+\ii\sk B)^{-1}\| \leq C_\rho
\end{equation}
for all $\sk\in\C$ with $|\sk|>\rho$.
\end{proposition}

\begin{proof}
Assume that $\det(A + \ii\sk B)=0$ for some $\sk\in\C$ with $\Re \sk\neq 0$.
Then also
\begin{equation*}
\det(A^\dagger - \ii\overline{\sk} B^\dagger)= \overline{\det(A + \ii\sk
B)}=0.
\end{equation*}
Therefore, there is a $\chi\neq 0$ such that
\begin{equation}\label{neu:1}
(A^\dagger - \ii\overline{\sk} B^\dagger)\chi=0.
\end{equation}
In particular, we have $(B A^\dagger - \ii\overline{\sk} B
B^\dagger)\chi=0$. Therefore, since $B A^\dagger$ is self-adjoint, we get
\begin{equation*}
\langle\chi, BA^\dagger \chi\rangle = \langle\chi, BB^\dagger \chi\rangle\,
\Im\sk \qquad\text{and}\qquad   \langle\chi, BB^\dagger \chi\rangle\,
\Re\sk =0.
\end{equation*}
The second equality implies that $\chi\in\Ker B^\dagger$. Then, by
\eqref{neu:1}, $\chi\in\Ker A^\dagger$. By Lemma \ref{kerne} we have $\Ker
A^\dagger\cap \Ker B^\dagger=\{0\}$. Thus, $\chi=0$ which contradicts the
assumption and, hence, $\Re \sk=0$.

Since $\det(A+\ii\sk B)$ is a polynomial in $\sk$, it has a finite number of
zeroes. Take an arbitrary $\rho>0$ such that all its zeroes lie in the disk
$|\sk|<\rho$. Using the matrix inverse formula we represent any element of
$(A+\ii\sk B)^{-1}$ as a quotient of two polynomials. Obviously, the degree
of the nominator does not exceed the degree of the denominator. In turn,
this implies the estimate \eqref{absch}.
\end{proof}

\begin{theorem}\label{thm:cont}
Let the boundary conditions $(A,B)$ satisfy \eqref{abcond}. The scattering
matrix $S(\sk)=S(\sk;A,B,\underline{a})$ is a meromorphic function in the
complex $\sk$-plane. In the upper half-plane $\Im\sk>0$ it has at most a
finite number of poles which are located on the imaginary semiaxis
$\Re\sk=0$.
\end{theorem}

\begin{proof}
Assume that $\det K(\sk;A,B,\underline{a})=0$ for some $\sk\in\C$ with
$\Im\sk>0$ and $\Re\sk\neq 0$. This implies that the homogeneous equation
\begin{equation*}
K(\sk;A,B,\underline{a})\begin{pmatrix} s \\ \alpha \\ \beta \end{pmatrix}
= 0
\end{equation*}
has a nontrivial solution with $s\in\cK_{\cE}$, $\alpha\in\cK_{\cI}^{(-)}$,
and $\beta\in\cK_{\cI}^{(+)}$. Consider the function
$\psi(x)=\{\psi_j(x)\}_{j\in\cI\cup\cE}$ defined by
\begin{equation*}
\psi_{j}(x)=\begin{cases} s_{j} \e^{\ii\sk x} &\text{for}
                                          \;j\in\cE, \\
                                  \alpha_{j} \e^{\ii\sk x}+
        \beta_{j} \e^{\ii\sk a_j} \e^{-\ii\sk x} & \text{for}\; j\in\cI. \end{cases}
\end{equation*}
Obviously, $\psi(x)$ satisfies the boundary conditions
\eqref{Randbedingungen}. Moreover, $\psi\in L^2(\cG)$ since $\Im\sk>0$.
Hence, $\sk^2\in\C$ with $\Im\sk^2\neq 0$ is an eigenvalue of the operator
$\Delta(A,B,\underline{a})$ which contradicts the self-adjointness of
$\Delta(A,B,\underline{a})$.

Since $\det K(\sk;A,B,\underline{a})$ is an entire function in $\sk$ which
does not vanish identically, from \eqref{11} it follows that the scattering
matrix $S(\sk)$ is a meromorphic function in the complex $\sk$-plane. To
prove that the scattering matrix $S(\sk)$ has at most a finite number of
poles on the imaginary semiaxis $\{\sk\in\C|\;\Re\sk=0,\; \Im\sk>0\}$ it
suffices to show that the determinant $\det K(\sk;A,B,\underline{a})$ does
not vanish for all sufficiently large $\Im\sk>0$. From Proposition
\ref{proposition:2.3} it follows that there is $\rho>0$ and a constant
$C>0$ such that $\|\mathfrak{S}(\ii\kappa;A,B)\| \leq C$ for all
$\kappa>\rho$. Observing that $\|T(\ii\kappa;\underline{a})\|\leq
\e^{-a\kappa}$ with $\displaystyle a:=\min_{j\in\cI} a_j$ for all
$\kappa>0$, we obtain that $\|\mathfrak{S}(\ii\kappa;A,B)
T(\ii\kappa;\underline{a})\|<1$ for all $\kappa> \max\{\rho,a^{-1}\log C\}$.
\end{proof}

\begin{remark}\label{rem:3.10}
The positive eigenvalues of the operator $-\Delta(A,B,\underline{a})$ on a
compact metric graph $(\cG,\underline{a})$ are determined by the equation
\begin{equation}\label{eq:det}
\det[\1 - \mathfrak{S}(\sk; A,B) T(\sk;\underline{a})] = 0
\end{equation}
with $T(\sk;\underline{a})$ defined in \eqref{T:def:neu}.
This fact has first been observed by Carlson in \cite[Section 3.2]{Carlson}.

Observing that $\mathfrak{S}(\sk; A,B) T(\sk;\underline{a})$ is strictly
contractive for all $\sk\in\C$ with $\Im\sk>0$ and sufficiently large
$\Re\sk>0$ one obtains an absolutely convergent expansion
\begin{equation*}
\log\det[\1 - \mathfrak{S}(\sk; A,B) T(\sk;\underline{a})] = 1 +
\sum_{n=1}^\infty\frac{1}{n} \tr [\mathfrak{S}(\sk; A,B)
T(\sk;\underline{a})]^n.
\end{equation*}
Starting from this observation Kottos and Smilansky presented in
\cite{Kottos} a derivation of a representation for the eigenvalues counting
function in terms of closed paths on the graph. This representation
strongly resembles the celebrated Selberg trace formula for compact Riemann
manifolds \cite{Selberg}, \cite{McKean} (a generalization of the Poisson
summation formula). The Selberg trace formula establishes a relation
between the length spectrum of prime geodesics and the spectrum of the
Laplace operator on the manifold.

A mathematically rigorous version of the Kottos-Smilansky representation
for the Laplace transform of the eigenvalues counting function has been
proven earlier by Roth \cite{Roth:1}, \cite{Roth:2} for the case of
standard boundary conditions (see Example \ref{3:ex:3} below). Some related
results can also be found in \cite{Gaveau:1}, \cite{Gaveau:2},
\cite{Naimark:Solomyak}, \cite{Nicaise}, \cite{Nicaise:2},
\cite{Solomyak:1}. The Weyl asymptotics for fourth-order differential
operators on compact graphs has been proven in \cite{Dekoninck:Nicaise}.

The representation from \cite{Kottos} was used by Gutkin and Smilansky in
\cite{GS} to determine the lengths of internal edges of a compact metric
graph from the spectrum of the associated Laplace operator. A rigorous
proof of this result has been recently given by Kurasov and Nowaczyk in
\cite{KuNo} for standard boundary conditions.
\end{remark}

\subsection{Symmetry Properties of the Scattering Matrices}

We say that the boundary conditions $(A,B)$ satisfying \eqref{abcond} are
\emph{real} if the self-adjoint operator $\Delta(A,B)$ is real, that is,
its domain $\Dom(\Delta(A,B,\underline{a}))$ is invariant with respect to
the complex conjugation. This implies that
$\overline{\Delta(A,B,\underline{a})\psi} =
\Delta(A,B,\underline{a})\overline{\psi}$ for all
$\psi\in\Dom(\Delta(A,B,\underline{a}))$.

By \eqref{thru} the operator $\Delta(A,B,\underline{a})$ is real if and only
if the maximal isotropic subspace $\cM(A,B)$ defined by \eqref{M:def} is
invariant with respect to complex conjugation. This immediately implies that
$\Delta(A,B,\underline{a})$ is real if and only if the boundary conditions
$(A,B)$ and $(\overline{A},\overline{B})$ are equivalent.

\begin{lemma}\label{lemma:3:14:real}
Boundary conditions $(A,B)$ satisfying \eqref{abcond} are real if and only
if there are real matrices $(A^\prime, B^\prime)$ and an invertible matrix
$C$ such that $A^\prime=CA$ and $B^\prime=CB$.
\end{lemma}

\begin{proof}
The ``if'' part is obvious. To prove the ``only if'' part we assume that
boundary conditions $(A,B)$ satisfying \eqref{abcond} are real such that
$\overline{A}=CA$ and $\overline{B}=CB$ for some invertible $C$. Therefore,
\begin{equation}\label{bar}
\overline{A}+\kappa\overline{B} = C(A+\kappa B)
\end{equation}
for all $\kappa\in\C$. Choose $\kappa>0$ sufficiently large such that
$A+\kappa B$ is invertible, which is possible by Proposition
\ref{proposition:2.3}. Equation \eqref{bar} implies that
\begin{equation*}
C=(\overline{A}+\kappa\overline{B})(A+\kappa B)^{-1}.
\end{equation*}
Now set $A^{\prime}=(A+\kappa B)^{-1}A$ and $B^{\prime}=(A+\kappa
B)^{-1}B$. The matrices $A^{\prime}$ and $B^{\prime}$ are real since
$\overline{A^{\prime}}=(\overline{A}+\kappa\overline{B})^{-1}\overline{A}=
(A+\kappa B)^{-1}C^{-1}\overline{A}=(A+\kappa B)^{-1}A=A^{\prime}$ and
similarly for $B^{\prime}$.
\end{proof}

\begin{lemma}\label{lem:trans}
Boundary conditions $(A,B)$ satisfying \eqref{abcond} are real if and only
if the unitary matrix $\mathfrak{S}(\sk;A,B)$ defined by \eqref{uuu:def} is
symmetric for at least one $\sk>0$ and, thus, for all $\sk>0$, that is
\begin{equation}\label{TTT}
\mathfrak{S}(\sk;A,B)^T = \mathfrak{S}(\sk;A,B)
\end{equation}
with ``$T$'' denoting the transpose.
\end{lemma}

\begin{proof}
By Corollary 2.1 in \cite{KS1} one has the relation
\begin{equation}\label{real}
\mathfrak{S}(\sk;A,B)^T = \mathfrak{S}(\sk;\overline{A},\overline{B}).
\end{equation}
If $\mathfrak{S}(\sk;A,B)$ is symmetric, then $\mathfrak{S}(\sk;A,B) =
\mathfrak{S}(\sk;\overline{A},\overline{B})$.  By Proposition \ref{unit:neu}
the boundary conditions $(A,B)$ and $(\overline{A},\overline{B})$ are
equivalent, which implies that they are real.

Conversely, if the boundary conditions $(A,B)$ are real, then by Lemma
\ref{lemma:3:14:real} both matrices $A$ and $B$ may be chosen real. Hence,
$\mathfrak{S}(\sk;A,B) = \mathfrak{S}(\sk;\overline{A},\overline{B})$,
which by \eqref{real} implies \eqref{TTT}.
\end{proof}

Recall that the scattering matrix $S(\sk;A,B,\underline{a})$ possesses a
property similar to \eqref{real} (see Corollary 3.2 in \cite{KS1} and
Theorem 2.2 in \cite{KS3}):
\begin{equation*}
S(\sk;A,B,\underline{a})^T = S(\sk;\overline{A},\overline{B},\underline{a}).
\end{equation*}

Observe that for all $\sk>0$
\begin{equation}\label{hanal}
\mathfrak{S}(\sk; A,B)^\dagger = \mathfrak{S}(\sk; A,B)^{-1} =
\mathfrak{S}(-\sk; A,B).
\end{equation}

\begin{corollary}\label{cor:hanal} Let the boundary conditions $(A,B)$ satisfy \eqref{abcond}.
For all complex $\sk$ the relation
\begin{equation}\label{hanal1}
S(-\sk;A,B,\underline{a})=S(\sk;A,B,\underline{a})^{-1}
\end{equation}
is valid.
\end{corollary}

\begin{proof}
By Theorem \ref{thm:cont} it suffices to prove \eqref{hanal1} for almost all
$\sk>0$. Assume that $\det[\1-\mathfrak{S}(\sk;A,B)T(\sk;\underline{a})]\neq
0$. Taking the adjoint of \eqref{S-matrix} and using relations
\eqref{hanal} and $T(\sk;\underline{a})^{\dagger}=T(-\sk;\underline{a})$,
we obtain
\begin{equation*}
S(\sk;A,B,\underline{a})^{\dagger} =
\begin{pmatrix} \1 & 0 & 0
\end{pmatrix}\mathfrak{S}(-\sk;
A,B)(\1-T(-\sk;\underline{a})\mathfrak{S}(-\sk; A,B))^{-1}\begin{pmatrix}
\1 \\ 0 \\ 0
\end{pmatrix}.
\end{equation*}
Note that $K_1(\1+K_2 K_1)^{-1} = (\1+K_1 K_2)^{-1} K_1$ if at least one of
the inverses $(\1+K_2 K_1)^{-1}$ and $(\1+K_1 K_2)^{-1}$ exists. Thus, we
obtain
\begin{equation*}
\begin{aligned} S(\sk;A,B,\underline{a})^{\dagger} &=
\begin{pmatrix} \1 & 0 & 0
\end{pmatrix}(\1-\mathfrak{S}(-\sk; A,B)T(-\sk;\underline{a}))^{-1}\mathfrak{S}(-\sk;
A,B)\begin{pmatrix} \1 \\ 0 \\ 0
\end{pmatrix}\\
&= S(-\sk;A,B,\underline{a}).
\end{aligned}
\end{equation*}
\end{proof}

In the context of analytic S-matrix theory (see, e.g., \cite{ELOP},
\cite{Olive}) relation \eqref{hanal1} is called Hermitian analyticity.

\subsection{Graphs With No Internal Edges.}\label{subsec:3.2}

The case with no internal edges $(\cI=\emptyset)$ deserves special
attention. The relation \eqref{11} for the scattering matrix simplifies to
\begin{equation}
\label{svertex} S(\sk;A,B) = \mathfrak{S}(\sk;A,B),
\end{equation}
where $\mathfrak{S}(\sk;A,B)$ is defined in \eqref{uuu:def}. As special
cases we obtain $\sk$-independent scattering matrices for Dirichlet and
Neumann boundary conditions
\begin{equation}
\label{dirne}
\begin{aligned}
S(\sk;A=\1,B=0)&=-&\1\quad &\mbox{(Dirichlet boundary conditions),}\\
S(\sk;A=0,B=\1)&=&\1\quad &\mbox{(Neumann boundary conditions)}.
\end{aligned}
\end{equation}
Since $\cM(A=\1,B=0)^\perp = \cM(A=0,B=\1)$, equations \eqref{dirne}
provide an example to \eqref{sperp}.

By Proposition \ref{unit:neu}, with condition \eqref{abcond} being
satisfied, both matrices $A\pm \ii\sk B$ are invertible for all real
$\sk\neq 0$. Now by the representation \eqref{svertex} the scattering
matrix $S(\sk; A, B)$ is a rational function in $\sk$. By Proposition
\ref{proposition:2.3}, all possible poles of the scattering matrix are
located on the imaginary axis.

We collect some properties of the single-vertex scattering matrix
\eqref{svertex} associated to $(A,B)$. By Proposition \ref{unit:neu} we
conclude that the scattering matrix $S(\sk;A,B)$ is uniquely determined by
the maximal isotropic space $\cM(A,B)$, i.e., the equality
\begin{equation*}
S(\sk; A,B)=S(\sk; A^\prime,B^\prime)
\end{equation*}
holds for all $\sk\neq 0$ if and only if $\cM(A,B) = \cM(A',B')$.
Conversely, again by Proposition \ref{unit:neu}, the scattering matrix
$S(\sk)$ for arbitrarily fixed $\sk>0$ uniquely fixes the boundary
conditions,
\begin{equation}
\label{bdinv} A(\sk)=-\frac{1}{2}(S(\sk)-\1),\qquad
B(\sk)=\frac{1}{2\ii\sk}(S(\sk)+\1).
\end{equation}

Relation \eqref{svertex} and Proposition \ref{proposition:2.3} imply the
following result.

\begin{lemma}\label{3:lem:1}
For any boundary conditions $(A,B)$ satisfying \eqref{abcond} the
scattering matrix $S(\sk)$ $:= S(\sk; A,B)$ possesses the following
properties for all $\sk\in\R\setminus\{0\}$:
\begin{equation}
\label{ker1} \Ker B=\Ker\left(S(\sk)+\1\right),\qquad \Ker
A=\Ker\left(S(\sk)-\1\right).
\end{equation}
Therefore, $-1$ is an eigenvalue of $S(\sk)$ for all $\sk \in\R$ if and only
if $\det B=0$ and $1$ is an eigenvalue of $S(\sk)$ for all $\sk\in\R$ if and
only if $\det A=0$.
\end{lemma}

Let $\cM_{\mathrm{D}}=\cM(A=\1, B=0)$ and $\cM_{\mathrm{N}}=\cM(A=0,B=\1)$
denote the maximal isotropic subspaces corresponding to Dirichlet and
Neumann boundary conditions, respectively. As an immediate consequence of
definition \eqref{M:def} of $\cM(A,B)$ we obtain that
\begin{equation*}
\cM \cap \cM_{\mathrm{D}} = \{0\} \oplus \Ker B\qquad\text{and}\qquad \cM
\cap \cM_{\mathrm{N}} =  \Ker A \oplus \{0\}
\end{equation*}
with respect to the orthogonal decomposition ${}^d\cK=\cK\oplus\cK$. Thus,
if $(A,B)$ satisfies \eqref{abcond}, then
\begin{equation*}
\dim \Ker A= \dim\left(\cM(A,B)\cap\cM_{\rm N}\right)\quad\text{and}\quad
\dim \Ker B= \dim\left(\cM(A,B)\cap\cM_{\rm D}\right).
\end{equation*}

Let $P_{\Ker A}$ and $P_{\Ker B}$ be the orthogonal projections in $\cK$
onto the kernel of $A$ and of $B$, respectively. Note that since $\Ker A =
\Ker CA$ and $\Ker B = \Ker CB$ for any invertible $C:\,\cK\rightarrow\cK$,
the projections $P_{\Ker A}$ and $P_{\Ker B}$ depend on $\cM(A,B)$ only.

\begin{corollary}\label{3:cor:2}
Any single vertex scattering matrix satisfies the limit relations
\begin{equation}\label{3:cor:1:eq}
\begin{split}
\lim_{\sk\;\rightarrow\;
  +\infty}S(\sk;A,B)& = \1-2P_{\Ker B},\\
\lim_{\sk\;\rightarrow\;
  0}S(\sk;A,B)& = -\1+2P_{\Ker A}.
\end{split}
\end{equation}
Moreover, the scattering matrix admits an expansion
\begin{equation}\label{expansion}
S(\sk; A,B) = \sum_{m=0}^\infty S^{(m)}(A,B) \sk^{-m},
\end{equation}
absolutely converging for all sufficiently large $\sk\in\C$.
\end{corollary}

\begin{proof}
Fix an arbitrary $\sk_0>0$. For arbitrary $\sk>\sk_0$ relations
\eqref{sinv} and \eqref{svertex} imply that
\begin{equation*}
S(\sk;A,B) = \frac{\sk+\sk_0}{\sk-\sk_0}\1 - \frac{4\sk\sk_0}{(\sk-\sk_0)^2}
\left[S(\sk_0;A,B) +
 \frac{\sk+\sk_0}{\sk-\sk_0}\1\right]^{-1}.
\end{equation*}
Observe that for large $\sk$
\begin{equation*}
\left[S(\sk_0;A,B) +
 \frac{\sk+\sk_0}{\sk-\sk_0}\1\right]^{-1} = \frac{\sk-\sk_0}{2\sk_0} P_{\Ker(S(\sk_0;A,B) +
 \1)}+ O(1),
\end{equation*}
where $P_{\Ker(S(\sk_0;A,B) + \1)}$ is the orthogonal projection in $\cK$
onto $\Ker(S(\sk_0;A,B) + \1)$. By Lemma \ref{3:lem:1} this leads to the
first relation in \eqref{3:cor:1:eq}. The second relation in
\eqref{3:cor:1:eq} follows from the first one using \eqref{sperp} and
\eqref{perp}. The expansion \eqref{expansion} can be proved by applying
standard perturbation theory \cite{Kato}.
\end{proof}

\begin{proposition}\label{3:prop:4:neu}
For any boundary conditions $(A,B)$ satisfying \eqref{abcond} with $\Ker
B=\{0\}$ the resulting scattering matrix $S(\sk,A,B)$ has a representation
\begin{equation}\label{srep:neu}
S(\sk;A,B)=\frac{\sk + \ii H/2}{\sk - \ii H/2},
\end{equation}
where $H$ is an $|\cE|\times |\cE|$ Hermitian matrix.
\end{proposition}

\begin{proof}
Note that $B^{-1} A$ is self-adjoint since $A B^{\dagger}$ is:
\begin{equation*}
B^{-1} A - A^\dagger {B^\dagger}^{-1}= B^{-1}(A- B A^\dagger
{B^\dagger}^{-1})=B^{-1}(A- A B^\dagger {B^\dagger}^{-1})=0.
\end{equation*}
{}From \eqref{svertex} it follows that
\begin{equation}\label{help}
\begin{split}
S(\sk;A,B) &= - (A+\ii\sk B)^{-1} B B^{-1} (A-\ii\sk B)\\
&=-(B^{-1}A + \ii \sk)^{-1} (B^{-1}A - \ii \sk),
\end{split}
\end{equation}
which implies \eqref{srep:neu} with $H = 2 B^{-1} A$.
\end{proof}

\begin{lemma}\label{lem:3.13}
Given an arbitrary Hermitian $|\cE|\times|\cE|$ matrix $H$ there are
boundary conditions $(A,B)$ satisfying \eqref{abcond} and $\Ker B = \{0\}$
such that
\begin{equation*}
\lim_{\sk\rightarrow+\infty} \sk(S(\sk;A,B)-\1)=\ii H.
\end{equation*}
These boundary conditions are unique up to the equivalence in the sense of
Definition \ref{def:equiv}.
\end{lemma}

\begin{proof}
Fix $\sk_0>0$ arbitrarily. Obviously, the matrix
\begin{equation*}
R=\frac{\sk_0 + \ii H/2}{\sk_0 - \ii H/2}
\end{equation*}
is unitary. Now set
\begin{equation*}
A=-\frac{1}{2}(R-\1)\qquad\text{and}\qquad B=\frac{1}{2\ii \sk_0}(R+\1).
\end{equation*}
It is easy to check that $(A,B)$ satisfies \eqref{abcond} and $\Ker B =
\{0\}$. Furthermore, for $\sk=\sk_0$ the scattering matrix $S(\sk;A,B)$
coincides with $R$. Proposition \ref{3:prop:4:neu} implies the claim.

To prove the uniqueness assume that there are boundary conditions
$(A^\prime,B^\prime)$ satisfying \eqref{abcond} such that
\begin{equation}\label{eq:uniq:1}
\lim_{\sk\rightarrow+\infty} \sk(S(\sk;A^\prime,B^\prime)-\1)=\ii H.
\end{equation}
By Corollary \ref{3:cor:2}, $\Ker B^\prime=\{0\}$ such that, by Proposition
\ref{3:prop:4:neu},
\begin{equation*}
S(\sk;A^\prime, B^\prime) = \frac{\sk+\ii H^\prime/2}{\sk-\ii
H^\prime/2}\quad\text{with}\quad H^\prime = 2 {B^\prime}^{-1} A^\prime.
\end{equation*}
Comparing this with \eqref{eq:uniq:1} shows that $H=H^\prime$ and,
therefore, $S(\sk;A,B)=S(\sk;A^\prime,B^\prime)$ for all $\sk>0$. Applying
Proposition \ref{unit:neu} completes the proof.
\end{proof}

To proceed further we introduce some new notions. Recall that the set of
all maximal isotropic subspaces (the Lagrangian Grassmannian) is isomorphic
to the group $\mathsf{U}(|\cE|)$ of all $|\cE|\times |\cE|$ unitary
matrices. Proposition \ref{unit:neu} provides an explicit parametrization
$\cM(U)$ of the maximal isotropic subspaces by the elements of the group
$\mathsf{U}(|\cE|)$. Let $\mu$ be the normalized Haar measure on
$\mathsf{U}(|\cE|)$.

\begin{definition}\label{generic-bc}
A property $P$ holds for Haar almost all boundary conditions if
\begin{equation*}
\begin{split}
&\mu\{U\in\mathsf{U}(|\cE|)|\ P\ \text{holds for the boundary conditions}\\
&\qquad\qquad\qquad\text{defined by the maximal isotropic subspace}\
\cM(U)\}=\mu(\mathsf{U}(|\cE|))=1.
\end{split}
\end{equation*}
is valid.
\end{definition}

We recall the following well-known result:

\begin{lemma}\label{generic:ev}
For arbitrary $N\in\N$ the set of all unitary matrices $U\in\mathsf{U}(N)$
with simple spectrum is of full Haar measure. Moreover,
\begin{itemize}
\item[(i)]{for any given $\lambda\in\C$ such that $|\lambda|=1$ the set
\begin{equation}\label{set:U}
\{U\in \mathsf{U}(N)|\ \lambda\ \text{is an eigenvalue of}\ U\}
\end{equation}
is of zero Haar measure;}
\item[(ii)]{for any given $\chi\in\C^N$ both set
\begin{equation*}
\{U\in \mathsf{U}(N)|\ \chi\ \text{is an eigenvector of}\ U\}
\end{equation*}
is of zero Haar measure;}
\item[(iii)]{for any given $\chi_1,\chi_2\in\C^N$ the sets
\begin{equation}\label{U:1}
\{U\in \mathsf{U}(N)|\ \langle\chi_1, U\chi_2\rangle = 0\}
\end{equation}
and
\begin{equation}\label{U:2}
\Big\{U\in \mathsf{U}(N)|\,\Big\langle\chi_1,\,
\frac{U-\1}{U+\1}\chi_2\Big\rangle=0\Big\}
\end{equation}
are of zero Haar measure.}
\end{itemize}
\end{lemma}

The proof of Lemma \ref{generic:ev} is based on the fact that any proper
analytic subvariety of $\mathsf{U}(N)$ has Haar measure zero
\cite{Varadarajan}. The proof of the main claim can be found in
\cite{Varadarajan}. The details of the proof of claims (i) -- (iii) are
left to the reader.

\begin{proposition}\label{propo:3.21}
For Haar almost all boundary conditions $\Ker A=\{0\}$ and $\Ker B=\{0\}$.
\end{proposition}

\begin{proof}
By Lemma \ref{generic:ev}, for an arbitrary $\sk>0$ the numbers $\pm 1$ are
not eigenvalues of $S(\sk;A,B)$ for Haar almost all boundary conditions
$(A,B)$. By Lemma \ref{3:lem:1} this implies the claim.
\end{proof}

We obtain the following consequence of Proposition \ref{3:prop:4:neu} and
Lemma \ref{generic:ev}.

\begin{theorem}\label{3:prop:4}
{}For Haar almost all boundary conditions $(A,B)$ the associated scattering
matrix $S(\sk;A,B)$ has a representation in the form
\begin{equation}
\label{srep} S(\sk;A,B)=\frac{\sk+\ii H/2}{\sk - \ii H/2},
\end{equation}
where $H$ is a Hermitian matrix with non-vanishing entries and simple
eigenvalues. Furthermore, for Haar almost all boundary conditions $(A,B)$
\begin{itemize}
\item[(i)]{given an arbitrary $\chi\in\cK$ neither of the eigenvectors of
$S(\sk;A,B)$ is orthogonal to $\chi$;}
\item[(ii)]{there are no
$\chi_1,\chi_2\in\cK$, $\langle\chi_1,\chi_2\rangle=0$ and $\chi_{1,2}\neq
0$ such that the inner product $\langle\chi_1, S(\sk;A,B)\chi_2\rangle$
vanishes identically for all $\sk>0$.}
\end{itemize}
\end{theorem}

\begin{proof}
Representation \eqref{srep} follows from Propositions \ref{3:prop:4:neu} and
\ref{propo:3.21}. Observe that $\lambda\in\R$ is a degenerated eigenvalue
of $H$ if and only if $(\sk+\ii \lambda/2)(\sk-\ii \lambda/2)^{-1}$ is a
degenerated eigenvalue of $S(\sk;A,B)$ for all $\sk>0$. Thus, by Lemma
\ref{generic:ev} all eigenvalues of $H$ are simple.

To prove that all entries of the matrix $H$ are non-vanishing we note that
by Lemma \ref{3:lem:1} and Proposition \ref{propo:3.21} $\1+S(\sk; A,B)$ is
invertible for Haar almost all boundary conditions. Thus, \eqref{srep}
implies that
\begin{equation}\label{will:be:used}
H = -2\ii\sk\,\frac{S(\sk;A,B)-\1}{S(\sk;A,B)+\1}
\end{equation}
for all $\sk>0$. By (iii) of Proposition \ref{generic:ev} the claim follows.

To prove the claim (i) we note that the eigenvectors of $S(\sk;A,B)$ are
those of the Hermitian matrix $H$ and, thus, are independent of $\sk>0$.
Therefore, (i) follows from Lemma \ref{generic:ev}. The claim (ii) also
follows from Lemma \ref{generic:ev}.
\end{proof}

Let $\fM_N$ be the set of all Hermitian $N\times N$ matrices endowed with
the metric topology induced by the operator norm. Consider the probability
measure $\varkappa$ on $\fM_N$ defined for any Borel set $M\subset\fM_N$ via
\begin{equation}\label{kappa:def}
\varkappa(M)=\mu\Big(\Big\{\frac{\1+\ii H/2}{\1-\ii H/2}\Big|\, H\in M\Big\}\Big),
\end{equation}
where $\mu$ is the normalized Haar measure on $\sU(N)$.
For example, if $N=1$, then
\begin{equation*}
\varkappa(M) = \frac{1}{\pi} \int_M \frac{dh}{1+h^2},\qquad M\subset\R.
\end{equation*}
On $\fM_N$ we also consider the measure
\begin{equation}\label{nu:def}
\begin{split}
d \nu & = \prod_{k=1}^N dH_{kk} \prod_{1\leq m<k\leq N} d\,\Re H_{mk}\,\,
d\,\Im H_{mk} \\ & = \prod_{k=1}^N dH_{kk} \prod_{1\leq m<k\leq N}
d|H_{mk}|\, d(\arg H_{mk})
\end{split}
\end{equation}

Recall (see, e.g., \cite{Varadarajan}) that on any chart of the manifold
$\sU(N)$ the Haar measure is absolutely continuous with respect the measure
induced by the $N(N+1)/2$-dimensional Lebesgue measure via a coordinate
map. In particular, this implies the following

\begin{lemma}\label{lem:nu}
The measure $\varkappa$ is absolutely continuous with respect to the
measure $\nu$.
\end{lemma}

\section{Local Boundary Conditions}\label{sec:reconstruction}

The notion of local boundary conditions has been introduced in our paper
\cite{KS1} and discussed in more detail in \cite{KS5}. Local boundary
conditions couple only those boundary values of $\psi$ and of its
derivative $\psi^\prime$ which belong to the same vertex.

In this section we show that given boundary conditions $(A,B)$ satisfying
\eqref{abcond} and the sets $\cE$ and $\cI$ there exists a graph $\cG$ for
which these boundary conditions are \emph{local}. Moreover, among all such
graphs there exists a unique maximal graph.

The precise definition of local boundary conditions is as follows. With
respect to the orthogonal decomposition $\cK =
\cK_{\cE}=\oplus\cK_{\cI}^{(-)}\oplus\cK_{\cI}^{(+)}$ any element $\chi$ of
$\cK$ can be represented as a vector
\begin{equation}\label{elements}
\chi=\begin{pmatrix}\{\chi_e\}_{e\in\cE}\\ \{\chi^{(-)}_i\}_{i\in\cI}\\
\{\chi^{(+)}_i\}_{i\in\cI}\end{pmatrix}.
\end{equation}
Consider the orthogonal decomposition
\begin{equation}\label{K:ortho}
\cK = \bigoplus_{v\in V} \cL_{v}
\end{equation}
with $\cL_{v}$ the linear subspace of dimension $\deg(v)$ spanned by those
elements \eqref{elements} of $\cK$ which satisfy
\begin{equation}
\label{decomp}
\begin{split}
\chi_e=0 &\quad \text{if}\quad e\in \cE\quad\text{is not incident with the vertex}\quad v,\\
\chi^{(-)}_i=0 &\quad \text{if}\quad v\quad\text{is not an initial vertex of}\quad i\in \cI,\\
\chi^{(+)}_i=0 &\quad \text{if}\quad v\quad\text{is not a terminal vertex
of}\quad i\in \cI.
\end{split}
\end{equation}
Obviously, the subspaces $\cL_{v_1}$ and $\cL_{v_2}$ are orthogonal if
$v_1\neq v_2$.

Set ${^d}\cL_v:=\cL_v\oplus\cL_v\cong\C^{2\deg(v)}$. Obviously, each
${}^d\cL_v$ inherits a symplectic structure from ${}^d\cK$ in a canonical
way, such that the orthogonal decomposition
\begin{equation*}
\bigoplus_{v\in V} {^d}\cL_v = {}^d\cK
\end{equation*}
holds.

\begin{definition}\label{propo}
Given the graph $\cG=\cG(V,\cI,\cE,\partial)$, boundary conditions $(A,B)$
satisfying \eqref{abcond} are called \emph{local on} $\cG$ if the maximal
isotropic subspace $\cM(A,B)$ of $\cK$ has an orthogonal decomposition
\begin{equation*}
\cM(A,B)=\bigoplus_{v\in V}\;\cM(v),
\end{equation*}
with $\cM(v)$ maximal isotropic subspaces of ${^d}\cL_v$.

Otherwise the boundary conditions are called \emph{non-local}.
\end{definition}

For instance, for a single-vertex graph any boundary conditions are local. The boundary conditions
considered in Example 3.4 of \cite{KS5} are non-local.

Recall (see Section \ref{sec:2}) that $\cM(A,B)$ denotes the maximal
isotropic subspace of $\cK$ corresponding to the boundary conditions
$(A,B)$. It is defined through \eqref{M:def}.

The following proposition gives another characterization of local boundary
conditions (cf.\, \cite{KS5}).

\begin{proposition}\label{def:local}
Given the graph $\cG=\cG(V,\cI,\cE,\partial)$, the boundary conditions
$(A,B)$ satisfying \eqref{abcond} are local on $\cG$ if and only if there
is an invertible map $C:\, \cK\rightarrow\cK$ and linear transformations
$A(v)$ and $B(v)$ in $\cL_{v}$ such that the simultaneous orthogonal
decompositions
\begin{equation}\label{permut}
CA= \bigoplus_{v\in V} A(v)\quad \text{and}\quad CB= \bigoplus_{v\in V} B(v)
\end{equation}
are valid.
\end{proposition}

\begin{proof}
By Theorem \ref{thm:3.1} the boundary conditions $(A,B)$ are local if and
only if there exist linear maps $(A(v),B(v)):\;
{}^d\cL_{v}\rightarrow\cL_{v}$ such that $\cM(v)=\Ker\, (A(v), B(v))$ for
all $v\in V$ and
\begin{equation*}
\Ker\, (A,B) = \bigoplus_{v\in V} \Ker\, (A(v), B(v)).
\end{equation*}
In turn, the last condition holds if and only if there is an invertible map
$C:\, \cK\rightarrow\cK$ such that \eqref{permut} holds.
\end{proof}

\begin{remark}
Let $P_{\cL_v}$ be the orthogonal projection in $\cK$ onto $\cL_v$. For an
arbitrary graph $\cG$ the number
\begin{equation*}
A_{v,v^\prime} := \tr P_{\cL_v} T(0;\underline{a}) P_{\cL_{v^\prime}}
\end{equation*}
is the $(v,v^\prime)$-entry of the adjacency matrix of $\cG$. In
particular, for simple graphs $A_{v,v^\prime}=1$ if and only if the
vertices $v\neq v^\prime$ are adjacent and $A_{v,v^\prime}=0$ otherwise.
\end{remark}

\begin{definition}\label{def:7.4}
Given a $\cG=(V,\cI,\cE,\partial)$ to any vertex $v\in V=V$ we associate
the graph $\cG_v=(\{v\},\cI_v,\cE_v,\partial_v)$ with the following
properties
\begin{itemize}
\item[(i)]{$\cI_v=\emptyset$,}
\item[(ii)]{$\partial_v(e)=v$ for all $e\in\cE_v$,}
\item[(iii)]{$|\cE_v|=\deg_{\cG}(v)$, the degree of the vertex $v$ in the graph $\cG$,}
\item[(iv)]{there is an injective map $\Psi_v:\; \cE_v\rightarrow\cE\cup\cI$ such that
$v\in\partial\circ\Psi_v(e)$ for all $e\in\cE_v$.}
\end{itemize}
\end{definition}

Boundary conditions $(A(v), B(v))$ on each of the graphs $\cG_v$ induce
local boundary conditions $(A,B)$ on the graph $\cG$ with
\begin{equation*}
A= \bigoplus_{v\in V} A(v)\quad \text{and}\quad B= \bigoplus_{v\in V} B(v).
\end{equation*}

\begin{example}[Standard boundary conditions]\label{3:ex:3}
Given a graph $\cG$ with $|V|\geq 2$ and minimum degree not less than two,
define the boundary conditions $(A(v),B(v))$ on $\cG_v$ for every $v\in V$
by the $\deg(v)\times\deg(v)$ matrices
\begin{equation}
\label{ABspecial:neu}
\begin{aligned}
A(v)= \begin{pmatrix}
    1&-1&0&\ldots&&0&0\\
    0&1&-1&\ldots&&0 &0\\
    0&0&1&\ldots &&0 &0\\
    \vdots&\vdots&\vdots&&&\vdots&\vdots\\
    0&0&0&\ldots&&1&-1\\
    0&0&0&\ldots&&0&0
     \end{pmatrix},\qquad
B(v)= \begin{pmatrix}
    0&0&0&\ldots&&0&0\\
    0&0&0&\ldots&&0&0\\
    0&0&0&\ldots&&0&0\\
    \vdots&\vdots&\vdots&&&\vdots&\vdots\\
    0&0&0&\ldots&&0&0\\
    1&1&1&\ldots&&1&1
\end{pmatrix}.
\end{aligned}
\end{equation}
Obviously, $A(v)B(v)^\dagger=0$ and $(A(v),B(v))$ has maximal rank. Thus,
for every $v\in V$ the boundary conditions \eqref{ABspecial:neu} define
self-adjoint Laplace operators $\Delta(A(v),B(v))$ on $L^2(\cG_v)$. By
Lemma \ref{lemma:3:14:real} they are real. The corresponding scattering
matrices are given by
\begin{equation*}
[S_v(\sk)]_{e,e^\prime}= -\frac{\deg(v)-2}{\deg(v)}\delta_{e,e^\prime}
+\frac{2}{\deg(v)} (1-\delta_{e,e^\prime})
\end{equation*}
with $\delta_{e,e^\prime}$ Kronecker symbol. They do not depend on $\sk$
and are permutation invariant and symmetric. Except for the case $\deg(v)=2$
all their entries are non-vanishing.

The boundary conditions \eqref{ABspecial:neu} induce local boundary
conditions $(A,B)$ on the graph $\cG$. The resulting boundary conditions (under various names)
are widely used in different models (see, e.g., the review \cite{Kuchment:0}
and references quoted there).
\end{example}

Now we are in position to discuss the question how to reconstruct the graph
from the boundary conditions. First we present the main result of this
section.

\begin{theorem}\label{reconstruct}
Given the sets $\cE$ and $\cI$ let the boundary conditions $(A,B)$ satisfy
\eqref{abcond}. Then, there is a graph $\cG=\cG(V,\cI,\cE,\partial)$ with
$|\cI|$ internal and $|\cE|$ external edges for which these boundary
conditions are local. This graph is unique under the requirement that the
number of vertices $|V|$ is maximal.
\end{theorem}

Theorem \ref{reconstruct} as well as a sketch of its proof was presented by
the authors in \cite{KS1}. Below we will give a complete proof. We emphasize
that the graph referred to in this theorem need not satisfy Assumption
\ref{con:graph}, that is, it may be disconnected and may have tadpoles.

Consider the completely disjoint graph $\widehat{\cG}$ with $|\cE|$
external and $|\cI|$ internal edges having $|\cI|+|\cE|$ disjoint
components, i.e., any vertex of this graph is of degree 1. Let
$\widehat{V}$ be the vertex set of this graph,
\begin{equation}
\label{equiv1}
\widehat{V}=\{\partial(e)\}_{e\in\cE}\cup\{\partial^-(i)\}_{i\in\cI}\cup\{\partial^+(i)\}_{i\in\cI}.
\end{equation}
Now consider an arbitrary partition $\underline{\cO}$ of the set
$\widehat{V}$,
\begin{equation*}
\underline{\cO} = \{\cO(v)\}_{v\in V},
\end{equation*}
where $\cO(v)$ are subsets of $\widehat{V}$ satisfying
\begin{equation*}
\qquad\bigcup_{v\in V} \cO(v)=\widehat{V}\qquad\text{and}\qquad
\cO(v)\cap\cO(v')=\emptyset\quad\text{for}\quad v\neq v'.
\end{equation*}
Here $V(\underline{\cO})$ is a set labeling the blocks $\cO(v)$ of the
partition $\underline{\cO}$. Let $\cG(\underline{\cO})$ be the graph
obtained from $\widehat{\cG}$ by identifying points in $\widehat{V}$
whenever they lie in the same $\cO(v)$ for some $v\in V(\underline{\cO})$.
The vertex set $V(\cG(\underline{\cO}))$ of the graph
$\cG(\underline{\cO})$ is then, obviously, equipotent to
$V(\underline{\cO})$ and, thus, can be identified with
$V(\underline{\cO})$. Since the property for two points to belong to the
same $\cO(v)$ is an equivalence relation, we can write
\begin{equation}
\label{equiv2} \cG(\underline{\cO})=\widehat{\cG}/\underline{\cO}.
\end{equation}

We will call the graph $\cG(\underline{\cO})$ the \emph{contraction} of the
graph $\widehat{\cG}$ associated with the partition $\underline{\cO}$. In
the case of the trivial partition $\underline{\cO}_{\rm
trivial}=\{\widehat{V}\}$, the resulting graph has exactly one vertex.

We will say that a partition $\underline{\cO}^{\prime}$ of $\widehat{V}$ is
\emph{finer} than $\underline{\cO}$ and write
$\underline{\cO}^{\prime}\succeq\underline{\cO}$ if to any
$\cO^{\prime}(v^{\prime})\in\underline{\cO}^{\prime}$ there is
$\cO(v)\in\underline{\cO}$ with $\cO^{\prime}(v^{\prime})\subseteq \cO(v)$.
We will also write $\underline{\cO}^{\prime}\succ\underline{\cO}$ if
$\underline{\cO}^{\prime}\succeq\underline{\cO}$ and
$\underline{\cO}^{\prime}\neq\underline{\cO}$.

The finest partition $\underline{\cO}_{\rm finest}$ is the partition which
is formed by all one-element subsets of $\widehat{V}$. Note that
$\widehat{\cG}=\cG(\underline{\cO}_{\rm finest})$.

For a given partition $\underline{\cO}$ the condition that the graph
$\cG(\underline{\cO})$ has no tadpoles, i.e. for no edge its initial and
terminal vertices coincide, is equivalent to the condition that all
$\partial^-(i)$ and $\partial^+(i)$ lie in different $\cO(v)$. The
condition that the graph $\cG(\underline{\cO})$ is connected is equivalent
to the statement that to each pair of vertices $v,v^{\prime}\in
V(\cG(\underline{\cO}))$ there is a sequence $i_{1},i_{2},\cdots,i_{K}$ of
elements in $\cI$ and vertices $v=v_{1},\cdots, v_{K+1}=v^{\prime}$ such
that $i_{k}\in \mathcal{S}(v_{k})\cap \mathcal{S}(v_{k+1})$ for all $1\leq
k\leq K$.

\begin{definition}\label{admin:part}
A partition $\underline{\cO}$ of the set $\widehat{V}$ (defined in
\eqref{equiv1}) is \emph{admissible} for the boundary conditions $(A,B)$ if
the maximal isotropic subspace $\cM(A,B)$ of\,\, ${}^d\cK$ has an orthogonal
decomposition
\begin{equation*}
\cM(A,B)=\bigoplus_{v\in V(\underline{\cO})}\;\cM(v),
\end{equation*}
with $\cM(v)$ maximal isotropic subspaces of ${^d}\cL_v$.
\end{definition}

By definition, if a partition $\underline{\cO}$ is admissible for boundary
conditions $(A,B)$, then it is also admissible for all boundary conditions
$(A^\prime,B^\prime)$ such that $\cM(A^\prime,B^\prime)=\cM(A,B)$ and it
also makes sense to speak of admissibility for a maximal isotropic subspace
$\cM$. Furthermore, a partition $\underline{\cO}$ is admissible for the
boundary conditions $(A,B)$ if and only if $^{d}\cL_v\cap\cM(A,B)$ is a
maximal isotropic subspace of $^{d}\cL_v$ for each $v\in
V(\underline{\cO})$. For each $\cM$ there is at least one admissible
partition for each $\cM$, namely the trivial one, where
$\underline{\cO}=\{\widehat{V}\}$. In this case the resulting graph
$\cG(\underline{\cO})$ has exactly one vertex.

Definitions \ref{propo} and \ref{admin:part} immediately imply the
following proposition.

\begin{proposition}\label{def:2.1}
Let $\underline{\cO}$ be an arbitrary partition of the set $\widehat{V}$.
The boundary conditions $(A,B)$ are local on the graph
$\cG(\underline{\cO})$ if and only if the partition $\underline{\cO}$ is
admissible for these boundary conditions.
\end{proposition}

\begin{definition}\label{def:4:5}
An admissible partition $\underline{\cO}$ for boundary conditions $(A,B)$
is \emph{maximal} if there is no admissible partition
$\underline{\cO}^\prime$ such that
$\underline{\cO}^\prime\succ\underline{\cO}$.
\end{definition}

\begin{proposition}\label{lem:maximality}
For any boundary conditions $(A,B)$ satisfying \eqref{abcond} there is a
unique maximal admissible partition $\underline{\cO}(A,B)$ of the set
$\widehat{V}$.
\end{proposition}

We start the proof of the proposition with the following observation.

Let $\underline{\cO}_1=\{\cO_1(v_1)\}_{v_1\in V_1}$ and
$\underline{\cO}_2=\{\cO_2(v_2)\}_{v_2\in V_2}$ be two partitions of
$\widehat{V}$. By $\underline{\cO}_1\cap\underline{\cO}_2$ we denote the set
of all nonempty intersections $\cO_1(v_1)\cap\cO_2(v_2)\neq \emptyset$.
Obviously, any two distinct elements of
$\underline{\cO}_1\cap\underline{\cO}_2$ have empty intersection.
Furthermore, the union of all elements of $\cO_1(v_1)\cap\cO_2(v_2)$ gives
the whole $\widehat{V}$. Thus, $\underline{\cO}_1\cap\underline{\cO}_2$ is
also a partition of $\widehat{V}$. Obviously,
$\underline{\cO}_1\cap\underline{\cO}_2\succeq\underline{\cO}_1$ and
$\underline{\cO}_1\cap\underline{\cO}_2\succeq\underline{\cO}_2$.

For any $\mathcal{O}\subset\underline{\mathcal{O}}$ we set
\begin{equation}\label{c:O}
{^d}\cL_{\mathcal{O}} := \bigoplus_{v\in\cO} {^d}\cL_v.
\end{equation}

\begin{lemma}\label{3:lem:5:o}
Let $\underline{\cO}_1$ and $\underline{\cO}_2$ be two admissible partitions
for the maximal isotropic subspace $\cM$. Then the partition
$\underline{\cO}_1\cap\underline{\cO}_2$ is also admissible for $\cM$.
\end{lemma}

\begin{proof}
Let $\cO_{1}$ and $\cO_{2}$ be any two subsets of $\widehat{V}$ with empty
intersection and such that
\begin{equation*}
\cM_{\cO_{q}} := \cM \cap {^d}\cL_{\mathcal{O}_q}
\end{equation*}
is maximal isotropic in $\cL_{\mathcal{O}_q}$ for $q=1,2$. Then, obviously,
$\cM_{\cO_{1}}\oplus\cM_{\cO_{2}}$ is maximal isotropic in
$^{d}\cL_{\cO_{1}\cup\cO_{2}}=^{d}\cL_{\cO_{1}}\oplus ^{d}\cL_{\cO_{2}}$
such that the relation
$\cM_{\cO_{1}\cup\cO_{2}}=\cM_{\cO_{1}}\oplus\cM_{\cO_{2}}$ holds.

With this observation it suffices to prove the following: Given a maximal
isotropic subspace $\cM$, let $\cO$ and $\cO^{\prime}$ be two subsets of
$\widehat{V}$ having nonempty intersection. Assume that both subspaces
$\cM_{\cO}:=\cM\cap\,^{d}\cL_{\cO}$ and $\cM_{\cO^\prime}:=
\cM\cap\,^{d}\cL_{\cO^\prime}$ are maximal isotropic in $^{d}\cL_{\cO}$ and
$^{d}\cL_{\cO^\prime}$, respectively. Then
$\cM_{\cO\cap\cO^{\prime}}:=\cM\cap\,^{d}\cL_{\cO\cap\cO^{\prime}}$ is
maximal isotropic in $^{d}\cL_{\cO\cap\cO^{\prime}}$.

The isotropy is obvious. By Lemma \ref{ortho:lemma} to prove the maximality
it suffices to show the relation
\begin{equation}\label{dimen:o}
\cM_{\cO\cap\,\cO^{\prime}}^\perp = J \cM_{\cO\cap\,\cO^{\prime}}
\end{equation}
with $J$ the canonical symplectic matrix on $^{d}\cL_{\cO\cap\cO^{\prime}}$.
{}From the definition \eqref{c:O} of $^{d}\cL_\cO$ it follows that
$^{d}\cL_{\cO\cap\cO'}=^{d}\cL_{\cO}\cap\,^{d}\cL_{\cO^\prime}$. Thus,
\begin{equation*}
\cM_{\cO\cap\cO^{\prime}} = \cM \cap\, ^{d}\cL_\cO\cap\,^{d}\cL_{\cO^\prime}
= \cM_\cO \cap \cM_{\cO^\prime}.
\end{equation*}
Since $\cM_\cO$ and $\cM_{\cO^\prime}$ are maximal isotropic, by Lemma
\ref{ortho:lemma} we have $\cM_\cO^\perp = J \cM_\cO$ and
$\cM_{\cO^\prime}^\perp = J \cM_{\cO^\prime}$. Therefore, for any
$\chi\in{^{d}\cL_{\cO\cap\,\cO^{\prime}}}$ there are $\chi_1,
\chi_2\in\cM_{\cO}$ and $\chi_1^\prime, \chi_2^\prime\in\cM_{\cO^{\prime}}$
such that
\begin{equation*}
\chi = \chi_1 + J\chi_2 = \chi_1^\prime + J\chi_2^\prime.
\end{equation*}
This implies that $\chi_1-\chi_1^\prime = -J(\chi_2-\chi_2^{\prime})$.
Since both $\chi_1-\chi_1^\prime$ and $\chi_2-\chi_2^{\prime}$ are elements
of $\cM$ from Lemma \ref{ortho:lemma} again it follows that they both are
null vectors. Therefore,
$\chi_1=\chi_1^\prime\in\cM_{\cO\cap\,\cO^{\prime}}$ and
$\chi_2=\chi_2^\prime\in\cM_{\cO\cap\,\cO^{\prime}}$. Thus, we obtain
$^{d}\cL_{\cO\cap\cO'}=\cM_{\cO\cap\,\cO^{\prime}}\oplus J
\cM_{\cO\cap\,\cO^{\prime}}$ which implies \eqref{dimen:o}.
\end{proof}

\begin{proof}[Proof of Proposition \ref{lem:maximality}]
The relation $\succeq$ is a partial order on the set of all admissible
partitions of $\widehat{V}$. Thus, by the Zorn lemma there exists an
admissible partition $\underline{O}_{\mathrm{max}}$ such that
\begin{equation}\label{Zorn}
\text{there is no admissible partition}\quad \underline{\cO}\quad \text{such
that} \quad \underline{\cO}\succ\underline{O}_{\mathrm{max}}.
\end{equation}

Assume there is an admissible partition $\underline{\cO}$ incomparable with
$\underline{O}_{\mathrm{max}}$, that is, neither $\underline{O}\succeq
\underline{O}_{\mathrm{max}}$ nor $\underline{O}_{\mathrm{max}}\succeq
\underline{O}$ is valid. By Lemma \ref{3:lem:5:o} the partition
$\underline{\cO}\cap\underline{O}_{\mathrm{max}}$ is admissible. Moreover,
$\underline{\cO}\cap\underline{O}_{\mathrm{max}}\succeq\underline{\cO}$ and
$\underline{\cO}\cap\underline{O}_{\mathrm{max}}\succeq\underline{O}_{\mathrm{max}}$,
which contradicts \eqref{Zorn}. Thus, $\underline{O}_{\mathrm{max}}$ is a
maximal admissible partition in the sense of Definition \ref{def:4:5}. Its
uniqueness is now obvious.
\end{proof}

\begin{proof}[Proof of Theorem \ref{reconstruct}]
The claim follows immediately from Propositions \ref{def:2.1} and
\ref{lem:maximality}.
\end{proof}

We conclude this section with two comments. First, an admissible partition
$\underline{\cO}$ for the maximal isotropic subspace $\cM$ is also
admissible for $\cM^{\perp}$. If $\underline{\cO}$ is maximal for $\cM$,
then it is also maximal for $\cM^\perp$. Indeed, let
$\underline{\mathcal{O}}$ be an admissible partition for the maximal
isotropic subspace $\cM$. Observe that the canonical symplectic matrix $J$
\eqref{J:canon} leaves the subspace $^{d}\cL_{\mathcal{O}}$ invariant for
any $\mathcal{O}\in\underline{\mathcal{O}}$. Hence,
\begin{equation*}
(\cM^{\perp})_{\mathcal{O}}=\cM^{\perp}\cap\,^{d}\cL_{\mathcal{O}}=J\cM\cap\,^{d}\cL_{\mathcal{O}}=
J(\cM\cap\,^{d}\cL_{\mathcal{O}})=J\cM_{\mathcal{O}}.
\end{equation*}
But then we obtain
\begin{equation*}
\bigoplus_{\mathcal{O}\in\underline{\mathcal{O}}}
(\cM^{\perp})_{\mathcal{O}}=
\bigoplus_{\mathcal{O}\in\underline{\mathcal{O}}}
J\cM_{\mathcal{O}}=J\cM=\cM^{\perp},
\end{equation*}
which implies that the partition $\underline{\mathcal{O}}$ is admissible
for $\cM^\perp$.

Second, there is an alternative way of proving Theorem 4.6 based on the
concept of reducible matrices (see \cite{GHJ} and the literature quoted
therein) which goes back to Frobenius. An $n\times n$ matrix $M$ is called
\emph{reducible} (by Frobenius \emph{zerlegbar} or \emph{zerfallend}) if
there is a permutation matrix $\Pi$ such that $\Pi M \Pi^{-1}$ has
triangular block form
\begin{equation*}
\Pi M \Pi^{-1}=\begin{pmatrix} M_{00} & M_{01} \\ 0 & M_{11}
\end{pmatrix}.
\end{equation*}
Otherwise $Z$ is called \emph{irreducible}. Recall that any permutation
matrix is unitary. Hence, if $M$ is unitary, then necessary $M_{01}=0$ and
both $M_{00}$ and $M_{11}$ are also unitary. If for a unitary matrix
$U\in\sU(n)$ there is a permutation matrix $\Pi$ such that
\begin{equation*}
\Pi U\Pi^{-1}=\begin{pmatrix}
U_1&0&\cdots&0\\
0&U_2&\cdots&0\\
\vdots&\vdots&\ddots&\vdots\\
0&0&\cdots&U_m
\end{pmatrix},\qquad m\leq n,
\end{equation*}
where every $U_k$ is irreducible, then $\Pi$ is called \emph{maximal}. One
can prove that a maximal permutation matrix does exist and it is unique up
to a permutation of the blocks and permutations within each of the blocks.

Now fix $\sk_0>0$ arbitrarily and determine a maximal permutation matrix
$\Pi$ for $\mathfrak{S}({\sk}_0;A,B)$. Then by relation \eqref{sinv} this
gives
\begin{equation}\label{block2}
\Pi \mathfrak{S}(\sk)\Pi^{-1} =
\begin{pmatrix}
\mathfrak{S}_1(\sk)&0&\cdots&0\\
0&\mathfrak{S}_2(\sk)&\cdots&0\\
\vdots&\vdots&\ddots &\vdots\\
0&0&\cdots&\mathfrak{S}_m(\sk)
\end{pmatrix}
\end{equation}
for all $\sk\in\R$. Obviously, $\Pi$ and $m$ determine uniquely the graph
for which the boundary conditions $(A,B)$ are local. The number $m$ in
\eqref{block2} is the number of vertices of this graph, the blocks
$\mathfrak{S}_1(\sk),\ldots,\mathfrak{S}_m(\sk)$ represent the matrices
\eqref{uuu:def} associated with maximal isotropic subspaces $\cM(v)$ from
Definition \ref{propo}.

\section{Harmonic Analysis of Scattering Matrices}\label{sec:harmony}

In this section we perform harmonic analysis of the scattering matrix with
respect to the lengths $\underline{a}=\{a_i\}_{i\in\cI}\in(\R_+)^{|\cI|}$ of
the internal edges of the graph $\cG$. Throughout the whole section we will
assume that the (topological) graph $\cG$ as well as the boundary
conditions $(A,B)$ are fixed. To carry out the analysis we will now treat
$\underline{a}$ as a parameter which may belong to $\R^{|\cI|}$ or even
$\C^{|\cI|}$.

We start with the following simple but important observation.

\begin{lemma}\label{periodic}
For arbitrary $\sk>0$ the scattering matrix $S(\sk;A,B,\underline{a})$ is
uniquely defined as a solution of \eqref{11} for all
$\underline{a}\in\R^{|\cI|}$. Moreover, the scattering matrix is periodic
with respect to $\underline{a}$,
\begin{equation*}
S\left(\sk;A,B,\underline{a}+\frac{2\pi}{\sk}\underline{\ell}\right)=S(\sk;A,B,\underline{a}),
\qquad \underline{a}\in\R^{|\cI|}
\end{equation*}
for arbitrary $\underline{\ell}\in\Z^{|\cI|}$.
\end{lemma}

\begin{proof}
It suffices to consider those $\underline{a}\in\R^{|\cI|}$ for which $\det
K(\sk; A, B, \underline{a})=0$, since the claim is obvious when the
determinant is non-vanishing. For $\underline{a}\in(\R_+)^{|\cI|}$ the fact
that $S(\sk; A,B,\underline{a})$ is uniquely defined as a solution of
\eqref{11} is guaranteed by Theorem \ref{3.2inKS1}. The case of arbitrary
$\underline{a}\in\R^{|\cI|}$ can be treated in exactly the same way (see
the proof of Theorem 3.2 in \cite{KS1}).

The periodicity follows immediately from \eqref{12} and the fact that the
matrix $T(\sk;\underline{a})$ defined in \eqref{T:def:neu} is
$\frac{2\pi}{\sk}\Z^{|\cI|}$-periodic.
\end{proof}

Lemma \ref{periodic} suggests to consider a Fourier expansion of the
scattering matrix. The following theorem ensures the absolute convergence
of the corresponding Fourier series.

\begin{theorem}\label{thm:main:harmony}
Let $\sk>0$ be arbitrary. For all $\underline{a}\in\R^{|\cI|}$ the Fourier
expansion of the scattering matrix
\begin{equation}\label{Fourier:exp}
S(\sk;A,B,\underline{a})=\sum_{\underline{n}\in\Z^{|\cI|}}
\widehat{S}_{\underline{n}}(\sk;A,B)\,
\e^{\ii\sk\langle\underline{n},\underline{a}\rangle}
\end{equation}
with
\begin{equation}\label{fourier:coef}
\widehat{S}_{\underline{n}}(\sk;A,B) =
\left(\frac{\sk}{2\pi}\right)^{\mid\cI\mid}\;
\int\limits_{[0,2\pi/\sk]^{|\cI|}}d
\underline{a}\:S(\sk;A,B,\underline{a})\:
\e^{-\ii\sk\langle\underline{n},\underline{a}\rangle}
\end{equation}
converges absolutely and uniformly on compact subsets of $\R^{|\cI|}$. The
Fourier coefficients \eqref{fourier:coef} vanish for all
$\underline{n}=\{n_i\}_{i\in\cI}\in\Z^{|\cI|}$ for which $n_i<0$ for at
least one $i\in\cI$.
\end{theorem}

For the proof we need a couple of auxiliary results. Set
\begin{equation*}
\cA=\left\{\underline{a}=\{a_i\}_{i\in\cI}\big|\: \Re a_i\in\R,\:\Im a_i
>0\right\}\subset\C^{|\cI|}.
\end{equation*}

\begin{lemma}\label{lem:det:neq:0}
For any $\sk>0$ the determinant $\det K(\sk;A,B,\underline{a})$ does not
vanish for all $\underline{a}\in\cA$.
\end{lemma}

\begin{proof}
If $\underline{a}\in \cA$, then $\|\mathfrak{S}(\sk;A,B)
T(\sk;\underline{a})\| < 1$. Thus, all eigenvalues of $\mathfrak{S}(\sk;A,B)
T(\sk;\underline{a})$ are strictly less than $1$. Therefore, $\det
K(\sk;A,B,\underline{a})\neq 0$ for all $\underline{a}\in \cA$.
\end{proof}

Set
\begin{equation*}
\Sigma(\sk;A,B) := \{\underline{a}\in\mathrm{clos}(\cA)|\, \det K(\sk;
A,B,\underline{a})=0\}.
\end{equation*}
By Lemma \ref{lem:det:neq:0} we have the inclusion
$\Sigma(\sk;A,B)\subset\partial\cA$. If $\Sigma(\sk;A,B)\neq\emptyset$,
then it is a real analytic subvariety of $\cA$ of codimension at least 1.
The following example shows that the set $\Sigma(\sk;A,B)$ in general need
not to belong to $\R^{|\cI|}\subset\partial\cA$.

\begin{example}\label{ex:Sigma}
Consider the graph $\cG$ with $\cI=\{i_1,i_2,i_3\}$ and $\cE=\{e_1,e_2\}$
depicted in Fig.~\ref{fig:example}. The local boundary conditions $(A,B)$
are defined by the standard boundary conditions (see Example \ref{3:ex:3})
at the vertices of the graph.
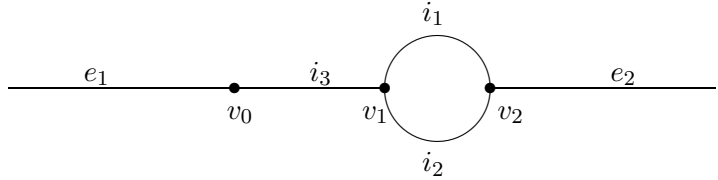
\begin{figure}[htb]
\centerline{ \unitlength1mm
\begin{picture}(120,40)
\put(20,20){\line(1,0){50}} \put(50,20){\circle*{1.5}}
\put(70,20){\circle*{1.5}}  \put(30,21){$e_1$} \put(60,21){$i_3$}
\put(49,16){$v_0$}  \put(67,16){$v_1$} \put(85,16){$v_2$}
\put(77,20){\circle{20}} \put(84,20){\circle*{1.5}} \put(75,29){$i_1$}
\put(75,9){$i_2$} \put(84,20){\line(1,0){30}} \put(100,21){$e_2$}
\end{picture}}
\caption{\label{fig:example} The graph from Example \ref{ex:Sigma}. \newline
\hfill }
\end{figure}
\newline A simple calculation shows that
\begin{equation*}
\Sigma(\sk;A,B)=\left\{a_{i_1}=\frac{\pi n_1}{\sk},\, a_{i_2}=\frac{\pi
n_2}{\sk},\, a_{i_3}\in\C_+\Big|\, n_1,n_2\in\Z,\, n_1+n_2\in 2\Z\right\},
\end{equation*}
which is a subset of the boundary $\partial\cA$ but not of $\R^{|\cI|}$.
Here $\C_+$ denotes the closed upper complex half-plane.
\end{example}

\begin{proposition}\label{thm:6.1:harmo}
Let $\sk>0$ be arbitrary. For all
\begin{equation*}
\underline{a}\in\mathrm{clos}(\cA):=\left\{\underline{a}=\{a_i\}_{i\in\cI}\big|\:
\Re a_i\in\R,\: \Im a_i \geq 0\right\}
\end{equation*}
the scattering matrix $S(\sk;A,B,\underline{a})$ is uniquely defined as a
solution of \eqref{11} and satisfies the bound
\begin{equation}\label{S:leq:1}
\|S(\sk;A,B,\underline{a})\| \leq 1.
\end{equation}
Moreover, it is a rational function of
$\underline{t}=\{t_i\}_{i\in\cI}\in\D^{|\cI|}=\{\zeta\in\C|\,
|\zeta|<1\}^{|\cI|}$ with $t_i:=\e^{\ii\sk a_i}$, i.e.\ a quotient of a
$\cB(\cK_{\cE})$-valued polynomial and a $\C$-valued polynomial in the
variables $t_i$. Thus, for all $\underline{a}\in\mathrm{clos}(\cA)$ the
scattering matrix is $\frac{2\pi}{\sk}\Z^{|\cI|}$-periodic,
\begin{equation*}
S\left(\sk;A,B,\underline{a}+\frac{2\pi}{\sk}\underline{\ell}\right)=S(\sk,A,B,\underline{a}),\qquad
\underline{\ell}\in\Z^{|\cI|}.
\end{equation*}
\end{proposition}

\begin{proof}
By Lemma \ref{lem:det:neq:0} equation \eqref{12} has a unique solution for
all $\underline{a}\in\cA$. Equations \eqref{K:def:neu} and \eqref{zet} imply
that $K(\sk;A,B,\underline{a})$ is a polynomial function of the components
of $\underline{t}$. Obviously, $K(\sk;A,B,\underline{a})^{-1}$ is a
rational function of $\underline{t}$. Thus, by \eqref{12} the scattering
matrix $S(\sk;A,B,\underline{a})$ is a rational function of
$\underline{t}$. Thus, it is $\frac{2\pi}{\sk}\Z^{|\cI|}$-periodic.

Using \eqref{12} it is easy to check that this solution satisfies the
relation
\begin{equation*}
\begin{pmatrix} S(\sk;A,B,\underline{a})\\ \alpha(\sk;A,B,\underline{a})\\
\e^{-\ii\sk\underline{a}}\beta(\sk;A,B,\underline{a}) \end{pmatrix} =
\mathfrak{S}(\sk;A,B) \begin{pmatrix} \1\\ \beta(\sk;A,B,\underline{a})\\
\e^{\ii\sk\underline{a}}\alpha(\sk;A,B,\underline{a}) \end{pmatrix},
\end{equation*}
where $\mathfrak{S}(\sk;A,B)$ is defined in \eqref{uuu:def}. Since
$\mathfrak{S}(\sk;A,B)$ is unitary we obtain
\begin{equation}\label{unitaer}
\begin{split}
& S(\sk;A,B,\underline{a})^\dagger S(\sk;A,B,\underline{a}) +
\alpha(\sk;A,B,\underline{a})^\dagger (\1-\e^{-2\sk \Im\underline{a}})
\alpha(\sk;A,B,\underline{a})\\
& \qquad + \beta(\sk;A,B,\underline{a})^\dagger (\e^{2\sk
\Im\underline{a}}-\1) \beta(\sk;A,B,\underline{a})=\1,
\end{split}
\end{equation}
where $\e^{\pm 2\sk\Im\underline{a}} := \diag\{\e^{\pm 2 \sk \Im
a_i}\}_{i\in\cI}$. {}From \eqref{unitaer} it follows immediately that
\begin{equation*}
0\leq S(\sk;A,B,\underline{a})^\dagger S(\sk;A,B,\underline{a}) \leq \1
\end{equation*}
in the sense of quadratic forms. This proves the bound \eqref{S:leq:1} for
all $\underline{a}\in\cA$. Recalling Lemma \ref{periodic} completes the
proof.
\end{proof}

By Proposition \ref{thm:6.1:harmo} the scattering matrix $S(\sk;
A,B,\underline{a})$ is continuous at all points $\underline{a}\in\cA$. A
priori it is not clear whether its boundary values coincide with those
given by equation \eqref{S-matrix} for all $\underline{a}$ in the subset
$\Sigma(\sk;A,B)$ of $\partial\cA$ where $\det K(\sk; A,B,\underline{a})=0$.

The following result shows the continuity of the scattering matrix
$S(\sk;A, B, \underline{a})$ at points $\underline{a}\in\Sigma(\sk;A,B)$
along some non-tangential directions. Actually this result and an
analyticity argument presented below will imply that the scattering matrix
is continuous at all points $\underline{a}\in\mathrm{clos}(\cA)$.

\begin{proposition}\label{Grenzwert}
Let $\underline{a}\in\partial\cA$ and $\sk>0$ be arbitrary. For any sequence
$\{\underline{a}^{(n)}\}_{n\in\N}$, $\underline{a}^{(n)}\in\cA$ converging
to $\underline{a}\in\partial\cA$ and for all $i\in\cI$ satisfying
\begin{equation}\label{nontangential}
\lim_{n\rightarrow\infty}\frac{a_i^{(n)}-a_i}{|\underline{a}^{(n)}-\underline{a}|}=\gamma
\in\C\setminus\{0\},
\end{equation}
where $\gamma$ is independent of $i$, the relation
\begin{equation}\label{S:convergence}
\lim_{n\rightarrow\infty} S(\sk;A, B, \underline{a}^{(n)}) = S(\sk;A, B,
\underline{a})
\end{equation}
holds.
\end{proposition}

For the proof we need the following lemma, for which we introduce the
shorthand notation
\begin{equation*}
K(\underline{a})\equiv K(\sk; A, B,\underline{a}),\quad
S(\underline{a})\equiv S(\sk; A, B,\underline{a}),\quad\Sigma\equiv
\Sigma(\sk;A,B),\quad \mathfrak{S}\equiv\mathfrak{S}(\sk;A,B) .
\end{equation*}

\begin{lemma}\label{Grenzwert:Hilfe}
For arbitrary $\sk>0$ and arbitrary
$\underline{a}\in\partial\cA\setminus\Sigma$ let
$\underline{a}^{(n)}\in\cA$ be a sequence converging to
$\underline{a}\in\R$ and satisfying \eqref{nontangential}. Then
\begin{equation*}
\lim_{n\rightarrow\infty} P_{\Ran K(\underline{a})^\dagger}
K(\underline{a}^{(n)})^{-1} P_{\Ran
K(\underline{a})}=\left(K(\underline{a})|_{\Ran
K(\underline{a})^\dagger}\right)^{-1} P_{\Ran K(\underline{a})}.
\end{equation*}
\end{lemma}

\begin{proof}
It suffices to show that
\begin{equation}\label{Ziel:1}
\lim_{n\rightarrow\infty} P_{\Ran K(\underline{a})^\dagger}
K(\underline{a}^{(n)})^{-1} K(\underline{a}) = P_{\Ran
K(\underline{a})^\dagger}.
\end{equation}
Indeed, for arbitrary $\widetilde{\chi}\in\Ran K(\underline{a})$ there is a
unique $\chi\in\Ran K(\underline{a})^\dagger$ such that
$K(\underline{a})\chi = \widetilde{\chi}$. Applying \eqref{Ziel:1} to the
vector $\chi\in\Ran K(\underline{a})$ we obtain
\begin{equation*}
\lim_{n\rightarrow\infty} P_{\Ran K(\underline{a})^\dagger}
K(\underline{a}^{(n)})^{-1} \widetilde{\chi} = \chi \equiv
\left(K(\underline{a})|_{\Ran K(\underline{a})^\dagger}\right)^{-1}
\widetilde{\chi},
\end{equation*}
from which the claim follows.

Since $T(\sk;\underline{a})$ is a contraction for all
$\underline{a}\in\partial\cA$ and $\mathfrak{S}$ is unitary, similarly to
the observation made in the proof of Theorem \ref{3.2inKS1} we conclude that
$\Ran K(\underline{a})^\dagger=\Ran K(\underline{a})$. Thus, \eqref{Ziel:1}
is equivalent to the condition
\begin{equation}\label{Ziel:0}
\lim_{n\rightarrow\infty} P_{\Ran K(\underline{a})}
K(\underline{a}^{(n)})^{-1} K(\underline{a}) = P_{\Ran K(\underline{a})}.
\end{equation}
We represent $K(\underline{a})$ and $K(\underline{a}^{(n)})$ as block
matrices
\begin{equation*}
K(\underline{a}) = \begin{pmatrix} K_{00}(\underline{a}) & 0 \\ 0 & 0
\end{pmatrix},\qquad K(\underline{a}^{(n)}) = \begin{pmatrix} K_{00}(\underline{a}^{(n)}) &
K_{01}(\underline{a}^{(n)}) \\ K_{10}(\underline{a}^{(n)}) &
K_{11}(\underline{a}^{(n)})
\end{pmatrix}
\end{equation*}
with respect to the orthogonal decomposition $\cK = \Ran K(\underline{a})
\oplus \Ker K(\underline{a})$.

For arbitrary $\widetilde{\chi}_0\in\Ran K(\underline{a})$ we consider the
linear equation
\begin{equation*}
\begin{split}
K_{00}(\underline{a}^{(n)}) \chi_0^{(n)} +
K_{01}(\underline{a}^{(n)})\chi_1^{(n)} & = \widetilde{\chi}_0,\\
K_{10}(\underline{a}^{(n)}) \chi_0^{(n)} +
K_{11}(\underline{a}^{(n)})\chi_1^{(n)} & = 0.
\end{split}
\end{equation*}
By the assumption that $K(\underline{a}^{(n)})$ is invertible, this equation
is uniquely solvable for all $n\in\N$. Since $K_{00}(\underline{a})$ is
invertible and $K_{00}(\underline{a}^{(n)})\rightarrow
K_{00}(\underline{a})$ as $n\rightarrow\infty$, we have that
$K_{00}(\underline{a}^{(n)})$ is invertible for all sufficiently large
$n\in\N$ and $\|[K_{00}(\underline{a}^{(n)})]^{-1}\|$ is uniformly bounded.
In particular, we have
\begin{equation}\label{Ziel:3}
\left[K_{11}(\underline{a}^{(n)}) -
K_{10}(\underline{a}^{(n)})[K_{00}(\underline{a}^{(n)})]^{-1}
K_{01}(\underline{a}^{(n)})\right] \chi_1^{(n)} = -
K_{10}(\underline{a}^{(n)})[K_{00}(\underline{a}^{(n)})]^{-1}\widetilde{\chi}_0
\end{equation}
and
\begin{equation}\label{Ziel:4}
\chi_0^{(n)} = [K_{00}(\underline{a}^{(n)})]^{-1}\widetilde{\chi}_0 -
[K_{00}(\underline{a}^{(n)})]^{-1} K_{01}(\underline{a}^{(n)}) \chi_1^{(n)}.
\end{equation}

Observe now that
\begin{equation*}
\begin{split}
K_{01}(\underline{a}^{(n)}) & = - P_0
\mathfrak{S}[T(\sk;\underline{a}^{(n)})-T(\sk;\underline{a})] P_1,\\
K_{10}(\underline{a}^{(n)}) & = - P_1
\mathfrak{S}[T(\sk;\underline{a}^{(n)})-T(\sk;\underline{a})] P_0,\\
K_{11}(\underline{a}^{(n)}) & = - P_1
\mathfrak{S}[T(\sk;\underline{a}^{(n)})-T(\sk;\underline{a})] P_1,
\end{split}
\end{equation*}
where $P_0$ and $P_1$ abbreviate the orthogonal projections onto $\Ran
K(\underline{a})$ and $\Ker K(\underline{a})$, respectively. Therefore, by
\eqref{nontangential} the limits
\begin{equation*}
\begin{split}
\widetilde{K}_{01} & :=\lim_{n\rightarrow\infty}
|\underline{a}-\underline{a}^{(n)}|^{-1} K_{01}(\underline{a}^{(n)}),\\
\widetilde{K}_{10} & :=\lim_{n\rightarrow\infty}
|\underline{a}-\underline{a}^{(n)}|^{-1} K_{10}(\underline{a}^{(n)}),\\
\widetilde{K}_{11} & :=\lim_{n\rightarrow\infty}
|\underline{a}-\underline{a}^{(n)}|^{-1} K_{11}(\underline{a}^{(n)})
\end{split}
\end{equation*}
exist. The matrix $\widetilde{K}_{11}$ is invertible. Indeed, observe that
\begin{equation}\label{spaet:ref}
\Ker K(\underline{a})\subset \cK_{\cI}.
\end{equation}
By condition \eqref{nontangential}
\begin{equation*}
\lim_{n\rightarrow\infty} |\underline{a}-\underline{a}^{(n)}|^{-1}
\left[T(\sk;\underline{a}^{(n)}) - T(\sk;\underline{a})\right]
\end{equation*}
is equal to $\ii\sk\gamma T(\sk;\underline{a})$. Therefore,
\begin{equation*}
\displaystyle\lim_{n\rightarrow\infty}
|\underline{a}-\underline{a}^{(n)}|^{-1}\mathfrak{S}[T(\sk;\underline{a}^{(n)})-T(\sk;\underline{a})]\chi_1
=\ii\sk\gamma \mathfrak{S} T(\sk;\underline{a})\chi_1 =\ii\sk\gamma \chi_1
\end{equation*}
for all $\chi_1\in\Ker K(\underline{a})$, which implies that
$\widetilde{K}_{11}$ is invertible.

Divide equation \eqref{Ziel:4} by $|\underline{a}-\underline{a}^{(n)}|$.
Since $\widetilde{K}_{11}$ is invertible, we have that
$|\underline{a}-\underline{a}^{(n)}|^{-1} K_{11}(\underline{a}^{(n)})$ is
invertible for all sufficiently large $n\in\N$ and its inverse is uniformly
bounded. Since
\begin{equation*}
\lim_{n\rightarrow\infty}
K_{01}(\underline{a}^{(n)})=0\qquad\text{and}\qquad
\lim_{n\rightarrow\infty} K_{10}(\underline{a}^{(n)})=0,
\end{equation*}
the matrix
\begin{equation*}
|\underline{a}-\underline{a}^{(n)}|^{-1}
K_{10}(\underline{a}^{(n)})[K_{00}(\underline{a}^{(n)})]^{-1}
K_{01}(\underline{a}^{(n)})
\end{equation*}
vanishes in the limit $n\rightarrow\infty$. Thus, taking the limit
$n\rightarrow\infty$ we obtain that the solutions $\chi_1^{(n)}$ of
equation \eqref{Ziel:3} are uniformly bounded in $n$ and, thus, by
\eqref{Ziel:4}
\begin{equation*}
\lim_{n\rightarrow\infty} \chi_0^{(n)} =
[K_{00}(\underline{a})]^{-1}\widetilde{\chi}_0.
\end{equation*}
This equality implies \eqref{Ziel:0}.
\end{proof}

\begin{proof}[Proof of Proposition \ref{Grenzwert}]
{}From Theorem \ref{3.2inKS1} and Lemma \ref{lem:det:neq:0} it follows that
\begin{equation*}
S(\underline{a}) = - \begin{pmatrix}\1 & 0 & 0 \end{pmatrix}
\left(K(\underline{a})|_{\Ran K(\underline{a})^\dagger}\right)^{-1} P_{\Ran
K(\underline{a})} \mathfrak{S}
\begin{pmatrix}\1 \\ 0 \\ 0 \end{pmatrix}
\end{equation*}
and
\begin{equation*}
\begin{split}
S(\underline{a}^{(n)}) & = - \begin{pmatrix}\1 & 0 & 0 \end{pmatrix}
K(\underline{a}^{(n)})^{-1}  \mathfrak{S}
\begin{pmatrix}\1 \\ 0 \\ 0 \end{pmatrix}\\
& = - \begin{pmatrix}\1 & 0 & 0 \end{pmatrix} K(\underline{a}^{(n)})^{-1}
P_{\Ran K(\underline{a})} \mathfrak{S}
\begin{pmatrix}\1 \\ 0 \\ 0 \end{pmatrix}.
\end{split}
\end{equation*}
Thus, to prove the claim it suffices to show that
\begin{equation}\label{Ziel:Grenzwert}
\lim_{n\rightarrow\infty}\begin{pmatrix}\1 & 0 & 0 \end{pmatrix}
K(\underline{a}^{(n)})^{-1} P_{\Ran K(\underline{a})} =
\begin{pmatrix}\1 & 0 & 0 \end{pmatrix}
\left(K(\underline{a})|_{\Ran K(\underline{a})^\dagger}\right)^{-1} P_{\Ran
K(\underline{a})}.
\end{equation}

{}From \eqref{spaet:ref} it follows that $\Ker K(\underline{a})$ is
orthogonal to $\cK_{\cE}$. Thus,
\begin{equation*}
\begin{pmatrix}\1 & 0 & 0 \end{pmatrix} K(\underline{a}^{(n)})^{-1} P_{\Ran K(\underline{a})} =
\begin{pmatrix}\1 & 0 & 0 \end{pmatrix} P_{\Ran K(\underline{a})}
K(\underline{a}^{(n)})^{-1} P_{\Ran K(\underline{a})}
\end{equation*}
for any $n\in\N$. Applying Lemma \ref{Grenzwert:Hilfe} we obtain
\eqref{Ziel:Grenzwert}.
\end{proof}

For fixed $\sk>0$, consider
\begin{equation}\label{F:def}
F(\underline{t}) := S(\sk; A,B,\underline{a})\quad \text{with}\quad
\underline{t}=\{t_i\}_{i\in\cI},\quad t_i=\e^{\ii a_i \sk}.
\end{equation}
The map $\underline{a}\mapsto \underline{t}$ maps the set
\begin{equation*}
 \left\{\underline{a}\in\C^{|\cI|}\big|\:
\underline{a}=\{a_i\}_{i\in\cI}\; \text{with}\; 0<\Re a_i\leq
2\pi/\sk\;\text{and}\; \Im a_i > 0\; \text{for all}\; i\in\cI \right\}.
\end{equation*}
bijectively onto the polydisc $\D^{|\cI|} = \{\zeta\in\C|\: |\zeta| <
1\}^{|\cI|}$. The interval $(0,2\pi/\sk]$ is mapped onto the torus
$\T^{|\cI|} = \{\zeta\in\C|\: |\zeta| = 1\}^{|\cI|}$.

\begin{lemma}\label{lem:Hardy}
The function $F$ is inner.
\end{lemma}

\begin{remark}\label{Blaschke}
We recall that an operator-valued function on a polydisc $\D^d$ is said to
be inner if it is holomorphic in $\D^d$ and takes unitary values for almost
all points of $\T^d\subset\partial(\D^d)$ (the so called distinguished
boundary of $\D^d$ \cite{Hoermander}). For $d=1$ matrix-valued inner
functions are studied, e.g., in \cite{Potapov}. In particular, an analog of
the canonical factorization theorem for matrix-valued inner functions has
been proven there. For the case $|\cE|\geq 1$ arbitrary but
$|\cI|=1$ by this result one can obtain a representation of the scattering
matrix as a finite Blaschke product.
\end{remark}

\begin{proof}[Proof of Lemma \ref{lem:Hardy}]
{}From Proposition \ref{thm:6.1:harmo} it follows that $F$ is holomorphic
in the punctured open polydisc $\D^{|\cI|}\setminus\{0\}$. By
\eqref{S:leq:1} we have $\|F(\underline{t})\|\leq 1$ for all
$\underline{t}\in\D^{|\cI|}\setminus\{0\}$. Therefore, the Laurent
expansion of $F$ contains no terms with negative powers. Thus, $F$ is
holomorphic in $\D^{|\cI|}$. For every $\underline{t}\in\T^{|\cI|}$ the
operator $F(\underline{t})$ is unitary, which means that $F$ is an inner
function.
\end{proof}

\begin{proof}[Proof of Theorem \ref{thm:main:harmony}]
Since by Proposition \ref{thm:6.1:harmo} $F(\underline{t})$ is a rational
function, it can be analytically continued on all of
$\underline{t}\in\C^{|\cI|}$. Moreover, it is holomorphic in the polydisc
$\D_{1+\epsilon}^{|\cI|} = \{\zeta\in\C|\: |\zeta|<1+\epsilon\}^{|\cI|}$
for some $\epsilon>0$. To show this, by Hartogs' theorem it suffices to
consider the analytic continuation with respect to a single variable
$t_i\in\C$ keeping all other variables fixed. By the bound \eqref{S:leq:1}
all possible poles of this continuation lie outside a disc
$\{t_i\in\C|\;|t_i|<r\}$ with $r>1$.

In turn, this implies (see, e.g., Theorem 2.4.5 in \cite{Hoermander}) that
the Taylor series of the function $F(\underline{t})$ converges absolutely
and uniformly for all $\underline{t}\in\T^{|\cI|}$. Combining this with
Lemma \ref{Grenzwert} proves the absolute and uniform convergence of the
Fourier expansion \eqref{Fourier:exp}.

By Lemma \ref{lem:Hardy} the Fourier coefficients \eqref{fourier:coef}
satisfy $\widehat{S}_{\underline{n}}(\sk;A,B)=0$ for any $\underline{n}\in
\Z^{|\cI|}$ with $n_i<0$ for at least one $i\in\cI$.
\end{proof}

\begin{remark}\label{Taylor}
Lemma \ref{lem:Hardy} reduces the calculation of the Fourier coefficients
\eqref{fourier:coef} to the determination of the coefficients of the Taylor
series for the function \eqref{F:def}.
\end{remark}

\section{Walks on Graphs}\label{sec:4}

This section provides concepts and results which will be needed for the
discussion of the inverse problem as well as of the combinatorial problems
in Section \ref{sec:5}. Throughout the entire section we will assume that
the graph $\cG$ is connected.

A nontrivial \emph{walk} $\bw$ on the graph $\cG$ from $e^\prime\in\cE$ to
$e\in\cE$ is a sequence
\begin{equation*}
\{e^\prime,i_1,\ldots,i_N,e\}
\end{equation*}
of edges such that
\begin{itemize}
\item[(i)]{$i_1,\ldots,i_N\in\cI$,}
\item[(ii)]{$v_0 := \partial(e^\prime)\in\partial(i_1)$ and $v_N :=
\partial(e)\in\partial(i_N)$;}
\item[(iii)]{for any $k\in\{1,\ldots,N-1\}$ there is a
vertex $v_k\in V$ such that $v_k\in\partial(i_k)$ and
$v_k\in\partial(i_{k+1})$};
\item[(iv)]{$v_k = v_{k+1}$ for some
$k\in\{0,\ldots,N-1\}$ if and only if $i_k$ is a
tadpole.}
\end{itemize}

\begin{remark}\label{richtung}
We emphasize that the definition of walks on the graph $\cG$ is independent
of the particular choice of the orientation of the edges. In particular,
the vertex $v_k\in\partial(i_k)$, $v_k\in\partial(i_{k+1})$ need not to be
the terminal vertex of the edge $i_k$ and the initial vertex of the edge
$i_{k+1}$.

More general walks than those considered here are used in \cite{KS7} to
study positivity preserving semigroups generated by Laplace operators on
metric graphs.
\end{remark}

The number $N$ is the \emph{combinatorial length}
$|\bw|_{\mathrm{comb}}\in\N$ and the number
\begin{equation*}
|\bw| = \sum_{k=1}^N a_{i_k} > 0
\end{equation*}
is the \emph{metric length} of the walk $\bw$.

\begin{example}
Let $\cG=(V,\cI,\cE,\partial)$ with $V=\{v_0,v_1\}$, $\cI=\{i\}$,
$\cE=\{e\}$, $\partial(e)=v_0$, and $\partial(i)=\{v_0,v_1\}$. Then the
sequence \{e,i,e\} is not a walk, whereas \{e,i,i,e\} is a walk from $e$ to
$e$.
\end{example}

\begin{proposition}\label{prop:2.1}
Given an arbitrary nontrivial walk $\bw=\{e^\prime,i_1,\ldots,i_N,e\}$
there is a unique sequence $\{v_k\}_{k=0}^N$ of vertices such that $v_0 =
\partial(e^\prime)\in\partial(i_1)$, $v_N =
\partial(e)\in\partial(i_N)$, $v_k\in\partial(i_k)$, and
$v_k\in\partial(i_{k+1})$.
\end{proposition}

\begin{proof}
Assume on the contrary that there are two different sequences
$\{v_k\}_{k=0}^N$ and $\{v^\prime_k\}_{k=0}^N$ satisfying the assumption of
the proposition. This implies that there is a number $\ell\in\{0,\ldots,N-2\}$
such that $v_k=v^\prime_k$ for all $k\in\{0,\ldots,\ell\}$ but $v_{\ell+1}\neq
v^\prime_{\ell+1}$. Obviously, the vertices $v_\ell$, $v_{\ell+1}$, and
$v^\prime_{\ell+1}$ are incident with the same edge. Thus, either
$v_\ell=v_{\ell+1}$ or $v_\ell=v^\prime_{\ell+1}$, which is a contradiction.
\end{proof}

A \emph{trivial} walk $\bw$ on the graph $\cG$ from $e^\prime\in\cE$ to
$e\in\cE$ is a tuple $\{e^\prime,e\}$ with
$\partial(e)=\partial(e^\prime)$. Both the combinatorial and the metric
length of a trivial walk are zero.

A \emph{chain} $\bc$ on the graph $\cG$ is an ordered sequence
$\{v_0,\ldots,v_N\}$ of vertices of the graph $\cG$ such that any two
consecutive vertices are adjacent. Any vertex of $\cG$ may appear more than
once in this sequence. By Proposition \ref{prop:2.1} to any nontrivial walk
$\bw=\{e^\prime,i_1,\ldots,i_N,e\}$ we can uniquely associate the chain
\begin{equation*}
\bc(\bw)=\big\{v_{0}=\partial(e^{\prime}),v_{1},\cdots,v_{N}=\partial(e)\big\},\qquad
N=|\bw|_{\mathrm{comb}}
\end{equation*}
such that $v_k\in\partial(i_k)\cap\partial(i_{k+1})$ for all
$k\in\{1,\ldots,N-1\}$. The chain of a trivial walk is a one element set.
Conversely, a chain $\bc$ specifies a walk on $\cG$. This walk is unique if
any two vertices are adjacent by no more than one internal edge.

A walk $\bw=\{e^\prime,i_1,\ldots,i_N,e\}$ \emph{traverses} an internal edge
$i\in\cI$ if $i_k=i$ for some $1\leq k \leq N$. It \emph{visits} the vertex
$v$ if $v\in\bc(\bw)$.

A \emph{tour} $\fT$ is an unordered set $\{i_1,\ldots,i_M\}$ of pairwise
different internal edges such that there is a walk $\bw$ which traverses
every internal edge $i_k$ of the tour at least once and no other internal
edge. The tour associated with a trivial walk is the empty tour $\fT^{(0)}$.

The \emph{score} $\underline{n}(\bw)$ of a walk $\bw$ is the set
$\{n_i(\bw)\}_{i\in\cI}$ with $n_i(\bw)\geq 0$ the number of times the walk
$\bw$ traverses the internal edge $i\in\cI$. In particular, the tour
associated with a walk $\bw$ is the set
\begin{equation*}
\fT(\bw):=\big\{i\in\cI|\,\, n_i(\bw)>0\}.
\end{equation*}
Any trivial walk has the score
$\underline{n}=\underline{0}:=\{0,\ldots,0\}$.

Let $\cW_{e,e^{\prime}}$, $e,e^{\prime}\in\cE$ be the (infinite if
$\cI\neq\emptyset$) set of all walks $\bw$ on $\cG$ from $e^{\prime}$ to
$e$. By reversing a walk $\bw$ from $e^\prime$ to $e$ into a walk
$\bw_{\mathrm{rev}}$ from $e$ to $e^\prime$ we obtain a natural one-to-one
correspondence between $\cW_{e,e^{\prime}}$ and $\cW_{e^{\prime},e}$.
Obviously, $|\bw|=|\bw_{\mathrm{rev}}|$ and
$\underline{n}(\bw)=\underline{n}(\bw_{\mathrm{rev}})$.

Obviously, the combinatorial length of a walk $\bw$ is given by
\begin{equation*}\label{comb:length}
|\bw|_{\mathrm{comb}}=\sum_{i\in\cI}n_{i}(\bw)
\end{equation*}
and the metric length of a walk $\bw$ can be represented as a sum
\begin{equation}\label{length}
|\bw|=\sum_{i\in\cI}n_{i}(\bw)
a_{i}=\langle\underline{n}(\bw), \underline{a}\rangle,
\end{equation}
where $\langle\cdot,\cdot\rangle$ is the scalar product in $\R^{|\cI|}$ and
the sets $\underline{a}$ and $\underline{n}$ are understood as elements of
$\R^{|\cI|}$.

We have the following obvious result.

\begin{lemma}\label{4:lem:0}
If the Assumption \ref{2cond} holds, i.e., the lengths $\{a_i\}_{i\in\cI}$
are rationally independent, then the metric length $|\bw|$ of a walk $\bw$
uniquely determines its score $\underline{n}(\bw)$.
\end{lemma}

In other words once we know the length of a walk, we know how often any
internal edge has been traversed during that walk.

Let $\cG_v$ be the graph without internal edges associated with a vertex
$v$ (see Definition \ref{def:7.4}). Since the boundary conditions $(A,B)$
on $\cG$ are assumed to be local, we can consider the Laplace operator
$\Delta(A(v),B(v))$ on $L^2(\cG_v)$ associated with the boundary conditions
$(A(v),B(v))$ induced by $(A,B)$ (see Definition \ref{def:local}). By
\eqref{svertex} the scattering matrix for $\Delta(A_v, B_v )$ is given by
\begin{equation*}
S_v(\sk) = - (A(v)+\ii\sk B(v))^{-1} (A(v)-\ii\sk B(v)).
\end{equation*}

Now to each walk $\bw=\{e^\prime,i_1,\ldots,i_N,e\}$ from $e^\prime\in\cE$
to $e\in\cE$ on the graph $\cG$ we associate a weight $W(\bw;\sk)$ by
\begin{equation}\label{weight}
W(\bw;\sk) = \e^{\ii\sk\langle\underline{n}(\bw),
  \underline{a}\rangle}\,\widetilde{W}(\bw;\sk)
\end{equation}
with
\begin{equation}\label{weight:2}
\widetilde{W}(\bw;\sk)=\prod_{k=0}^{|\bw|_{\mathrm{comb}}}
\left[S_{v_{k}}(\sk)\right]_{e_{k}^{(+)}, e_{k}^{(-)}}.
\end{equation}
Here $e_k^{(\pm)}\in\cE_{v_k}$ are defined as
\begin{equation*}
e_k^{(-)}=\begin{cases}\Psi_{v_k}^{-1}(i_k), & \text{if}\quad 1\leq k\leq |\bw|_{\mathrm{comb}},\\
\Psi^{-1}_{v_k}(e^\prime), & \text{if}\quad k=0,
\end{cases}
\end{equation*}
and
\begin{equation*}
e_k^{(+)}=\begin{cases}\Psi_{v_k}^{-1}(i_{k+1}), & \text{if}\quad 0\leq k\leq |\bw|_{\mathrm{comb}}-1,\\
\Psi^{-1}_{v_k}(e), & \text{if}\quad k=|\bw|_{\mathrm{comb}}+1,
\end{cases}
\end{equation*}
where the map $\Psi_v$ is defined in Definition \ref{def:7.4}. Note that
$\widetilde{W}(\bw;\sk)$ is independent of the metric properties of the
graph. Obviously, $\widetilde{W}(\bw;\sk)=[S_{\partial(e)}(\sk)]_{e^\prime,
e}$ if the walk $\bw$ is trivial.

We write $\cW_{e,e^{\prime}}$ as an infinite union of disjoint, non-empty
sets by grouping together all walks $\bw$ with the same score
$\underline{n}(\bw)$,
\begin{equation*}
\cW_{e,e^{\prime}}(\underline{n})=\left\{{\bf w}\in
\cW_{e,e^{\prime}}\left.\right|\underline{n}({\bf w})=\underline{n}\right\}
\end{equation*}
such that
\begin{equation}
\label{union}
\cW_{e,e^{\prime}}=\bigcup_{\underline{n}\in(\N_0)^{|\cI|}}\cW_{e,e^{\prime}}(\underline{n}).
\end{equation}
Note that these sets depend only on topology of the graph $\cG$ and are
independent of its metric properties. Also if $\bw\in
\cW_{e,e^{\prime}}(\underline{n})$ then $\bw_{\mathrm{rev}}\in
\cW_{e^{\prime},e}(\underline{n})$.
$\cW_{e,e^{\prime}}(\underline{0})=\emptyset$ if and only if $\partial(e)
\neq \partial(e^{\prime})$.

We have the following obvious result.

\begin{lemma}\label{4:lem:2}
The sets $\cW_{e,e^{\prime}}(\underline{n})$ are at most finite. The number
of different walks in $\cW_{e,e^{\prime}}(\underline{n})$ satisfies the
bound
\begin{equation}
\label{nnumber} |\cW_{e,e^{\prime}}(\underline{n})|\leq
\frac{|\underline{n}|!} {\displaystyle\prod_{i\in\cI}n_{i}!},
\end{equation}
where
\begin{equation*}
|\underline{n}| =\sum_{i\in\cI}n_{i}
\end{equation*}
is the combinatorial length of any walk
$\bw\in\cW_{e,e^{\prime}}(\underline{n})$.
\end{lemma}

\begin{theorem}\label{verloren}
For any local boundary conditions $(A,B)$ satisfying \eqref{abcond} the
matrix elements of the $\underline{n}$-th Fourier coefficients
\eqref{fourier:coef} are given by the sum over the walks with the score
$\underline{n}$,
\begin{equation}\label{verloren:2}
[\widehat{S}_{\underline{n}}(\sk;A,B)]_{e,e^\prime} =
\sum_{\bw\in\cW_{e,e^\prime}(\underline{n})} \widetilde{W}(\bw;\sk)
\end{equation}
whenever $\cW_{e,e^{\prime}}(\underline{n})$ is nonempty and
$[\widehat{S}_{\underline{n}}(\sk;A,B)]_{e,e^\prime}=0$ if
$\cW_{e,e^{\prime}}(\underline{n})=\emptyset$.
\end{theorem}

\begin{proof}
Obviously, it suffices to show that the $\underline{n}$-th coefficient of
the multi-dimensional Taylor expansion of the function $F(\underline{t})$
(defined in \eqref{F:def}) with respect to $\underline{t}\in\D^{|\cI|}$
coincides with the r.h.s.\ of \eqref{verloren:2}. Recall that by Theorem
\ref{3.2inKS1} and Lemma \ref{lem:det:neq:0} for all $\underline{a}\in\cA$
the scattering matrix is given by
\begin{equation}\label{S:verloren}
S(\sk; A,B,\underline{a})=-\begin{pmatrix} \1 & 0 & 0 \end{pmatrix} \left[\1
-\mathfrak{S}(\sk;A,B) T(\sk;\underline{a})\right]^{-1}\mathfrak{S}(\sk;A,B)
\begin{pmatrix} \1 \\ 0 \\ 0 \end{pmatrix},
\end{equation}
where $\mathfrak{S}(\sk;A,B)$ is given by \eqref{svertex} and
$T(\sk;\underline{a})$ by \eqref{T:def:neu}. Since
$\|T(\sk;\underline{a})\|<1$ for all $\underline{a}\in\cA$ we obtain the
series
\begin{equation}\label{Reihe}
\begin{split}
& S(\sk; A,B,\underline{a})=\\ & = \sum_{n=0}^\infty
\begin{pmatrix} \1 & 0 & 0
\end{pmatrix} \left[\mathfrak{S}(\sk; A,B) T(\sk;\underline{a}) \right]^n \mathfrak{S}(\sk; A,B)
\begin{pmatrix} \1 \\ 0 \\ 0
\end{pmatrix}
\end{split}
\end{equation}
converging in norm for all $\underline{a}\in\cA$.

Recall that
\begin{equation*}
\mathfrak{S}(\sk; A,B) = \mathfrak{S}(\sk; CA, CB)
\end{equation*}
for every invertible $C$. It follows directly from Definition
\ref{def:local} that
\begin{equation*}
\mathfrak{S}(\sk; A,B) = \bigoplus_{v\in V} S(\sk; A(v),B(v)).
\end{equation*}
Plugging this equality into \eqref{Reihe} proves that
\begin{equation}\label{sseries}
[S(\sk;A,B,\underline{a})]_{e,e^{\prime}}=\sum_{\bw\in \cW_{e,e^{\prime}}}
W(\bw; \sk) \equiv \sum_{\bw\in \cW_{e,e^{\prime}}}\widetilde{W}(\bw; \sk)\,
\e^{\ii\sk\langle\underline{n}(\bw),\underline{a}\rangle}
\end{equation}
for all $\sk>0$ and $\underline{a}\in\cA$. By uniqueness of the Fourier
expansion Theorem \ref{verloren} implies the claim.
\end{proof}

\begin{remark}
Theorem \ref{verloren} implies that the scattering matrix of the metric
graph $(\cG, \underline{a})$ is determined by the scattering matrices
associated with all its single vertex subgraphs. This result can also be
obtained by applying the factorization formula \cite{KS3}.
\end{remark}

Since $\widetilde{W}(\bw;\sk)$ is independent of the metric properties of
the graph, it is natural to view \eqref{sseries} as a \emph{combinatorial
Fourier expansion} of the scattering matrix $S(\sk;A,B,\underline{a})$. We
will show that \eqref{sseries} actually coincides with the Fourier expansion
\eqref{Fourier:exp} in Theorem \ref{thm:main:harmony}.

For given $e,e^{\prime}\in\cE$ consider the set of scores of all walks from
$e^{\prime}$ to $e$,
\begin{equation}\label{enn}
\cN_{e,e^{\prime}}=\left\{\underline{n}\in(\N_0)^{|\cI|}\left.\right|
\text{there is a walk}\, \bw\in \cW_{e,e^{\prime}}(\underline{n})\right\}.
\end{equation}
Since $\underline{n}(\bw)=\underline{n}(\bw_{\mathrm{rev}})$, we have
$\cN_{e,e^{\prime}}=\cN_{e^{\prime},e}$. By Theorems \ref{thm:main:harmony}
and \ref{verloren} we can rewrite \eqref{Fourier:exp} as
\begin{equation}\label{ssum}
[S(\sk;A,B,\underline{a})]_{e,e^{\prime}} =
\sum_{\underline{n}\in\cN_{e,e^{\prime}}}
[\widehat{S}_{\underline{n}}(\sk;A,B)]_{e,e^{\prime}}\, \e^{\ii\sk
\langle\underline{n},\underline{a}\rangle}.
\end{equation}

By Lemma \ref{4:lem:2}, \eqref{verloren:2} is a finite sum. Hence,
$[\widehat{S}_{\underline{n}}(\sk;A,B)]_{e,e^{\prime}}$ is a rational
function of $\sk$. This function may vanish identically even if
$\cW_{e,e^{\prime}}(\underline{n})\neq\emptyset$. This is well illustrated
by the following simple example.

\begin{figure}[htb]
\centerline{ \unitlength1mm
\begin{picture}(120,40)
\put(20,20){\line(1,0){80}} \put(50,20){\circle*{1.5}}
\put(70,20){\circle*{1.5}}  \put(30,21){$e$} \put(60,21){$i$}
\put(90,21){$e^\prime$}  \put(49,16){$v_0$}  \put(69,16){$v_1$}
\put(51,20){\vector(1,0){10}}
\end{picture}}
\caption{\label{graph:fig:line} The graph from Example
\ref{vanish:identically}. The length of the internal edge $i$ is arbitrary.
The arrow shows the orientation of the internal edge. \hfill }
\end{figure}
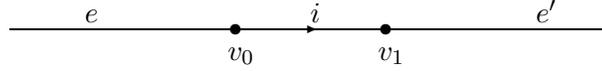

\begin{example}\label{vanish:identically}
Consider the graph $\cG$ depicted in Fig.\ \ref{graph:fig:line}. Choose the
boundary conditions $A\underline{\psi} + B\underline{\psi}' = 0$ with
\begin{equation*}
A=\begin{pmatrix} 1 & -1 & 0 & 0 \\ 0 & 0 & 1 & -1 \\ 0 & 0 & 0 & 0 \\ 0 & 0
& 0 & 0
\end{pmatrix},\qquad
B=\begin{pmatrix} 0 & 0 & 0 & 0 \\ 0 & 0 & 0 & 0 \\ 1 & 1 & 0 & 0 \\ 0 & 0
& 1 & 1
\end{pmatrix}
\end{equation*}
and
\begin{equation*}
\underline{\psi} = \begin{pmatrix} \psi_e(0) \\
                                   \psi_i(0) \\
                                   \psi_i(a_i) \\
                                    \psi_{e^\prime}(0)
                                     \end{pmatrix},\qquad
\underline{\psi}' = \begin{pmatrix} \psi_e'(0) \\
                                   \psi_i'(0) \\
                                   -\psi_i'(a_i) \\
                                   \psi'_{e^\prime}(0)
                                     \end{pmatrix}.
\end{equation*}
Obviously, these boundary conditions are local on $\cG$. It is easy to see
that
\begin{equation*}
S_{v_0}(\sk)= \begin{pmatrix} 1 & 0 \\ 0 & 1
\end{pmatrix}\qquad\text{and}\qquad S_{v_1}(\sk)= \begin{pmatrix} 1 & 0 \\ 0 & 1
\end{pmatrix}.
\end{equation*}
Thus, for any walk from $e$ to $e$ the factor $\widetilde{W}(\bw;\sk)$
equals zero. Therefore, $S(\sk;\underline{n})_{e,e}=0$ for all
$\underline{n}\in\N_0$. Note that $\cW_{e,e}(\underline{n})\neq\emptyset$
for all $\underline{n}\in 2\N_0$.
\end{example}

Below we will show that in the generic case the matrix elements vanishes
identically if and only if $\cW_{e,e^{\prime}}(\underline{n})=\emptyset$.
First we will prove this statement for a particular choice of boundary
conditions. These boundary conditions will be used for the discussion of
some combinatorial problems in Section \ref{sec:5}.

\begin{theorem}\label{4:theo:2}
Let $\{H(v)\}_{v\in V}$ be a collection of $\deg(v)\times\deg(v)$
self-adjoint matrices with strictly positive entries. Let $(A(v),B(v))$ be
the boundary conditions associated with $H(v)$ by Lemma \ref{lem:3.13}.
Define the local boundary conditions $(A,B)$ on the graph $\cG$ given by
\begin{equation*}
A=\bigoplus_{v\in V}A(v)\qquad\text{and}\qquad B=\bigoplus_{v\in V}B(v)
\end{equation*}
with respect to the orthogonal decomposition \eqref{K:ortho}. Then for any
$e,e^\prime\in\cE$ the matrix element
$[\widehat{S}_{\underline{n}}(\sk;A,B)]_{e,e^\prime}$ of the Fourier
coefficient \eqref{fourier:coef} of the scattering matrix
$S(\sk;A,B,\underline{a})$ vanishes identically for all $\sk>0$ if and only
if $\cW_{e,e^{\prime}}(\underline{n})=\emptyset$.
\end{theorem}

We introduce some notion which will be used in the proof. Consider an
arbitrary nontrivial walk $\bw=\{e^\prime,i_1,\ldots,i_N,e\}$. We say that
the walk is \emph{transmitted} at the vertex
\begin{equation*}
v_k\in\bc(\bw) = \{v_0=\partial(e^\prime),\ldots,v_k,\ldots,v_N=\partial(e)\}
\end{equation*}
if either $v_k=\partial(e)$ or $v_k=\partial(e^\prime)$ or
$v_k\in\partial(i_k)$, $v_k\in\partial(i_{k+1})$, and $i_k\neq i_{k+1}$. We
say that a trivial walk from $e^\prime$ to $e$ is transmitted at the vertex
$v=\partial(e)=\partial(e^\prime)$ if $e\neq e^\prime$. Otherwise the walk
is said to be \emph{reflected}. Let $\sT(\bw)$ denote the ordered set of
vertices of the graph $\cG$ successively visited by the walk $\bw$, at which
the walk is transmitted. Thus, the set $\sT(\bw)$ can be obtained from the
chain $\bc(\bw)$ by deleting all vertices, at which the walk is reflected.

Set
\begin{equation*}
t_{e,e^\prime}(\underline{n}) =
\min_{\bw\in\cW_{e,e^\prime}(\underline{n})}|\sT(\bw)|.
\end{equation*}
Then
\begin{equation*}
\cW_{e,e^\prime}^{\mathrm{min}}(\underline{n})=\{\bw\in\cW_{e,e^\prime}(\underline{n})
\mid |\sT(\bw)|=t_{e,e^\prime}(\underline{n})\}
\end{equation*}
is the set of walks from $e^\prime$ to $e$ with the score $\underline{n}$
having the smallest number of transmissions.

\begin{proof}[Proof of Theorem \ref{4:theo:2}]
The ``if'' part of the statement follows from Theorem \ref{verloren}. To
prove the ``only if'' part, we assume that
$\cW_{e,e^{\prime}}(\underline{n})\neq\emptyset$. For an arbitrary walk
$\bw\in\cW_{e,e^{\prime}}(\underline{n})$ consider the weight
$\widetilde{W}(\bw;\sk)$ given by \eqref{weight:2}. For
$\sk\rightarrow+\infty$ from Lemma \ref{lem:3.13} we obtain the following
asymptotics of this weight
\begin{equation*}
\widetilde{W}(\bw;\sk) = \sk^{-|\sT(\bw)|}\, \ii^{|\sT(\bw)|} \prod_{v\in
\sT(\bw)} [H(v)]_{e^{(+)}(v),e^{(-)}(v)}[1+o(1)].
\end{equation*}
The product is understood in the ordered sense. By assumption this product
is strictly positive.

By Theorem \ref{verloren} the matrix element
$[\widehat{S}_{\underline{n}}(\sk;A,B)]_{e,e^\prime}$ has the following
asymptotics
\begin{equation*}
[\widehat{S}_{\underline{n}}(\sk;A,B)]_{e,e^\prime} =
\sk^{-t_{e,e^\prime}(\underline{n})}\, \ii^{t_{e,e^\prime}(\underline{n})}
\sum_{\bw\in\cW_{e,e^\prime}^{\mathrm{min}}(\underline{n})} \, \prod_{v\in
\sT(\bw)} [H(v)]_{e^{(+)}(v),e^{(-)}(v)}[1+o(1)].
\end{equation*}
Noting that
\begin{equation*}
\sum_{\bw\in\cW_{e,e^\prime}^{\mathrm{min}}(\underline{n})} \, \prod_{v\in
\sT(\bw)} [H(v)]_{e^{(+)}(v),e^{(-)}(v)}
\end{equation*}
is a sum of strictly positive numbers proves that
$[\widehat{S}_{\underline{n}}(\sk;A,B)]_{e,e^\prime}$ does not vanish
identically.
\end{proof}

We turn to the discussion of general local boundary conditions. The set of all local boundary conditions is isomorphic to the direct product
\begin{equation*}
\mathsf{U}_{\cG}:=\displaystyle\Bigtimes_{v\in V}\mathsf{U}(\deg(v))
\end{equation*}
of the groups of all unitary $\deg(v)\times\deg(v)$ matrices. Let $\mu$ be
the probability measure on $\mathsf{U}_{\cG}$ defined as a product measure
by
\begin{equation*}
\mu(\sigma)=\bigotimes_{v\in V}\mu_v(\sigma_v),\qquad
\sigma=\Bigtimes_{v\in V} \sigma_v,
\end{equation*}
where $\mu_v$ is the normalized Haar measure on $\mathsf{U}(\deg(v))$ and
$\sigma_v$ is an open subset of $\mathsf{U}(\deg(v))$.

\begin{definition}\label{generic:alg}
A property $P$ holds for Haar almost all local boundary conditions if
\begin{equation*}
\begin{split}
&\mu\Bigl\{U\in\Bigtimes_{v\in V}\mathsf{U}(\deg(v))|\ P\ \text{holds for the boundary conditions}\\
&\qquad\qquad\text{defined by the maximal isotropic subspace}\
\cM(U)\Bigr\}=\mu\left(\Bigtimes_{v\in V}\mathsf{U}(\deg(v))\right)=1.
\end{split}
\end{equation*}
is valid.
\end{definition}

The following theorem shows that for Haar almost all local boundary conditions
the function $\widehat{S}_{\underline{n}}(\sk)_{e,e^{\prime}}$ does not vanish
identically whenever $\cW_{e,e^{\prime}}(\underline{n})\neq\emptyset$.

\begin{theorem}\label{4:theo:1}
For Haar almost all local boundary conditions $(A,B)$ satisfying
\eqref{abcond} the matrix element $[\widehat{S}_{\underline{n}}(\sk;
A,B)]_{e,e^{\prime}}$ of the Fourier coefficient \eqref{fourier:coef} of
the scattering matrix $S(\sk;A,B,\underline{a})$ vanishes identically for
all $\sk>0$ if and only if $\cW_{e,e^{\prime}}(\underline{n})= \emptyset$.
\end{theorem}

\begin{proof}
The group $\mathsf{U}_{\cG}$ is a real-analytic manifold of real dimension
$\dim_\R \mathsf{U}_{\cG} = N_{\cG}$ with
\begin{equation*}
N_{\cG}:= \sum_{v\in V} [\deg(v)]^2.
\end{equation*}
Let $\mathsf{U}^0(\deg(v))$ be the subset of $\mathsf{U}(\deg(v))$
consisting of those unitaries which do not have $-1$ as an eigenvalue. By
Lemma \ref{generic:ev} the set $\mathsf{U}^0(\deg(v))$ is of full Haar
measure. It is an open dense subset in $\mathsf{U}(\deg(v))$. Set
$\mathsf{U}_{\cG}^0 := \displaystyle\Bigtimes_{v\in
V}\mathsf{U}^0(\deg(v))$.

For any given score $\underline{n}\in\cN_{e,e^\prime}$ define the map
$\Phi_{\underline{n}}:\; \mathsf{U}_{\cG}^0 \rightarrow \C$ by
\begin{equation}\label{phin:def}
U \mapsto \Phi_{\underline{n}}(U) = \sum_{\bw\in
\cW_{e,e^\prime}^{\mathrm{min}}(\underline{n})}\; \prod_{v\in\sT(\bw)}
[(U(v)+\1)^{-1}(U(v)-\1)]_{e^{(+)}(v),e^{(-)}(v)},
\end{equation}
where
\begin{equation*}
U=\bigoplus_{v\in V} U(v) \in \mathsf{U}_{\cG}^0
\end{equation*}
with respect to the orthogonal decomposition \eqref{K:ortho}.

Observe that by Lemma \ref{lem:3.13} and equation \eqref{will:be:used} the
quantity $\Phi_{\underline{n}}(U)$ with $U\in\mathsf{U}_\cG^0$ determines
the asymptotics for $\sk\rightarrow\infty$ of the Fourier coefficient
$\widehat{S}_{\underline{n}}(\sk; A_U, B_U)$ of the scattering matrix
$S(\sk; A_U, B_U,\underline{a})$,
\begin{equation*}
[\widehat{S}_{\underline{n}}(\sk;A_U,B_U)]_{e,e^\prime} =
\sk^{-t_{e,e^\prime}(\underline{n})}\, \ii^{t_{e,e^\prime}(\underline{n})}
\Phi_{\underline{n}}(U) [1+o(1)],
\end{equation*}
where $A_U=-(U - \1)/2$, $B_U=(U + \1)/(2\ii\sk)$ (see Proposition
\ref{unit:neu}). Therefore, (see the proof of Theorem \ref{4:theo:2}) the
function $U\mapsto\Phi_{\underline{n}}(U)$ is not identically zero.

Let $\cU\subset\mathsf{U}_{\cG}^0$ be an arbitrary open set belonging to a
single chart of the manifold $\mathsf{U}_{\cG}$. Denote by $\varkappa:\,
\cU\rightarrow\R^{\N_{\cG}}$ a (real-analytic) coordinate map. Let
$\mathcal{Z}\subset\mathsf{U}_{\cG}^0$ be the set of all zeroes of the
function $\Phi_{\underline{n}}$. Assume that $\mathcal{Z}\cap\cU$ has a
positive Haar measure. Then the function
$\Phi_{\underline{n}}\circ\varkappa^{-1}$ vanishes on a subset of
$\R^{\N_{\cG}}$ of a positive Lebesgue measure. Observe that
$\Re\Phi_{\underline{n}}\circ\varkappa^{-1}$ and
$\Im\Phi_{\underline{n}}\circ\varkappa^{-1}$ are real-analytic functions.
Hence, by the Lojaciewicz theorem \cite{Krantz:Parks}, the function
$\Phi_{\underline{n}}\circ\varkappa^{-1}$ is identically zero on
$\varkappa^{-1}(\cU)\subset\R^{\N_{\cG}}$. Thus, $\Phi_{\underline{n}}$
vanishes identically on $\mathsf{U}_{\cG}^0$, a contradiction.
\end{proof}

\section{Proof of Theorem \ref{1:theo:1}}\label{sec:thm1}

Given a finite noncompact connected graph $\cG=\cG(V,\cI,\cE,\partial)$, for
arbitrary $e,e^\prime\in\cE$ consider the set of nonnegative real numbers
\begin{equation*}
\cL_{e,e^\prime} = \left\{\langle
\underline{n}(\bw),\underline{a}\rangle|\, \bw\in\cW_{e,e^\prime} \right\},
\end{equation*}
where $\underline{n}(\bw)$ is the score of the walk $\bw$. Obviously,
$\cL_{e,e^{\prime}}=\cL_{e^{\prime},e}$. Also, $\cI=\emptyset$ if and only
if $\cL_{e,e^{\prime}}=\{0\}$ for all $e,e^\prime\in\cE$. Furthermore,
$\cL_{e^{\prime\prime},e}=\cL_{e^{\prime\prime},e^{\prime}}$ for all
$e^{\prime\prime}\in\cE$ whenever $\partial(e)=\partial(e^{\prime})$.

\begin{lemma}\label{lem:8.1}
Assume that the graph $\cG=\cG(V,\cI,\cE,\partial)$ is connected,
$\cE\neq\emptyset$. Then for any metric $\underline{a}$ on $\cG$ the set of
internal edges $\cI$ is nonempty if and only if $\cL_{e,e^\prime}$ is an
infinite discrete set for all $e,e^\prime\in\cE$,
\begin{equation*}
\cL_{e,e^\prime} = \{\omega_{e,e^\prime}(k)\}_{k\in\N_0},\quad
\text{where}\quad \omega_{e,e^\prime}(k)<\omega_{e,e^\prime}(k+1)\quad
\text{for all}\quad k\in\N_0.
\end{equation*}
\end{lemma}

\begin{proof}
The ``if'' part is obvious. For the ``only if'' part assume that
$\cI\neq\emptyset$. Then the set $\cL_{e,e^\prime}$ is infinite. Choose an
arbitrary $w\in\cL_{e,e^\prime}$. Observe that for any $\epsilon>0$ the set
\begin{equation*}
\{\underline{x}\in(\R_+)^{|\cI|}|\;
|\langle\underline{x},\underline{a}\rangle-w| < \epsilon\}
\end{equation*}
is bounded. Therefore, the set
\begin{equation*}
\{\underline{n}\in(\N_0)^{|\cI|}|\;
|\langle\underline{n},\underline{a}\rangle-w| < \epsilon\}
\end{equation*}
is at most finite. Thus, $\cL_{e,e^\prime}$ has no finite accumulation
points.
\end{proof}

The following uniqueness result is a key technical tool to determine the
topology of the graph from its scattering matrix. In contrast to Theorem
\ref{thm:main:harmony} which implies the uniqueness of the Fourier
expansion \eqref{Fourier:exp} for fixed $\sk>0$, this result corresponds to
the case, where $\underline{a}\in (\R_+)^{|\cI|}$ is kept fixed.

\begin{theorem}\label{eindeutigkeit}
Assume that the graph $\cG=\cG(V,\cI,\cE,\partial)$ is connected,
$\cE\neq\emptyset$. For any fixed $\underline{a}\in(\R_+)^{|\cI|}$
satisfying Assumption \ref{2cond} the scattering matrix $S(\sk; A,B,
\underline{a})$ has a unique absolutely converging expansion
\begin{equation}\label{6:11:Fourier}
S(\sk; A,B, \underline{a}) = \sum_{\underline{n}\in(\N_0)^{|\cI|}}
\widehat{S}_{\underline{n}}(\sk; A,B)\,
\e^{\ii\sk\langle\underline{n},\underline{a} \rangle},\qquad \sk\in\R_+.
\end{equation}
In particular, for any $e,e^\prime\in\cE$ the Fourier coefficients
$\left[\widehat{S}_{\underline{n}}(\sk; A,B)\right]_{e,e^\prime}$ and the
set
\begin{equation*}
\widetilde{\cL}_{e,e^\prime}:=\left\{\langle\underline{n},\underline{a}\rangle\big|\,
\underline{n}\in(\N_0)^{|\cI|}\,\,\text{and}\,\,
\bigl[\widehat{S}_{\underline{n}}(\sk; A,B)\bigr]_{e,e^\prime}\,\,\text{does
not vanish identically} \right\} \subset \cL_{e,e^\prime}
\end{equation*}
are uniquely determined by the map $\R_+\ni\sk\mapsto
\left[S(\sk;A,B,\underline{a})\right]_{e,e^\prime}$.
\end{theorem}

Note that for local boundary conditions leading to $\sk$-independent single
vertex scattering matrices (like standard boundary conditions considered in
Example \ref{3:ex:3} or more general boundary conditions discussed in Remark
\ref{k-independent}) the statement follows from the uniqueness theorem for
almost periodic functions. For general boundary conditions Theorem
\ref{thm:main:harmony} implies that the scattering matrix $S(\sk; A,B,
\underline{a})$ admits an expansion of the form \eqref{6:11:Fourier}.
However, it does not imply that $\R_+\ni\sk\mapsto
S(\sk;A,B,\underline{a})$ uniquely determines the coefficients
$\widehat{S}_{\underline{n}}(\sk; A,B)$ in \eqref{6:11:Fourier}.

\begin{remark}\label{rem}
By Theorem \ref{4:theo:1} the sets $\widetilde{\cL}_{e,e^\prime}$ and
$\cL_{e,e^\prime}$ agree for Haar almost all local boundary conditions.
Similarly, $\widetilde{\cL}_{e,e^\prime} = \cL_{e,e^\prime}$ for the
boundary conditions referred to in Theorem \ref{4:theo:2}.
\end{remark}

\begin{proof}[Proof of Theorem \ref{eindeutigkeit}]
Fix $\underline{a}\in(\R_+)^{|\cI|}$ and boundary conditions $(A,B)$. Set
for brevity $S(\sk):= S(\sk; A, B, \underline{a})$ and
$\widehat{S}_{\underline{n}}(\sk) := \widehat{S}_{\underline{n}}(\sk; A,B)$.
By Lemma \ref{lem:8.1} the set $\widetilde{\cL}_{e,e^\prime}$ is discrete,
$\widetilde{\cL}_{e,e^\prime}=\{\widetilde{w}_{e,e^\prime}(k)\}_{k\in\N_0}$
with $\widetilde{w}_{e,e^\prime}(k)<\widetilde{w}_{e,e^\prime}(k+1)$.

{}From Corollary \ref{3:cor:2} and Theorem \ref{verloren} it follows that
for all sufficiently large $\sk\in\C$ the Fourier coefficients
$\widehat{S}_{\underline{n}}(\sk)$ admit an absolutely converging expansion
of the form
\begin{equation}\label{S:series}
\widehat{S}_{\underline{n}}(\sk)=\sum_{m=0}^\infty
\widehat{S}_{\underline{n}}^{(m)} \sk^{-m}.
\end{equation}
By Assumption \ref{2cond} there is a unique
$\underline{n}_0\in(\N_0)^{|\cI|}$ such that $\langle\underline{n}_0,
\underline{a}\rangle = w_{e,e^\prime}(0)$.

Obviously,
\begin{equation*}
\lim_{\Im\sk\rightarrow +\infty} \sk^{-1} \log [S(\sk)]_{e,e^\prime} = \ii
\widetilde{w}_{e,e^\prime}(0)
\end{equation*}
and
\begin{equation*}
\lim_{\Im\sk\rightarrow+\infty} \e^{-\ii\sk\widetilde{w}_{e,e^\prime}(0)}
[S(\sk)]_{e,e^\prime} = [\widehat{S}_{\underline{n}_0}^{(0)}]_{e,e^\prime}.
\end{equation*}
Successively we can determine all coefficients of the series
\eqref{S:series} with $\underline{n}=\underline{n}_0$,
\begin{equation*}
\lim_{\Im\sk\rightarrow+\infty}
\sk^{m+1}\left(\e^{-\ii\sk\widetilde{w}_{e,e^\prime}(0)}
[S(\sk)]_{e,e^\prime}-\sum_{s=0}^m
[\widehat{S}_{\underline{n}_0}^{(s)}]_{e,e^\prime} \sk^{-s} \right) =
[\widehat{S}_{\underline{n}_0}^{(m+1)}]_{e,e^\prime},\qquad m\in\N_0.
\end{equation*}
Further, we observe that
\begin{equation*}
\lim_{\Im\sk\rightarrow +\infty} \sk^{-1} \log
\left([S(\sk)]_{e,e^\prime}-[\widehat{S}_{\underline{n}_0}(\sk)]_{e,e^\prime}\e^{\ii\sk
w_{e,e^\prime}(0)}\right) = \ii \widetilde{w}_{e,e^\prime}(1).
\end{equation*}
Since $\widetilde{\cL}_{e,e^\prime}$ is discrete, proceeding inductively as
above we can determine this set and all non-vanishing Fourier coefficients
in \eqref{6:11:Fourier}.
\end{proof}

The number $\omega_{e,e^\prime}(0)\in\cL_{e,e^\prime}$ is the metric length
of a geodesic walk from $e^\prime$ to $e$, that is a walk with the shortest
metric length. Under Assumption \ref{2cond} the geodesic walk is unique.
Obviously, $\omega_{e,e^\prime}(0)=0$ if and only if
$\partial(e)=\partial(e^\prime)$.

For any subsets $M_1$ and $M_2$ of $\R_{+}$ define $M_1+M_2$ to be the set
$\{a+b|\; a\in M_1,\;b\in M_2\} \subset \R_{+}$. With this notation we have
the following result:

\begin{lemma}\label{5:lem:1}
For any $e,e^{\prime}\in\cE$ the set $\cL_{e,e^{\prime}}$ is invariant under
the additive action of both $\cL_{e,e}$ and $\cL_{e^{\prime},e^{\prime}}$
in the sense that
\begin{equation}
\label{lincl}
\cL_{e,e^{\prime}}+\cL_{e,e}=\cL_{e,e^{\prime}}+\cL_{e^{\prime},e^{\prime}}=
\cL_{e,e^{\prime}}
\end{equation}
holds.
\end{lemma}

Equation \eqref{lincl} implies that for every $e\in\cE$ the set $\cL_{e,e}$
is an Abelian monoid (that is, a commutative semigroup with identity
$w_{e,e}(0)=0$) with respect to the arithmetic addition. Below this observation
will be used extensively.

Let $\fT(\bw)$ be the tour associated with a walk $\bw$. Given a tour $\fT$
we set
\begin{equation}\label{LT:def}
\cL_{e,e^{\prime}}^{\fT} :=
\left\{\langle\underline{n}(\bw),\underline{a}\rangle|\,
\bw\in\cW_{e,e^\prime}\quad\text{such that}\quad \fT(\bw)= \fT\right\}
\end{equation}
if $\fT$ is associated with at least one walk from $e^\prime$ to $e$ and
$\cL_{e,e^{\prime}}^{\fT}=\emptyset$ otherwise.

Obviously, $\cL_{e,e^{\prime}}^{\fT}=\cL_{e^{\prime},e}^{\fT}$. By
$w_{\fT}=\min\, \cL_{e,e^{\prime}}^{\fT}$ we denote the smallest number in
the set $\cL_{e,e^{\prime}}^{\fT}$. Clearly, $w_{\fT}$ is the metric length
of a shortest walk $\bw$ from $e^\prime\in\cE$ to $e\in\cE$ traversing each
$i\in\fT$ at least once,
\begin{equation*}
w_{\fT}=\sum_{i\in\fT} a_i n_i(\bw).
\end{equation*}
In general, even under Assumption \ref{2cond} such walk need not be unique.

Let $\cG_{\fT}$ be a compact subgraph of $\cG=(V,\cI,\cE,\partial)$
associated with the tour $\fT$, that is,
$\cG_{\fT}=(V_{\fT},\fT,\emptyset,\partial|_{\fT})$ with $V_{\fT}\subset V$
the set of vertices incident with the internal edges $i\in\fT$.

Given $\fT$ we consider the (additive) Abelian groups $\cC_1$ and $\cC_0$ generated by
the (oriented) 1-simplices $\sigma_i$, $i\in\fT$ and the 0-simplices
$\sigma_{v}$, $v\in V_{\fT}$, respectively, i.e.,
\begin{equation*}
\cC_1 = \left\{ \sum_{i\in\fT} n_i \sigma_i\Big|
n_i\in\Z\right\}\cong\Z^{|\fT|},\qquad \cC_0 = \left\{\sum_{v\in
\partial(\fT)} n_{v} \sigma_{v}\Big| n_{v}\in\Z\right\}\cong\Z^{|V_{\fT}|}.
\end{equation*}
Define the boundary operator $\delta_1: \cC_1 \rightarrow \cC_0$
\begin{equation*}
\delta_{1}:\quad c = \sum_{i\in\fT}n_{i} \sigma_{i}\quad
\longmapsto\quad\delta_{1} c=\sum_{i\in\fT}n_{i}
(\sigma_{\partial^{+}(i)}-\sigma_{\partial^{-}(i)}).
\end{equation*}
Obviously, $\delta_1$ is a group homomorphism and we have
\begin{equation}
\label{number} \delta_{1} c=0\quad\Longleftrightarrow\quad
\sum_{i:\;\partial^{+}(i)=v}n_{i}\,=\sum_{i:\;\partial^{-}(i)=v}n_{i}\quad\text{for
all}\quad v\in V_{\fT}.
\end{equation}

We extend the map $\cC_{1}\xrightarrow{\delta_{1}}\cC_{0}$ to a chain
complex
\begin{equation*}
0\longleftarrow\Z\mathop{\longleftarrow}\limits^{\delta_{0}}
\cC_{0}\mathop{\longleftarrow}^{\delta_{1}}\cC_{1}
\end{equation*}
with
\begin{equation*}
\delta_{0}c=\sum_{v\in V_{\fT}}n_{v}\quad\mbox{for}\quad c=\sum_{v\in
V_{\fT}}n_{v}\sigma_{v}\in\cC_0.
\end{equation*}
Obviously, $\delta_{0}\delta_{1}=0$. We call $\Ker \delta_1$ the first
homology group $H_1(\fT,\Z)$ of the graph $\cG_{\fT}$ and we set
$H_{0}(\fT,\Z)= \Ker\;\delta_{0}/ \Ran\;\delta_{1}$ to be the zeroth
homology group of $\cG_{\fT}$. The dimension of $H_1(\fT,\Z)$ is equal to
$|\fT|-|V_{\fT}|+1$ (see \cite[Section 3]{KS5}).

We say that the walks $\bw_1\in\cW_{e,e^\prime}$ and
$\bw_2\in\cW_{e,e^\prime}$ are equivalent ($\bw_1\sim\bw_2$) if
$\fT(\bw_1)=\fT(\bw_2)$. The factor set
$\cT_{e,e^\prime}=\cW_{e,e^\prime}/\sim$ is the set of all different tours
associated with the walks from $e^\prime\in\cE$ to $e\in\cE$. Obviously,
the number of such tours is bounded by $2^{|\cI|}+1$. If $\partial(e)=\partial(e^\prime)$,
then the empty tour $\fT^{(0)}$ corresponding to the trivial walk $\{e,e^\prime\}$
is contained in $\cT_{e,e}$.

\begin{lemma}\label{5:lem:3}
Assume that the metric graph $(\cG,\underline{a})$ satisfies conditions
(i), (ii) of Assumption \ref{con:graph} as well as Assumption \ref{2cond}.
If $\cI\neq\emptyset$, then for arbitrary $e,e^{\prime}\in\cE$ the set
$\cL_{e,e^{\prime}}$ has the following decomposition into disjoint subsets
\begin{equation}\label{ldecomp}
\cL_{e,e^{\prime}}=\bigsqcup_{\fT\in\cT_{e,e^{\prime}}}\cL_{e,e^{\prime}}^{\fT}.
\end{equation}
Every set $\cL_{e,e^{\prime}}^{\fT}$ has the form
\begin{equation}\label{lrep}
\begin{split}
\cL_{e,e^{\prime}}^{\fT} &= \Bigl\{w_{\fT} + 2\sum_{i\in\fT}n_{i}a_{i} +
\sum_{i\in\fT} |m_{i}|\,a_{i}\Big|\;n_{i}\in\N_0,\\ & \qquad\qquad\qquad
m_i\in\Z\quad\text{such that}\quad \sum_{i\in\fT} m_{i}\sigma_i \in
H_1(\fT,\Z)\Bigr\},
\end{split}
\end{equation}
where $w_{\fT}$ is the length of a shortest walk from $e^\prime$ to $e$
traversing each edge $i\in\fT$ at least once.
\end{lemma}

In particular, \eqref{lrep} implies that for any tour $\fT\in\cT_{e,e}$ the set
$\cL_{e,e}^{\fT}$ is a semigroup with respect to the arithmetic
addition.

\begin{proof}
The inclusion
\begin{equation*}
\begin{split}
\cL_{e,e^{\prime}}^{\fT} &\supset \Bigl\{w_{\fT} +
2\sum_{i\in\fT}n_{i}a_{i} + \sum_{i\in\fT}
|m_{i}|\,a_{i}\Big|\;n_{i}\in\N_0,\\ & \qquad\qquad\qquad
m_i\in\Z\quad\text{such that}\quad \sum_{i\in\fT} m_{i}\sigma_i \in
H_1(\fT,\Z)\Bigr\}
\end{split}
\end{equation*}
is obvious. To prove the opposite inclusion choose an arbitrary number $w
> w_{\fT}$ belonging to $\cL_{e,e^\prime}^{\fT}$. By \eqref{LT:def} there is
a walk
\begin{equation*}
\bw=\{e^\prime,i_1,\ldots,i_N,e\}
\end{equation*}
from $e^\prime\in\cE$ to $e\in\cE$ associated with the tour $\fT$ such that
$|\bw|=w$. Let $\widehat{\bw}=\{e^\prime,\widehat{i}_1,\ldots,
\widehat{i}_N,e\}$ be a shortest walk from $e^\prime\in\cE$ to $e\in\cE$
associated with the tour $\fT$, $|\widehat{\bw}| = w_{\fT}$. Obviously,
\begin{equation}\label{w:gleich:w}
w = w_{\fT} + \sum_{i\in\fT} n_i a_i
\end{equation}
with some $n_i\in\N_0$.

Consider the 1-chain associated with a walk $\bw$
\begin{equation*}
\sigma_{\bw} = \sum_{p=1}^{|\bw|_{\mathrm{comb}}} \sigma_{i_p} \sign(i_p),
\end{equation*}
where $\sign(i_p)=+1$ if the $p$-th element $v_p$ of the sequence
$\bc(\bw)=\{v_0,\ldots,v_p,\ldots,v_N\}$ is the terminal vertex of the
internal edge $i_p$ and $\sign(i_p)=-1$ otherwise. Similarly, consider the
1-chain
\begin{equation*}
\sigma_{\widehat{\bw}} = \sum_{p=1}^{|\widehat{\bw}|_{\mathrm{comb}}}
\sigma_{\widehat{i}_p} \sign(\widehat{i}_p)
\end{equation*}
associated with the shortest walk $\widehat{\bw}$. Consider the walk from
$e^\prime$ to $e^\prime$ which follows the walk $\bw$ from $e^\prime$ to
$e$ and then the reversed walk $\widehat{\bw}_{\mathrm{rev}}$ from $e$ to
$e^\prime$. Obviously, we have
\begin{equation}\label{sigma:gleich:sigma}
\sigma_{\bw} - \sigma_{\widehat{\bw}} \in H_1(\fT,\Z).
\end{equation}
Observe that if $\bw$ is such that the equality
$\sigma_{\bw}=\sigma_{\widehat{\bw}}$ holds, then the numbers $n_i$ in
\eqref{w:gleich:w} satisfy $n_i\in 2\N_0$ for all $i\in\fT$. This
observation combined with \eqref{sigma:gleich:sigma} completes the proof of
the equality \eqref{lrep}.

To prove \eqref{ldecomp} it suffices to show that
$\cL_{e,e^\prime}^{\fT_1}\cap \cL_{e,e^\prime}^{\fT_2}=\emptyset$ if
$\fT_1\neq\fT_2$. Assume that there exists a number $w\geq 0$ such that
$w\in\cL_{e,e^\prime}^{\fT_1}\cap \cL_{e,e^\prime}^{\fT_2}$. Thus,
\begin{equation*}
w = \sum_{i\in\cI} n_i^{(1)} a_i\qquad\text{and}\qquad w = \sum_{i\in\cI}
n_i^{(2)} a_i
\end{equation*}
for some $n_i^{(1)}\in\N_0$ and $n_i^{(2)}\in\N_0$. Thus, by Assumption
\ref{2cond}, we have $n_i^{(1)}=n_i^{(2)}$ for all $i\in\cI$. Noting that
\begin{equation*}
\fT_k=\{i\in\cI|\; n_i^{(k)}\neq 0\},\qquad k=1,2,
\end{equation*}
we obtain the equality $\fT_1=\fT_2$, a contradiction.
\end{proof}

Further we will be mainly interested in the case $e=e^\prime$ for an
arbitrary external edge $e\in\cE$. However, the case $e\neq e^\prime$ will
be of importance in Section \ref{sec:5}. Recall that the empty tour
$\fT^{(0)}$ (with $w_{\fT^{(0)}}=0$ ) is contained in $\cT_{e,e}$. Lemma
\ref{5:lem:3} implies that $w_{\fT_1}\neq w_{\fT_2}$ whenever
$\fT_1\neq\fT_2$. Therefore, $\cT_{e,e}$ is a totally ordered set with
respect to the order relation $\fT_1\vartriangleleft\fT_2$ defined by
$w_{\fT_1} \leq w_{\fT_2}$. Hence, the elements of the set $\cT_{e,e}$ can
be written in increasing order $\fT^{(0)}\vartriangleleft
\fT^{(1)}\vartriangleleft \ldots\vartriangleleft \fT^{(q)}$ such that
$w_{\fT^{(l)}}< w_{\fT^{(m)}}$ for any $1\leq l<m\leq q$, where
$q=|\cT_{e,e}|-1$.

Define an integer valued function $\eta: \{0,1,\ldots,q\}\rightarrow\N_0$
recursively by $\eta(0)=0$, $\eta(1)=1$, and
\begin{equation*}
\eta(k) := \begin{cases} \displaystyle\eta(k-1) & \text{if}\quad
\displaystyle\fT^{(k)} \subset
\bigcup_{l=1}^{k-1} \fT^{(l)},\\[3ex]
\eta(k-1)+1 & \text{if}\quad\displaystyle \fT^{(k)} \not\subset
\bigcup_{l=1}^{k-1} \fT^{(l)}
\end{cases} \qquad \text{for all} \quad 2\leq k \leq q.
\end{equation*}
For any $0\leq k\leq q$ we set
\begin{equation}\label{Ip:def}
\cI^{(\eta(k))} := \bigcup_{l=0}^{k} \fT^{(l)}
\end{equation}
such that $\cI^{(0)}=\emptyset$ and $\cI^{(\eta(q))}=\cI$. Further, for any
$p\in\{0,1,\ldots,\eta(q)\}$ we set
\begin{equation}\label{Leep}
\cL_{e,e}^{(p)} := \bigsqcup_{\substack{0\leq k\leq q\\ \eta(k)\geq
p}}\cL_{e,e}^{(\fT^{(k)})},
\end{equation}
where the sets $\cL_{e,e}^{(\fT^{(k)})}$ are defined by \eqref{lrep}. In
particular, $\cL_{e,e}^{(0)}=\cL_{e,e}$ and $\cL_{e,e}^{(\eta(q))}=\emptyset$.

Let $v_0 := \partial(e)$. For any $p\in\{0,1,\ldots,\eta(q)\}$ we set
\begin{equation}\label{Vp:def}
V^{(p)} := \{v_0\}\cup \bigcup_{i\in\cI^{(p)}}
\{\partial^-(i)\}\cup\{\partial^+(i)\}.
\end{equation}

By construction, the set $\cL_{e,e}^{(p)}$ consists of the metric lengths
of all walks on the graph $\cG$ from $e$ to $e$ traversing at least one
edge not belonging to $\cI^{(p)}$. The number
\begin{equation*}
w^{(p)} := \min \cL_{e,e}^{(p)}
\end{equation*}
is the metric length of a shortest walk from $e$ to $e$ on the graph $\cG$
traversing at least one edge not belonging to $\cI^{(p)}$. The next result
shows that this walk traverses \emph{precisely} one edge not belonging to
$\cI^{(p)}$, that is, $\cI^{(p+1)}\setminus \cI^{(p)}$ consists of
precisely one element. Along with Theorem \ref{eindeutigkeit} this constitutes the
second tool to determine the topology of the graph from its scattering
matrix.

\begin{theorem}\label{lem:7:7}
Under Assumptions \ref{con:graph} and \ref{2cond} the walk from $e$ to $e$
on the graph $\cG$ having the metric length $w^{(p)}$ traverses precisely
one internal edge $i\notin\cI^{(p)}$.
\end{theorem}

\begin{proof}
Let $\bw$ be a shortest walk (with the metric length $w_p$) from $e$ to $e$
traversing at least one edge not belonging to $\cI^{(p)}$. Assume to the
contrary that the walk $\bw$ traverses at least two different internal edges
not belonging to $\cI^{(p)}$, that is, there are
$\widehat{i}_1,\ldots,\widehat{i}_q\in\cI\setminus\cI^{(p)}$, $q\geq 2$,
such that $\widehat{i}_k\in\fT(\bw)$ for all $k\in\{1,\ldots,q\}$.
Obviously, $\bw$ visits $v_0=\partial(e)$ precisely twice.

\emph{Claim 1:} The set
\begin{equation*}
V^{(p)}\cap\left(\{\partial^-(\widehat{i}_1)\}\cup\ldots\cup\{\partial^-(\widehat{i}_q)\}\cup
\{\partial^+(\widehat{i}_1)\}\cup\ldots\cup\{\partial^+(\widehat{i}_q)\}\right)
\end{equation*}
consists of at least two elements. Assume to the contrary that this set
consists of a single vertex $\widehat{v}\in V^{(p)}$. Assume first that
$\widehat{v}\neq v_0$. Let
\begin{equation*}
\bw^\prime = \{e,i^{(1)},\ldots, i^{(r)}, i^{(r)}, \ldots, i^{(1)},e\},
\end{equation*}
where $i^{(1)},\ldots,i^{(r)}\in\cI^{(p)}$,
$\widehat{v}\in\partial(i^{(r)})$, be a shortest walk visiting the vertex
$\widehat{v}$. Obviously, $n_i(\bw)=n_i(\bw^\prime)$ for all $i\in\cI^{(p)}$.

If there is only one edge
$\widehat{i}\in\{\widehat{i}_1,\ldots,\widehat{i}_q\}$ incident with the
vertex $\widehat{v}$, then the walk $\bw$ traverses $\widehat{i}$ twice.
Therefore, the walk
\begin{equation}\label{w:prime:prime}
\bw^{\prime\prime} = \{e,i^{(1)},\ldots, i^{(r)},\widehat{i},\widehat{i},
i^{(r)}, \ldots, i^{(1)},e\},
\end{equation}
traverses precisely one edge $\widehat{i}\notin\cI^{(p)}$ and its metric
length is, obviously, smaller than $|\bw|$, which is a contradiction. Assume
that the edges $\widehat{i_1},\ldots,\widehat{i}_k\notin\cI^{(p)}$, $2\leq
k\leq q$ are incident with the vertex $\widehat{v}$. Let $\widehat{i_1}$ be
the edge with the shortest length among
$\widehat{i_1},\ldots,\widehat{i}_k$. Then,
\begin{equation*}
\begin{split}
|\bw| & \geq |\bw^\prime| + a_{\widehat{i}_1}+\ldots+a_{\widehat{i}_k}\\
& > |\bw^\prime| + k a_{\widehat{i}_1} \geq |\bw^\prime| + 2
a_{\widehat{i}_1}=|\bw^{\prime\prime}|.
\end{split}
\end{equation*}
Thus, the walk $\bw^{\prime\prime}$ traverses precisely one edge
$\widehat{i}_1\notin\cI^{(p)}$ and its length is smaller than $|\bw|$, again a
contradiction.

If $\widehat{v}=v_0$, the same arguments show that the walk $\{e,
\widehat{i}_1, \widehat{i}_1, e\}$ traverses precisely one edge
$\widehat{i}_1\notin\cI^{(p)}$ and its metric length is smaller than
$|\bw|$. This completes the proof of Claim 1.

\emph{Claim 2:}
$v_0\notin\{\partial^-(\widehat{i}_1)\}\cup\ldots\cup\{\partial^-(\widehat{i}_q)\}\cup
\{\partial^+(\widehat{i}_1)\}\cup\ldots\cup\{\partial^+(\widehat{i}_q)\}$.
Assume to the contrary that
$v_0\in\{\partial^-(\widehat{i}_1)\}\cup\ldots\cup\{\partial^-(\widehat{i}_q)\}\cup
\{\partial^+(\widehat{i}_1)\}\cup\ldots\cup\{\partial^+(\widehat{i}_q)\}$.
Since $\bw$ visits $v_0$ precisely twice, there are at most two different
edges (say $\widehat{i}_1$ and $\widehat{i}_2$) incident with $v_0$.
Without loss of generality we can assume that
$a_{\widehat{i}_1}<a_{\widehat{i}_2}$. The walk $\{e, \widehat{i}_1,
\widehat{i}_1, e\}$ traverses precisely one edge
$\widehat{i}_1\notin\cI^{(p)}$ and its metric length is smaller than
$|\bw|$,
\begin{equation*}
2 a_{\widehat{i}_1} < a_{\widehat{i}_1}+a_{\widehat{i}_2} \leq |\bw|.
\end{equation*}
Thus, $\widehat{i}_1$ is the only edge among
$\widehat{i}_1,\ldots,\widehat{i}_q$ incident with the vertex $v_0$.

By the Claim 1 there is a vertex $\widehat{v}\in V^{(p)}$, $\widehat{v}\neq
v_0$ such that
\begin{equation*}
\widehat{v} \in
V^{(p)}\cap\left(\{\partial^-(\widehat{i}_1)\}\cup\ldots\cup\{\partial^-(\widehat{i}_q)\}\cup
\{\partial^+(\widehat{i}_1)\}\cup\ldots\cup\{\partial^+(\widehat{i}_q)\}\right)
\end{equation*}
and either the walk $\bw$ or its reversed $\bw_{\mathrm{rev}}$ does not
leave $\cI^{(p)}$ after visiting the vertex $\widehat{v}$. Let
\begin{equation*}
\bw^\prime = \{e,i^{(1)},\ldots, i^{(r)}, i^{(r)}, \ldots, i^{(1)},e\},
\end{equation*}
where $i^{(1)},\ldots,i^{(r)}\in\cI^{(p)}$,
$\widehat{v}\in\partial(i^{(r)})$, be a shortest walk visiting the vertex
$\widehat{v}$. It is unique up to reversing. Obviously, after visiting the
vertex $\widehat{v}$ the remainder of the walk $\bw$ (or its reversed) agrees with the remainder $(i^{(r)},\ldots,i^{(r)},e)$ of the walk $\bw^\prime$.

Assume that $\partial(\widehat{i}_1)\bumpeq\{v_0,\widehat{v}\}$, that is,
either $\partial(\widehat{i}_1)=\{v_0,\widehat{v}\}$ or
$\partial(\widehat{i}_1)=\{\widehat{v}, v_0\}$. Then the walk
\begin{equation*}
\{e, i^{(1)},\ldots,i^{(r)},\widehat{i}_1,e\}
\end{equation*}
traverses precisely one edge $\widehat{i}_1\in\cI_p$ and its metric length
is smaller than that of the walk $\bw$. Thus,
$\partial(\widehat{i}_1)\not\bumpeq \{v_0, \widehat{v}\}$ and, hence, there
is an edge $\widehat{i}\in\{\widehat{i}_2,\ldots,\widehat{i}_q\}$ incident
with the vertex $\widehat{v}$.

Let $\widehat{i}\in\{\widehat{i}_2,\ldots,\widehat{i}_q\}$ be an edge
incident with the vertex $\widehat{v}$. The metric length of the walk $\bw$
is given by
\begin{equation*}
|\bw| = a_{\widehat{i}_1} + a_{\widehat{i}} + |\bw^\prime|/2 + \xi,
\end{equation*}
where $\xi$ is a nonnegative number. {}From the assumption $|\bw|<2
a_{\widehat{i}_1}$ it follows that
\begin{equation}\label{contra:1}
a_{\widehat{i}} + |\bw^\prime|/2 + \xi < a_{\widehat{i}_1}.
\end{equation}
On the other hand, $|\bw|<|\bw^{\prime\prime}|=|\bw^\prime| + 2
a_{\widehat{i}}$, where $\bw^{\prime\prime}$ is defined by
\eqref{w:prime:prime}. Thus,
\begin{equation*}
a_{\widehat{i}_1} +  \xi < a_{\widehat{i}} + |\bw^\prime|/2 \leq
a_{\widehat{i}} + |\bw^\prime|/2 + 2\xi,
\end{equation*}
which implies the inequality
\begin{equation*}
a_{\widehat{i}} + |\bw^\prime|/2 + \xi > a_{\widehat{i}_1}
\end{equation*}
contradicting to \eqref{contra:1} and, thus, completing the proof of Claim
2.

\emph{Claim 3:} The walk $\bw$ traverses precisely one edge
$\widehat{i}\notin\cI^{(p)}$. We continue to assume that $\bw$ visits at
least two edges not belonging to $\cI^{(p)}$. {}From the Claims 1 and 2 it
follows that there are two different vertices
$\widehat{v},{\widetilde{v}}\in V^{(p)}$ such that the walk $\bw$ does not
leave $\cI^{(p)}$ before visiting $\widehat{v}$ and after visiting
${\widetilde{v}}$. Let $\widehat{i}\notin\cI^{(p)}$ and
${\widetilde{i}}\notin\cI^{(p)}$ be edges incident with $\widehat{v}$ and
${\widetilde{v}}$, respectively. It is easy to show that
$\widehat{i}\neq{\widetilde{i}}$, $\partial(\widehat{i})\not\bumpeq
\{\widehat{v},{\widetilde{v}}\}$, and $\partial({\widetilde{i}})\not\bumpeq
\{\widehat{v},{\widetilde{v}}\}$.

Set
\begin{equation*}
\widehat{\bw} = \{e,i^{(1)},\ldots, i^{(r)}, i^{(r)}, \ldots, i^{(1)},e\}
\end{equation*}
and
\begin{equation*}
{\widetilde{\bw}} = \{e,i^{(r+1)},\ldots, i^{(r+s)}, i^{(r+s)}, \ldots,
i^{(r+1)},e\},
\end{equation*}
where $i^{(1)},\ldots,i^{(r+s)}\in\cI^{(p)}$,
$\widehat{v}\in\partial(i^{(r)})$, ${\widetilde{v}}\in\partial(i^{(r+s)})$,
be shortest walks visiting the vertices $\widehat{v}$ and ${\widetilde{v}}$,
respectively. Consider the walks
\begin{equation*}
{\bw}^\prime = \{e,i^{(1)},\ldots, i^{(r)}, \widehat{i}, \widehat{i},
i^{(r)}, \ldots, i^{(1)},e\},
\end{equation*}
and
\begin{equation*}
{\bw}^{\prime\prime} = \{e,i^{(r+1)},\ldots, i^{(r+s)}, {\widetilde{i}},
{\widetilde{i}}, i^{(r+s)}, \ldots, i^{(r+1)},e\}.
\end{equation*}

The metric length of the walk $\bw$ satisfies
\begin{equation*}
|\bw|\geq |\widehat{\bw}|/2 + |{\widetilde{\bw}}|/2 + a_{\widehat{i}} +
a_{{\widetilde{i}}}.
\end{equation*}
By Assumption \ref{2cond} either
\begin{equation*}
a_{\widehat{i}} + \sum_{k=1}^r a_{i^{(k)}} < a_{\widetilde{i}} +
\sum_{k=1}^s a_{i^{(r+k)}}
\end{equation*}
or
\begin{equation*}
a_{\widehat{i}} + \sum_{k=1}^r a_{i^{(k)}} > a_{\widetilde{i}} +
\sum_{k=1}^s a_{i^{(r+k)}}.
\end{equation*}
Thus, either the walk $\bw^\prime$ of length $\displaystyle
2(a_{\widehat{i}} + \sum_{k=1}^r a_{i^{(k)}})$, or the walk
$\bw^{\prime\prime}$ of length $\displaystyle 2(a_{\widetilde{i}} +
\sum_{k=1}^s a_{i^{(r+k)}})$ is shorter than the walk $\bw$. Since both
these walks traverse precisely one edge not belonging to $\cI_p$, we arrive
at a contradiction. The theorem is proven.
\end{proof}

As an immediate consequence of Theorem \ref{lem:7:7} we obtain the
following result.

\begin{corollary}\label{property:p}
Under Assumptions \ref{con:graph} and \ref{2cond} for any $v\in V^{(p)}$
there is a unique $k\in\{0,1,\ldots,p\}$ such that $w^{(k)}$ is the length
of the shortest walk on the graph $\cG$ from $e$ to $e$ visiting the vertex
$v$ precisely once. The number $w^{(k)}$ is necessarily of the form
\begin{equation*}
w^{(k)} = \sum_{i\in\cI} n_i a_i\qquad\text{with}\qquad n_i\in\{0,2\}.
\end{equation*}
\end{corollary}

For an arbitrary $e\in\cE$ let $\cG_{e}$ denote the graph obtained from
$\cG=(V,\cE,\cI,\partial)$ by removing all external edges $e^{\prime}\neq
e$, $\cG_{e}=(V,\cI,\{e\},\partial|_{\{e\}\cup\cI})$. So $\cG_{e}$ has only
one external edge $e$. The graphs $\cG_{e}$ and $\cG$ have the same
interior.

The strategy to prove the first part of Theorem \ref{1:theo:1} is to
construct the graph $\cG_{e}$. The following result combined with Theorem
\ref{eindeutigkeit} and Remark \ref{rem} implies the first part of Theorem
\ref{1:theo:1}.

\begin{theorem}\label{1:theo:1:equiv}
Assume that the metric graph $(\cG,\underline{a})$,
$\cG=(V,\cI,\cE,\partial)$ satisfies Assumptions \ref{con:graph} and
\ref{2cond}. Let $e\in\cE$ be arbitrary. The set $\cL_{e,e}$ determines the
graph $\cG_{\mathrm{int}}=(V,\cI,\emptyset,\partial|_{\cI})$, the vertex
$v_0=\partial(e)$ and the metric $\underline{a}$ uniquely.
\end{theorem}

We start the proof with the following simple observation: $\cL_{e,e}=\{0\}$
if and only if $\cG_e$ is a single vertex graph with no internal edges such
that $\cG_e=(\{v\},\emptyset,\{e\},\partial)$. Therefore, from now on we
assume that the graph $\cG_e$ contains at least one internal edge. Thus, by
Lemma \ref{lem:8.1} the set $\cL_{e,e}$ is infinite.

To prove Theorem \ref{1:theo:1:equiv} we construct a sequence of metric
graphs $(\cG_{e}^{(p)},\underline{a}^{(p)})$ with $\cG_{e}^{(p)} :=
(V^{(p)},\cI^{(p)},\{e\},\partial^{(p)})$ and
$\underline{a}^{(p)}=\{a_i\}_{i\in\cI^{(p)}}$, where
$\partial^{(p)}=\partial|_{\cI^{(p)}\cup\{e\}}$, and the sets $\cI^{(p)}$
and $V^{(p)}$ are given by \eqref{Ip:def} and \eqref{Vp:def}, respectively.
The set of lengths of all walks on the graph $\cG_{e}^{(p)}$ from $e$ to $e$
agrees with $\cL_{e,e}\setminus\cL_{e,e}^{(p)}$. By Theorem \ref{lem:7:7}
the graph $\cG_e^{(p+1)}$ can be constructed by adding an internal edge
$i_{p+1}$ to the graph $\cG_e^{(p)}$. By the discussion above,
$\cL_{e,e}^{(r)}=\emptyset$ for a sufficiently large $r\in\N$ and,
therefore, $\cG_e^{(r)}=\cG_e$ with $|\cI|=r$. Below we will prove that
this sequence can be constructed in a unique way.

\begin{figure}[htb]
\setlength{\unitlength}{1.5cm}
\begin{picture}(4,1.5)
\put(0,1){\line(1,0){2}} \put(1,0.6){$e$} \put(2,1){\circle*{0.1}}
\put(4,1){\circle*{0.1}} \put(2,1){\line(1,0){2}} \put(2.9,0.6){$i_1$}
\put(1.9, 1.2){$v_0$} \put(3.9, 1.2){$v_{1}$}
\end{picture}
\caption{\label{fig5} The graph $\cG_e^{(1)}$. It has one external edge $e$,
one boundary vertex $v$, one internal vertex $v_{1}$, and one internal edge
$i_1$ with length $a_{i_1}$.}
\end{figure}
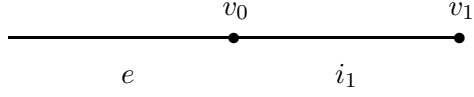

Before we turn to the general case, we will treat in detail the
construction of the graphs $\cG_e^{(1)}$ and $\cG_e^{(2)}$. Since
$V^{(0)}=\{v_0\}$, we have
$\cG_e^{(0)}=(\{v_0\},\emptyset,\{e\},\partial^{(0)})$ such that
$\partial^{(0)}(e)=v$. Now we construct
$\cG_e^{(1)}=(\{v_0,v_1\},\{i_1\},\{e\},\partial^{(1)})$ from
$\cG_e^{(0)}$. Here $\partial^{(1)}(e)=v_0$ and
$\partial^{(1)}(i_1)\bumpeq\{v_0,v_1\}$ (see Fig.~\ref{fig5}).
Since the shortest nontrivial walk from $e$ to $e$ on $\cG_e^{(1)}$ is
$\{e,i_1,i_1,e\}$ we necessarily have that the length of the internal edge
$i_1$ is equal to $a_{i_1}=w_{e,e}(1)/2>0$.

We turn to the construction of the graph $\cG_e^{(2)}$. By Lemma
\ref{5:lem:1} the set $\{2na_{i_1}\}_{n\in\N_0}$ is contained in
$\cL_{e,e}$. {}From \eqref{Leep} it follows that
\begin{equation}\label{k1}
\cL_{e,e}^{(1)}=\cL_{e,e}\setminus \{2na_{i_1}\}_{n\in\N_0}.
\end{equation}
If the set $\cL_{e,e}^{(1)}$ is empty, then we are finished:
$\cG_{e}=\cG_{e}^{(1)}$, $|V^{(1)}|=2$, and $|\cI^{(1)}|=1$.

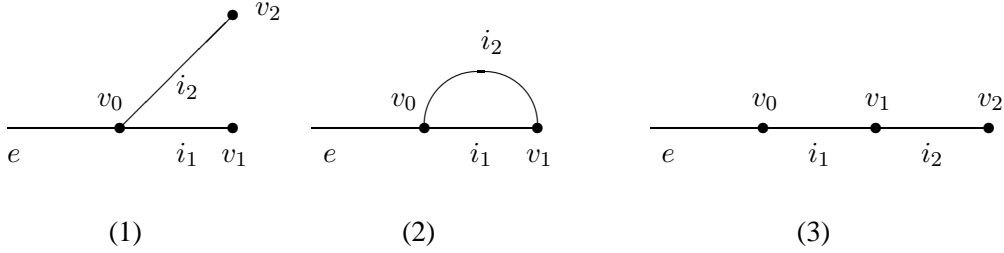
\begin{figure}[htb]
\setlength{\unitlength}{1.5cm}
\begin{picture}(9,2.5)
\put(0,1){\line(1,0){1}} \put(0,0.7){$e$} \put(1,1){\circle*{0.1}}

\put(0.8,1.2){$v_0$} \put(1.5,1.3){$i_{2}$} \put(1.5,0.7){$i_{1}$}
\put(1.9,0.7){$v_{1}$} \put(2,2){\circle*{0.1}} \put(2,1){\circle*{0.1}}
\put(1,1){\line(1,1){1}} \put(1,1){\line(1,0){1}} \put(2.2,2){$v_{2}$}

\put(0.9,0){(1)} \put(3.5,0){(2)} \put(7,0){(3)}

\put(2.7,1){\line(1,0){1}} \put(2.8,0.7){$e$} \put(3.7,1){\circle*{0.1}}

\put(3.4,1.2){$v_0$} \put(4.2,1.7){$i_{2}$} \put(4.1,0.7){$i_{1}$}
\put(4.6,0.7){$v_{1}$} \put(4.7,1){\circle*{0.1}}
\put(3.7,1){\line(1,0){1}} \put(4.2,1){\oval(1,1)[t]}

\put(5.7,1){\line(1,0){3}} \put(5.8,0.7){$e$} \put(6.7,1){\circle*{0.1}}
\put(7.7,1){\circle*{0.1}} \put(8.7,1){\circle*{0.1}} \put(6.6,1.2){$v_0$}
\put(7.6,1.2){$v_{1}$} \put(8.6,1.2){$v_{2}$} \put(7.1,0.7){$i_{1}$}
\put(8.1,0.7){$i_{2}$}
\end{picture}
\caption{\label{fig3} Three topologically different possibilities to
construct the graph $\cG_{e}^{(2)}$ from $\cG_{e}^{(1)}$.}
\end{figure}

Assume now that $\cL_{e,e}^{(1)}$ is nonempty and let $w^{(1)} > w_{e,e}(1)$
be the smallest number in $\cL_{e,e}^{(1)}$. By Lemma \ref{5:lem:3} from
Assumption \ref{2cond} it follows that $w^{(1)}$ and $w_{e,e}(1)$ are
rationally independent.

The graph $\cG_e^{(2)}$ will be constructed from $\cG_e^{(1)}$ by adding an
internal edge $i_2$. Ignoring the orientation of internal edges, there are
three topologically different possibilities to add this edge (see
Fig.~\ref{fig3}):
\begin{enumerate}
\item[(1)]{The edge $i_2$ has $v_0$ as the initial vertex. Its terminal vertex $v_2$ coincides neither
with $v_0$ nor with $v_1$. Thus, $V^{(2)}=\{v_0,v_1,v_2\}$.}
\item[(2)]{The edge $i_2$ has $v_0$ as the initial vertex and $v_{1}$ as the terminal one. Thus,
$V(\cG_e^{(2)})=\{v_0,v_1\}$.}
\item[(3)]{The edge $i_2$ has $v_1$ as the initial vertex. Its terminal vertex $v_2$ coincides neither
with $v_0$ nor with $v_1$. Thus, $V^{(2)}=\{v_0,v_1,v_2\}$.}
\end{enumerate}
There is only one choice of the length $a_{i_2}$ of the internal edge $i_2$
guaranteeing that the shortest walk from $e$ to $e$ traversing the edge
$i_2$ has the metric length $w^{(1)}$:
\renewcommand{\descriptionlabel}[1]%
{\hspace{\labelsep}\textrm{#1}}
\begin{description}
\item[Case (1):]{\quad $a_{i_2} = w^{(1)}/2$,}
\item[Case (2):]{\quad $a_{i_2} = w^{(1)}-a_{i_1}$,}
\item[Case (3):]{\quad $a_{i_2} = w^{(1)}/2 - a_{i_1}$.}
\end{description}
Note that in all three cases the lengths $a_{i_1}$ and $a_{i_2}$ of the
internal edges $i_2$ and $i_1$ are not only different but also rationally
independent.

Observe that the structure of the set $\cL_{e,e}^{(1)}$ uniquely fixes the
possibility to add the internal edge $i_2$ to the graph $\cG_e^{(1)}$.
Indeed, with the integer-valued function
\begin{equation}\label{num}
\nu(m):=\min\left\{n\in\Z\,\big|\, n a_{i_1} + m
w^{(1)}\in\cL_{e,e}^{(1)}\right\}, \qquad m\in\N.
\end{equation}
we have the following proposition.

\begin{proposition}
$\nu(m)=0$ for all $m\in\N$ if and only if the Case (1) holds.

$\displaystyle\nu(m)=\begin{cases} -m & \text{for all}\quad m\in 2\N,\\
1-m & \text{for all}\quad m\in 2\N - 1
\end{cases}$ \quad if and only if the Case (2) holds.

$\nu(m)= 2(1-m)$\,\, for all $m\in\N$ if and only if the Case (3) holds.
\end{proposition}

\begin{proof}
The ``if'' parts of all statements can be verified by direct calculations.
This immediately implies that the ``only if'' part of each statement holds.
\end{proof}

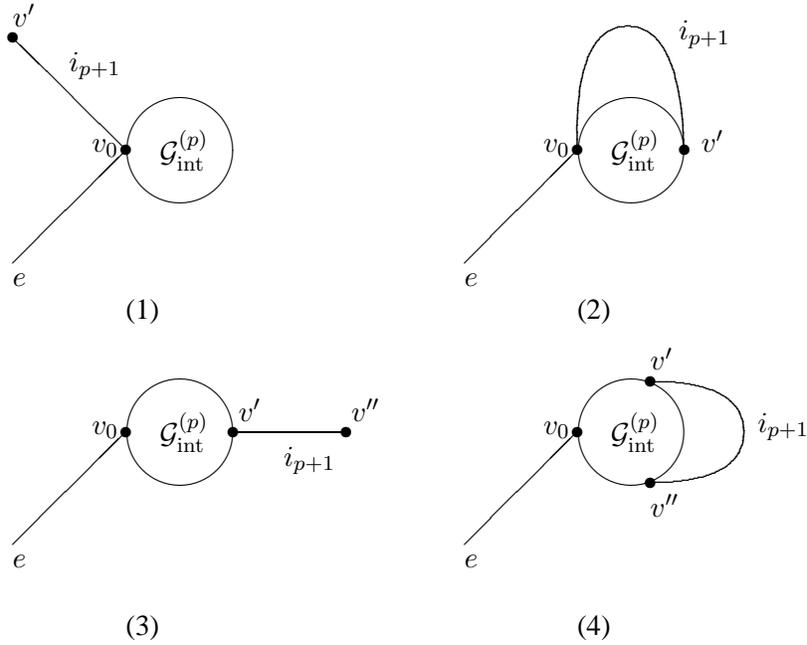
\begin{figure}[htb]
\setlength{\unitlength}{1.5cm}
\begin{picture}(9,6)
\put(1,3.5){\line(1,1){1}} \put(1,3.3){$e$} \put(2,4.5){\circle*{0.1}}

\put(2.48,4.5){\circle{3}} \put(2.3,4.4){$\cG_{\mathrm{int}}^{(p)}$}
\put(1.7,4.48){$v_0$}

\put(2,4.5){\line(-1,1){1}} \put(1,5.6){$v^\prime$} \put(1.5,5.2){$i_{p+1}$}
\put(1,5.5){\circle*{0.1}}

\put(2,3.0){(1)}


\put(5,3.5){\line(1,1){1}} \put(5,3.3){$e$} \put(6,4.5){\circle*{0.1}}

\put(6.48,4.5){\circle{3}} \put(6.3,4.4){$\cG_{\mathrm{int}}^{(p)}$}
\put(5.7,4.48){$v_0$} \put(7.1,4.48){$v^\prime$}

\put(6.95,4.5){\circle*{0.1}}

\tagcurve(6.48,4.5,6,4.5,6.45,5.6,6.95,4.5,6.48,4.5)


\put(6.9,5.48){$i_{p+1}$} \put(6,3.0){(2)}


\put(1,1.0){\line(1,1){1}} \put(1,0.8){$e$} \put(2,2.0){\circle*{0.1}}

\put(2.48,2.0){\circle{3}} \put(2.3,1.9){$\cG_{\mathrm{int}}^{(p)}$}
\put(1.7,1.98){$v_0$} \put(3.0,2.1){$v^\prime$}

\put(2.95,2.0){\circle*{0.1}} \put(2.95,2.0){\line(1,0){1}}
\put(3.95,2.0){\circle*{0.1}} \put(4.0,2.1){$v^{\prime\prime}$}
\put(3.4,1.7){$i_{p+1}$} \put(2,0.2){(3)}


\put(5,1.0){\line(1,1){1}} \put(5,0.8){$e$} \put(6,2.0){\circle*{0.1}}

\put(6.48,2.0){\circle{3}} \put(6.3,1.9){$\cG_{\mathrm{int}}^{(p)}$}
\put(5.7,1.98){$v_0$} \put(6.65,2.55){$v^\prime$}
\put(6.65,1.25){$v^{\prime\prime}$}

\put(6.65,2.45){\circle*{0.1}}

\put(6.65,1.55){\circle*{0.1}}

\tagcurve(6.48,2.0,6.65,2.45,7.48,2.0,6.65,1.55,6.48,2.0)
\put(7.6,2.0){$i_{p+1}$} \put(6,0.2){(4)}
\end{picture}
\caption{\label{fig3:bis} Four topologically different possibilities to
construct the graph $\cG_{e}^{(p+1)}$ from $\cG_{e}^{(p)}$. The Case (4)
may occur only if $|V^{(p)}|\geq 3$.}
\end{figure}

The construction of the graph $\cG_{e}^{(p+1)}$ from $\cG_{e}^{(p)}$ for
general $p\in\N$ is similar to the case $p=1$ already discussed. By Theorem
\ref{lem:7:7} the graph $\cG_{e}^{(p+1)}$ can be constructed by adding an
internal edge $i_{p+1}$ to the graph $\cG_{e}^{(p)}$. Ignoring the
orientation of internal edges, there are four topologically different
possibilities to add this edge (see Fig.~\ref{fig3:bis}):
\begin{enumerate}
\item[(1)]{The edge $i_{p+1}$ has $v_0$ as the initial vertex. Its terminal
vertex $v^\prime$ does not belong to $V^{(p)}$. Thus,
$V^{(p+1)}=V^{(p)}\cup\{v^\prime\}$.}
\item[(2)]{The edge $i_{p+1}$ has $v_0$ as the initial vertex. Its terminal vertex $v^\prime\neq v_0$
belongs to $V^{(p)}$. Thus, $V^{(p+1)} = V^{(p)}$.}
\item[(3)]{The edge $i_{p+1}$ has $v^\prime\in V^{(p)}$, $v^\prime\neq v_0$ as the
initial vertex. Its terminal vertex $v^{\prime\prime}$ does not belong to
$V^{(p)}$. Thus, $V^{(p+1)}=V^{(p)}\cup\{v^{\prime\prime}\}$.}
\item[(4)]{The edge $i_{p+1}$ has $v^\prime\in V^{(p)}$, $v^\prime\neq v_0$ as the
initial vertex. Its terminal vertex $v^{\prime\prime}\neq v^\prime$, $v^{\prime\prime}\neq v_0$ belongs
to $V^{(p)}$. Thus, $V^{(p+1)}=V^{(p)}$.}
\end{enumerate}

For every $m\in\N$ we define
\begin{equation*}
\begin{split}
\nu^{(p)}(m) & :=\min\Big\{n\in\Z\,\big|\, \exists\,
w\in\cL_{e,e}\setminus\cL_{e,e}^{(p)}\quad \text{such that}\quad m w^{(p)} +
n w/2\in\cL_{e,e}^{(p)}\Big\}.
\end{split}
\end{equation*}
For $p=1$ this definition agrees with \eqref{num}.

\begin{lemma}\label{lem:7:8}
Assume that the Case (2) holds. Then the shortest walk from $e$ to $e$
traversing the edge $i_{p+1}$ also traverses at least one edge
$i\in\cI^{(p)}$.
\end{lemma}

\begin{proof}
Assume to the contrary that the shortest walk from $e$ to $e$ traversing the
edge $i_{p+1}$ traverses none of the edges $i\in\cI^{(p)}$. By Lemma
\ref{lem:7:7} it is of the form
\begin{equation*}
\bw = \{e,i_{p+1},i_{p+1},e\}.
\end{equation*}
Let $v^\prime\neq v_0$ be the other vertex incident with $i_{p+1}$. By Corollary \ref{property:p} there is a walk
\begin{equation*}
\bw_0=\{e,j_1,\ldots,j_k,j_k,\ldots,j_1,e\},\qquad v^\prime\in\partial(j_k)
\end{equation*}
from $e$ to $e$ visiting the vertex $v^\prime$ with $j_1,\ldots,j_k\in\cI^{(p)}$. Its metric length is strictly smaller than $|\bw|=2 a_{i_{p+1}}$. But then the walk
\begin{equation*}
\bw^\prime = \{e,j_1,\ldots,j_k,i_{p+1},e\}
\end{equation*}
traversing the edge $i_{p+1}$ also traverses at least one edge in $\cI^{(p)}$. Its metric length satisfies the inequality
\begin{equation*}
|\bw^\prime| = \frac{1}{2}(|\bw|+|\bw_0|) < |\bw_0|,
\end{equation*}
which is a contradiction.
\end{proof}

\begin{lemma}\label{lem:case4}
Assume that the Case (4) holds. Let $\bw$ be the shortest walk from $e$ to
$e$ on the graph $\cG$ traversing the edge $i_{p+1}$ at least once. Let
$\bw^\prime$ be the shortest walk from $e$ to $e$ on the graph
$\cG_e^{(p)}$ visiting the vertex $v^\prime$ and $\bw^{\prime\prime}$ be the
shortest walk from $e$ to $e$ on the graph $\cG_e^{(p)}$ visiting the
vertex $v^{\prime\prime}$. Then
\begin{equation}\label{fT:cup}
\fT(\bw) = \fT(\bw^\prime)\cup \fT(\bw^{\prime\prime}) \cup\{i_{p+1}\}.
\end{equation}
\end{lemma}

\begin{proof}
A priori there are three possibilities:
\begin{equation*}
\begin{split}
(\alpha) &\qquad |\bw^\prime|/2 > |\bw^{\prime\prime}|/2 + a_{i_{p+1}},\\
(\beta) &\qquad |\bw^\prime|/2 = |\bw^{\prime\prime}|/2 + a_{i_{p+1}},\\
(\gamma) &\qquad |\bw^\prime|/2 < |\bw^{\prime\prime}|/2 + a_{i_{p+1}}.
\end{split}
\end{equation*}
Without loss of generality, by Assumption \ref{2cond} we may assume $|\bw^\prime|<|\bw^{\prime\prime}|$. Then,
inequality $(\alpha)$ contradicts Corollary \ref{property:p} for the vertex
$v^\prime$. Equality $(\beta)$ contradicts Assumption \ref{2cond}. Thus,
inequality $(\gamma)$ holds, which implies \eqref{fT:cup}.
\end{proof}

\begin{remark}\label{rem:7:12}
Lemma \ref{lem:case4} does not exclude the possibilities
$\fT(\bw^\prime)\subset\fT(\bw^{\prime\prime})$ and
$\fT(\bw^{\prime\prime})\subset\fT(\bw^\prime)$.
\end{remark}

\begin{proposition}\label{propo:cases}
(i) The Case (1) holds if and only if $\nu^{(p)}(m)=0$ for all $m\in\N$.

(ii) The Case (2) holds if and only if the following two conditions hold
simultaneously
\begin{equation}\label{Condition:1}
\displaystyle\nu^{(p)}(m)=\begin{cases} -m & \text{for all}\quad m\in 2\N,\\
1-m & \text{for all}\quad m\in 2\N - 1,
\end{cases}
\end{equation}
and
\begin{equation}\label{Condition:2}
\left\{\begin{aligned} & \text{For every}\quad w=\sum_{i\in\cI} n_i
a_i\in\cL_{e,e}\setminus\cL_{e,e}^{(p)}\quad\text{with}
\quad n_{i}\in\{0,2\}\quad \text{satisfying}\\
& m w^{(p)} + \nu^{(p)}(m) w/2 \in \cL_{e,e}^{(p)}\quad\text{for all}\quad
m\in\N\\ & \text{the equality}\quad w = w^{(k)}\quad \text{holds for
some}\quad k\in\{1,\ldots,p-1\}.
\end{aligned}\right.
\end{equation}

(iii) The Case (3) holds if and only if the following two conditions hold
simultaneously
\begin{equation}\label{Condition:3}
\nu^{(p)}(m)= 2(1-m)\quad \text{for all}\quad m\in\N
\end{equation}
and
\begin{equation}\label{Condition:4}
\left\{\begin{aligned}
& \text{For every}\quad w=\sum_{i\in\cI} n_i a_i\in\cL_{e,e}\setminus\cL_{e,e}^{(p)}\quad \text{satisfying}\\
& m w^{(p)} + \nu^{(p)}(m) w/2 \in \cL_{e,e}^{(p)}\quad\text{for all}\quad
m\in\N\\
& \text{the number}\quad w^\prime := \sum_{i\in\cI} n_i^\prime
a_i\quad\text{with}\quad n^\prime_i=\begin{cases} 2 & \text{if}\quad
n_i\neq 0\\ 0 & \text{if}\quad n_i=0
\end{cases}\\ & \text{satisfies}\quad m w^{(p)} + \nu^{(p)}(m) w^\prime/2 \in \cL_{e,e}^{(p)}\quad\text{for all}\quad
m\in\N.
\end{aligned}\right.
\end{equation}
\end{proposition}

Note that if the Case (4) holds, then either
\begin{equation*}
\displaystyle\nu^{(p)}(m)=\begin{cases} -m & \text{for all}\quad m\in 2\N\\
1-m & \text{for all}\quad m\in 2\N - 1
\end{cases}
\end{equation*}
or $\nu^{(p)}(m)= 2(1-m)$ for all $m\in\N$ depending on whether the
shortest walk traversing the edge $i_{p+1}\notin\cI^{(p)}$ traverses any
edge $i\in\cI^{(p)}$ at most once or there is an edge $i^\prime\in\cI^{(p)}$ traversed by this walk twice.

\begin{proof}[Proof of Proposition \ref{propo:cases}]
Observe that by Theorem \ref{lem:7:7} the number $w^{(p)}$ is the length of
the shortest walk from $e$ to $e$ traversing the edge $i_{p+1}$ at least
once and not traversing any edge $i\in\cI\setminus(\cI^{(p)}\cup\{i_{p+1}\})$.
With this observation the ``only if'' parts of all three statements can be
verified by direct calculations.

To prove the ``if'' part of the claim (i) we assume that $\nu^{(p)}(m)=0$
for all $m\in\N$. Then
\begin{equation}\label{assu}
\begin{split}
& \text{$m w^{(p)} - n w/2 \notin\cL_{e,e}^{(p)}$ for any
$w\in\cL_{e,e}\setminus\cL_{e,e}^{(p)}$ and for all $m,n\in\N$.}
\end{split}
\end{equation}

Assume that the Case (2) holds. Let $\bw$ be a shortest walk from $e$ to
$e$ on the graph $\cG_e^{(p)}$ visiting the vertex $v^\prime$. Therefore,
$|\bw|\in\cL_{e,e}\setminus\cL_{e,e}^{(p)}$. Then for arbitrary even $m\in\N$
\begin{equation*}
m w^{(p)} - \frac{m}{2} |\bw| \in \cL_{e,e}^{(p)},
\end{equation*}
which contradicts \eqref{assu}.

Assume that the Case (3) holds. Let $\bw$ be a shortest walk from $e$ to
$e$ on the graph $\cG_e^{(p)}$ visiting the vertex $v^\prime$. Then for any
$m\in\N$
\begin{equation*}
m w^{(p)} - (m-1) |\bw| \in \cL_{e,e}^{(p)},
\end{equation*}
which again contradicts \eqref{assu}.

Assume next that the Case (4) holds. Let $\bw$ be a shortest walk from $e$
to $e$ on the graph $\cG$ traversing the edge $i_{p+1}$ at least once. Let
$\bw^\prime$ be the shortest walk from $e$ to $e$ on the graph
$\cG_e^{(p)}$ visiting the vertex $v^\prime$ and let $\bw^{\prime\prime}$ be
the shortest walk from $e$ to $e$ on the graph $\cG_e^{(p)}$ visiting the
vertex $v^{\prime\prime}$. Observe that
\begin{equation*}
|\bw^\prime| + |\bw^{\prime\prime}|\in\cL_{e,e}\setminus\cL_{e,e}^{(p)}.
\end{equation*}
By Lemma \ref{lem:case4}
\begin{equation*}
\fT(\bw) = \fT(\bw^\prime)\cup \fT(\bw^{\prime\prime}) \cup\{i_{p+1}\}.
\end{equation*}
Therefore, for any odd $m>1$,
\begin{equation*}
m w^{(p)} - \frac{m-1}{2} (|\bw^\prime| + |\bw^{\prime\prime}|) \in
\cL_{e,e}^{(p)},
\end{equation*}
which once more contradicts \eqref{assu}. Thus, the claim (i) is proven.

We turn to the proof of the ``if'' part of the claim (ii). By (i) and by the
``only if'' part of (iii) the condition \eqref{Condition:1} implies that
either the Case (2) or the Case (4) holds. Assume that the Case (4) holds.
Let $v^\prime,v^{\prime\prime}\in V(\cG_e^{(p)})$, $v^\prime\neq
v^{\prime\prime}$ be vertices incident with the edge $i_{p+1}$. Let
$\bw^\prime$ and $\bw^{\prime\prime}$ be the shortest walks on the graph
$\cG_e^{(p)}$ from $e$ to $e$ visiting the vertices $v^\prime$ and
$v^{\prime\prime}$, respectively. There are no edges in $\cI^{(p)}$
traversed by both walks $\bw^\prime$ and $\bw^{\prime\prime}$. Indeed,
assume to the contrary that there is an edge $\widetilde{i}\in\cI^{(p)}$
traversed by both walks. Then, the shortest walk from $e$ to $e$ on the
graph $\cG_e^{(p+1)}$ traversing the edge $i_{p+1}$ traverses the edge
$\widetilde{i}$ twice. Therefore, $\nu^{(p)}(m)=2(1-m)$ for all $m\in\N$,
which contradicts \eqref{Condition:1}. Further, the scores of the walks
$\bw^\prime$ and $\bw^{\prime\prime}$ contain only $0$'s and $2$'s.

With these observations we obtain that the shortest walk from $e$ to $e$
traversing the edge $i_{p+1}$ (and, thus, visiting the vertices $v^\prime$
and $v^{\prime\prime}$) traverses every edge $i\in\cI^{(p)}$ at most once.
By Lemma \ref{lem:case4} the
number $w = |\bw^\prime| + |\bw^{\prime\prime}| \in
\cL_{e,e}\setminus\cL_{e,e}^{(p)}$ satisfies
\begin{equation*}
m w^{(p)} + \nu^{(p)}(m) w/2 \in \cL_{e,e}^{(p)}.
\end{equation*}
in particular for $m=3$ with $\nu^{(p)}(m) = -2$. By Corollary \ref{property:p} the equalities $|\bw^\prime|=w^{(k_1)}$ and
$|\bw^\prime|=w^{(k_2)}$ hold for some $k_1,k_2\in\{1,\ldots,p\}$. Therefore, $w
\neq w^{(k)}$ for all
$k\in\{1,\ldots,p\}$. Thus, \eqref{Condition:2} does not hold, which is a
contradiction. This proves the ``if'' part of the claim (ii).

Now we turn to the proof of the ``if'' part of the claim (iii). By what we
already proved only the Cases (3) or (4) may occur. Assume that the Case
(4) holds. Let $\bw$ be a shortest walk from $e$ to $e$ on the graph $\cG$
traversing the edge $i_{p+1}$ at least once. Recall that by Theorem \ref{lem:7:7} this walk does not traverse any edge $i\in\cI\setminus(\cI^{(p)}\cup\{i_{p+1}\})$. Let
$v^\prime,v^{\prime\prime}\in V(\cG_e^{(p)})$, $v^\prime\neq
v^{\prime\prime}$ be vertices incident with the edge $i_{p+1}$. Let
$\bw^\prime$ and $\bw^{\prime\prime}$ be the shortest walks on the graph
$\cG_e^{(p)}$ from $e$ to $e$ visiting the vertices $v^\prime$ and
$v^{\prime\prime}$, respectively. Recall (see Remark \ref{rem:7:12}) that it is possible that either
$\fT(\bw^\prime)\subset\fT(\bw^{\prime\prime})$ or
$\fT(\bw^{\prime\prime})\subset\fT(\bw^\prime)$.

The number $w=|\bw^\prime|+|\bw^{\prime\prime}|$ satisfies
\begin{equation*}
m w^{(p)} + \nu^{(p)}(m) w/2 \in\cL_{e,e}^{(p)}\quad\text{for all}\quad
m\in\N.
\end{equation*}
Let
\begin{equation*}
n_i^\prime := \begin{cases} 2 & \text{if}\quad n_i(\bw)\neq 0,\\ 0 &
\text{if}\quad n_i(\bw)= 0.
\end{cases}
\end{equation*}
Obviously, $\{n_i^\prime\}_{i\in\cI}$ is the score of a walk
on graph $\cG_e^{(p)}$ visiting both vertices $v^\prime$ and
$v^{\prime\prime}$.

Choose $m=3$ for which $\nu^{(p)}(m)=-4$. The walk $\bw$ traverses at least
one edge $\widehat{i}\in\cI^{(p)}$ precisely once and at least one edge in
$\cI^{(p)}$ twice. Let
\begin{equation*}
w^\prime := \sum_{i\in\cI} n_i^\prime a_i.
\end{equation*}
Therefore,
\begin{equation*}
m w^{(p)} + \nu^{(p)}(m) w^\prime/2 = \sum_{i\in\cI} \widetilde{n}_i a_i
\end{equation*}
with $\widetilde{n}_{\widehat{i}}=-1$. Hence, the condition
\begin{equation*}
m w^{(p)} + \nu^{(p)}(m) w^\prime/2 \in \cL^{(p)}_{e,e}
\end{equation*}
is not satisfied for $m=3$, which contradicts \eqref{Condition:4}.
\end{proof}

The following proposition ensures the uniqueness of the graph
$\cG_e^{(p+1)}$.

\begin{proposition}\label{lem:7:12}
In Cases (2), (3), and (4) there is a unique possibility to construct the
graph $\cG_e^{(p+1)} = (V^{(p+1)},\cI^{(p+1)},\{e\},\partial^{(p+1)})$ from
$\cG_e^{(p)}= (V^{(p)},\cI^{(p)},\{e\},\partial^{(p)})$. More precisely:

(i) Assume that the Case (2) or (3) holds. Then there is a unique vertex
$v^\prime\in V^{(p)}$, $v^\prime\neq v_0$ such that
\begin{itemize}
\item[(a)]{the set of lengths of
all walks from $e$ to $e$ on the graph $\cG_e^{(p+1)}$ obtained from
$\cG_e^{(p)}$ by adding the edge $i_{p+1}$ with
$v^\prime\in\partial(i_{p+1})$, is contained in $\cL_{e,e}$,}
\item[(b)]{the metric graph $\cG_e^{(p+1)}$ satisfies Assumptions \ref{con:graph} and \ref{2cond}.}
\end{itemize}
The metric length of the edge $i_{p+1}$ is given by
\begin{equation*}
a_{i_{p+1}} =
\begin{cases}
w^{(p)} - |\bw^\prime|/2 & \text{if the Case (2) holds},\\
(w^{(p)} - |\bw^\prime|)/2 & \text{if the Case (3) holds},
\end{cases}
\end{equation*}
where ${\bw}^\prime$ is a shortest walk on the graph $\cG_e^{(p)}$ from
$e$ to $e$ visiting the vertex $v^\prime$ precisely once.

(ii) Assume that the Case (4) holds. Then there is a unique pair of
vertices $v^\prime\neq v^{\prime\prime}\in V^{(p)}$ such that
\begin{itemize}
\item[(a)]{the set of lengths of all walks from $e$ to $e$ on the graph
$\cG_e^{(p+1)}$ obtained from $\cG_e^{(p)}$ by adding the edge $i_{p+1}$
with $\partial(i_{p+1})\bumpeq\{v^\prime, v^{\prime\prime}\}$, is contained
in $\cL_{e,e}$,}
\item[(b)]{the metric graph $\cG_e^{(p+1)}$ satisfies Assumptions \ref{con:graph} and \ref{2cond}.}
\end{itemize}
The metric length of the edge $i_{p+1}$ is given by
\begin{equation*}
a_{i_{p+1}} = w^{(p)} - (|\bw^\prime|+|\bw^{\prime\prime}|)/2,
\end{equation*}
where ${\bw}^\prime$ and $\bw^{\prime\prime}$ are the shortest walks on the
graph $\cG_e^{(p)}$ from $e$ to $e$ visiting the vertices $v^\prime$ and
$v^{\prime\prime}$, respectively.
\end{proposition}

\begin{proof}
The proofs in all three cases are rather similar. Therefore, we will deal with
the Case (4) only.

Let $v^\prime\neq v^{\prime\prime}$ and $\widetilde{v}^\prime\neq
\widetilde{v}^{\prime\prime}$ belong to $V^{(p)}$. Let $\bw^\prime$,
$\bw^{\prime\prime}$, $\widetilde{\bw}^\prime$, and
$\widetilde{\bw}^{\prime\prime}$ be the shortest walks from $e$ to $e$
visiting the vertices $v^\prime$, $v^{\prime\prime}$,
$\widetilde{v}^\prime$, and $\widetilde{v}^{\prime\prime}$, respectively.
Set
\begin{equation*}
a_{i_{p+1}} = w^{(p)} -
(|\bw^\prime|+|\bw^{\prime\prime}|)/2\quad\text{and}\quad
a_{\widetilde{i}_{p+1}} = w^{(p)} -
(|\widetilde{\bw}^\prime|+|\widetilde{\bw}^{\prime\prime}|)/2.
\end{equation*}
Assume that at least one of the inequalities
$v^\prime\neq\widetilde{v}^\prime$ and
$v^{\prime\prime}\neq\widetilde{v}^{\prime\prime}$ holds. Then
$a_{i_{p+1}}\neq {a}_{\widetilde{i}_{p+1}}$. Without loss of generality we
can assume that ${a}_{\widetilde{i}_{p+1}} < a_{i_{p+1}}$.

Let $\cG_e^{(p+1)}$ be obtained from $\cG_e^{(p)}$ by adding the edge
$i_{p+1}$ such that $\partial(i_{p+1})\bumpeq\{v^\prime, v^{\prime\prime}\}$
and correspondingly the graph $\widetilde{\cG}_e^{(p+1)}$ -- by adding the
edge $\widetilde{i}_{p+1}$ such that
$\partial(\widetilde{i}_{p+1})\bumpeq\{\widetilde{v}^\prime,
\widetilde{v}^{\prime\prime}\}$. Let $\ell$ be the set of lengths of all
walks from $e$ to $e$ on the graph $\cG_e^{(p+1)}$, and correspondingly
$\widetilde{\ell}$ -- on the graph $\widetilde{\cG}_e^{(p+1)}$.

Assume that both sets $\ell$ and $\widetilde{\ell}$ are contained in
$\cL_{e,e}$. In particular, we have that $2
w^{(p)}\in\ell\cap\widetilde{\ell}$, which implies
\begin{equation*}
a_{i_{p+1}} + {a}_{\widetilde{i}_{p+1}} +
(|\bw^\prime|+|\bw^{\prime\prime}|)/2 +
(|\widetilde{\bw}^\prime|+|\widetilde{\bw}^{\prime\prime}|)/2 \in\ell.
\end{equation*}
Since $\widetilde{i}_{p+1}\notin\cI^{(p+1)}$, from this equation it follows
that
\begin{equation*}
a_{\widetilde{i}_{p+1}} = \sum_{i\in\cI^{(p+1)}} n_i a_i
\end{equation*}
with some $n_i\in\N_0$. Since by assumption $a_{\widetilde{i}_{p+1}} <
a_{i_{p+1}}$, the coefficient $n_{i_{p+1}}$ has to be zero. Thus, the set
$\{a_{i_1},\ldots,a_{i_p},a_{\widetilde{i}_{p+1}}\}$ is not rationally
independent. Hence, the metric graph $\widetilde{\cG}_e^{(p+1)}$ does not
satisfy Assumption \ref{2cond}.
\end{proof}

Now we proceed by induction. By Lemma \ref{5:lem:3} after a finite number
of induction steps we get $\cL_{e,e}^{(r)}=\emptyset$ and, thus,
$\cG_{e}^{(r)}=\cG_e$. This completes the proof of Theorem
\ref{1:theo:1:equiv}.

\begin{proof}[Proof of the second part of Theorem \ref{1:theo:1}]
It suffices to show that whenever the interior $\cG_{\mathrm{int}}$ of the
graph $\cG$, the vertex $\partial(e)$, and the metric $\underline{a}$ are
known, the transmission amplitude \newline
$[S(\sk;A,B,\underline{a})]_{e,e^\prime}$, $e\neq e^\prime$ determines the
boundary vertex $\partial(e^\prime)$.

By Theorem \ref{eindeutigkeit} and Remark \ref{rem} the transmission
amplitude determines the set \hfill $\cL_{e,e^\prime} =$ $
\{w_{e,e^\prime}(k)\}_{k\in\N_0}$ uniquely. Since $e\neq e^\prime$ we have
that $w_{e,e^\prime}(0)>0$ is the metric length of the geodesic walk $\bw$
from $e^\prime$ to $e$. By Assumption \ref{2cond} the score
$\underline{n}(\bw)$ of this walk is uniquely determined. Since $\bw$ is
geodesic either $n_i=1$ or $n_i=0$ for all $i\in\cI$. Since the geodesic
walk is unique, the score determines the walk $\bw$ uniquely (up to
reversal). Thus, the vertex $\partial(e^\prime)$ is determined uniquely.
\end{proof}

\section{Proof of Theorem \ref{1:theo:2}}\label{sec:7}

In this section we will study local boundary conditions on a given metric
graph $\cG$ which lead to the same scattering matrix. More precisely, we
consider the following problem: Given local boundary conditions $(A,B)$ on
the metric graph $(\cG,\underline{a})$ find all local boundary conditions
$(A^\prime, B^\prime)$ on $\cG$ not equivalent to $(A,B)$ for which the
Laplace operators $\Delta(A^\prime, B^\prime; \underline{a})$ are
isoscattering with $\Delta(A, B; \underline{a})$, that is,
\begin{equation*}
S(\sk;A^\prime,B^\prime,\underline{a}) = S(\sk;A,B,\underline{a})
\end{equation*}
holds for all $\sk>0$. Recall (see Definition \ref{def:equiv}) that the
boundary conditions $(A,B)$ and $(A^\prime, B^\prime)$ are equivalent if
the corresponding maximal isotropic subspaces $\cM(A,B)$ and
$\cM(A^\prime,B^\prime)$ of ${}^d\cK$ coincide. The solution of this problem
is given by Theorem \ref{rb:main} below. This result immediately implies
Theorem \ref{1:theo:2}.

Let $\phi$ denote an arbitrary family $\{\phi_{j}\}_{j\in\cI\cup\cE}$ of
real numbers. To such $\phi$ we associate a unitary map $G(\phi)$ in $\cH$
defined by
\begin{equation*}
(G(\phi)\psi)_{j}(x)=\e^{\ii\phi_{j}}\psi_{j}(x),\qquad j\in\cI\cup\cE.
\end{equation*}
It is easy to verify that
\begin{equation*}
G(\phi)^\dagger \Delta(A,B;\underline{a})G(\phi)=\Delta(A U_\phi, B U_\phi;
\underline{a}),
\end{equation*}
where
\begin{equation}\label{u:form}
U_\phi = \left(\bigoplus_{e\in\cE}\e^{\ii \phi_e} \right) \oplus
\left(\bigoplus_{i\in\cI}\e^{\ii \phi_i} \right) \oplus
\left(\bigoplus_{i\in\cI}\e^{\ii \phi_i} \right)
\end{equation}
is a unitary transformation on
$\cK=\cK_{\cE}\oplus\cK_{\cI}^{(-)}\oplus\cK_{\cI}^{(+)}$. Motivated by our
study \cite{KS5} of magnetic Laplace operators on graphs we call such
transformations \emph{gauge transformations}. Transformations of the type
\eqref{u:form} correspond to the case of vanishing magnetic fields.

\begin{definition}\label{trivial:gauge}
A unitary transformation on $\cK$ of the form \eqref{u:form} is called a
\emph{trivial gauge transformation} if $\phi_e=0$ for all $e\in\cE$, that
is, if $U_\phi|_{\cK_{\cE}}=\1|_{\cK_{\cE}}$.
\end{definition}

The set of all trivial gauge transformations will be denoted by $\sU_0$.
Obviously, $\sU_0$ is an Abelian group isomorphic to the torus $\T^{|\cI|}$.

\begin{proposition}\label{6:prop:1}
Trivial gauge transformations do not change the scattering matrix, that is,
for arbitrary boundary conditions $(A,B)$ satisfying \eqref{abcond}
\begin{equation}\label{S=S}
S(\sk;AU, BU,\underline{a}) = S(\sk;A, B,\underline{a})\quad\text{for all}\quad \sk>0
\end{equation}
whenever $U\in\sU_0$.
\end{proposition}

\begin{proof}
Observe that
\begin{equation*}
\mathfrak{S}(\sk; AU, BU) = U^\dagger \mathfrak{S}(\sk; A, B) U
\end{equation*}
for all  unitary $U\in\sU(|\cE|+2|\cI|)$, where $\mathfrak{S}(\sk; A, B)$
is defined in \eqref{uuu:def}. Any gauge transformation $U$ commutes with
$T(\sk;\underline{a})$ defined in \eqref{T:def:neu}. Thus,
\begin{equation*}
K(\sk; AU, BU, \underline{a}) = U^\dagger K(\sk; A, B, \underline{a}) U.
\end{equation*}
Now relation \eqref{S-matrix} in Theorem \ref{3.2inKS1} implies equality
\eqref{S=S} for any $U\in\sU_0$.
\end{proof}

We note that from the proof it follows that the transmission probabilities
$|[S(\sk;A,B,\underline{a})]_{e,e^{\prime}}|^2$, $(e\neq e^{\prime})$ as
well as the reflection amplitudes $[S(\sk;A,B,\underline{a})]_{e,e}$ are
invariant under \emph{any} gauge transformation.

The following proposition shows that generically trivial gauge
transformations do not leave the Laplace operator unchanged.

\begin{proposition}\label{x:theorem:intro:1}
For Haar almost all local boundary conditions $(A,B)$ the equality
\begin{equation}\label{gleich}
\cM(A,B) = \cM(AU,BU)
\end{equation}
for some $U\in\sU_0$ implies that $U=\1$.
\end{proposition}

Using \eqref{muinv} we can rephrase this result as follows: For Haar almost
all local boundary conditions $(A,B)$ the invariance of the subspace
$\cM(A,B)\subset{}^d\cK$ with respect to ${}^dU:=\begin{pmatrix} U & 0 \\ 0
& U\end{pmatrix}$ for some $U\in\sU_0$, that is, the equality
\begin{equation*}
{}^dU \cM(A,B) = \cM(A,B),
\end{equation*}
implies that $U=\1$.

\begin{proof}
{}From Proposition \ref{propo:3.21} and Definition \ref{propo} it follows
that $A$ and $B$ are invertible for Haar almost all local boundary
conditions $(A,B)$. By Theorem 1 in \cite{KS5} equality \eqref{gleich}
implies that $U$ commutes with $A^{-1} B$. Recall that $A^{-1} B$ is
self-adjoint since $AB^\dagger$ is (see the proof of Proposition
\ref{3:prop:4:neu}). Therefore, $U$ and $\mathfrak{S}(\sk;A,B)$ (defined by
\eqref{uuu:def}) are commuting for all $\sk>0$.

Assume now that for some boundary conditions $(A,B)$ equality \eqref{gleich}
holds for some $U\neq\1$, $U\in\mathsf{U}_0$. Then there are two elements
$\chi_1\neq\chi_2$ of the standard basis in $\cK\cong\C^{|\cE|+2|\cI|}$,
which are eigenvectors of $U$ corresponding to different eigenvalues. Thus,
$\langle\chi_1, \mathfrak{S}(\sk;A,B)\chi_2\rangle=0$ for all $\sk>0$. But
by Lemma \ref{generic:ev} this can only hold for a set of boundary
conditions, which is of zero Haar measure.
\end{proof}

Let $\mathsf{F}(A,B)$ be the set of all local boundary conditions which are
isoscattering with $(A,B)$. {}From Proposition \ref{6:prop:1} it follows
that $\mathsf{U}_0\subset \mathsf{F}(A,B)$. Under additional assumptions
also the opposite inclusion holds for Haar almost all boundary conditions.
The following result implies Theorem \ref{1:theo:2}.

\begin{theorem}\label{rb:main}
Under Assumptions \ref{con:graph} and \ref{2cond} the equality
$\mathsf{F}(A,B)=\mathsf{U}_0$ holds for Haar almost all local boundary
conditions.
\end{theorem}

The remainder of this section is devoted to the proof of this theorem. Given
local boundary conditions $(A,B)$ satisfying \eqref{abcond} assume that
there are local boundary conditions $(A^\prime,B^\prime)$ such that
\begin{equation}\label{SS}
S(\sk; A,B,\underline{a})=S(\sk;
A^\prime,B^\prime,\underline{a})\qquad\text{for all}\quad \sk>0.
\end{equation}
By Proposition \ref{def:local} due to equation \eqref{svertex} the unitary
matrices $\mathfrak{S}(\sk;A,B)$ and $\mathfrak{S}(\sk;A^\prime,B^\prime)$
possess the orthogonal decompositions
\begin{equation}\label{deco}
\mathfrak{S}(\sk;A,B)=\bigoplus_{v\in V}
S_v(\sk;A_v,B_v)\qquad\text{and}\qquad
\mathfrak{S}(\sk;A^\prime,B^\prime)=\bigoplus_{v\in V}
S_v(\sk;A^\prime_v,B^\prime_v)
\end{equation}
with respect to decomposition \eqref{K:ortho}.

We start the proof of Theorem \ref{rb:main} with the following proposition.
Fix an arbitrary boundary vertex $v\in\partial V$ and consider the graph
$\cG_{v}=(\{v\},\cI_{v}=\emptyset,\cE_{v},\partial_{v})$ associated with the
vertex $v$ (see Definition \ref{def:7.4}).

Set $\sU_v=\{U\in\sU_0|\,U\chi=\chi\,\, \text{for all}\,\,
\chi\in\cK\ominus\cL_v\}$ such that
\begin{equation*}
\sU_0 = \Bigtimes_{v\in V} \sU_v.
\end{equation*}
Note that any $U\in\sU_v$ leaves all subspaces $\cL_v\subset\cK$ (see
Section \ref{sec:reconstruction}) invariant.

\begin{proposition}\label{rb:propo:1}
Under Assumptions \ref{con:graph} and \ref{2cond} for Haar almost all local
boundary conditions the scattering matrix $S(\sk; A, B, \underline{a})$
determines the maximal isotropic subspaces $\cM(A_v, B_v)\subset{}^d\cL_v$,
$v\in\partial V$ uniquely  up to trivial gauge transformations
$U\in\mathsf{U}_v$.
\end{proposition}

In other words, the scattering matrix $S(\sk; A, B, \underline{a})$
determines the family of maximal isotropic subspaces $\{\cM(A_v U, B_v
U)\}_{U\in\mathsf{U}_v}$ uniquely.

We split the proof into several lemmas.

By Theorem \ref{eindeutigkeit} and Remark \ref{rem} the scattering matrix
$S(\sk;A,B,\underline{a})$ on the metric graph $(\cG,\underline{a})$
uniquely determines all Fourier coefficients
$\widehat{S}_{\underline{n}}(\sk;A,B)$ in the Fourier expansion
\eqref{Fourier:exp}.

\begin{lemma}\label{lem:rb:1}
Let $e\in\cE$ be arbitrary, $v := \partial(e)$. For Haar almost all boundary
conditions $(A,B)$ the matrix element
$[\widehat{S}_{\underline{n}}(\sk;A,B)]_{e,e}$ with
$\underline{n}=\underline{0}$ determines the scattering matrix
$S_v(\sk)\equiv S(\sk;A_v,B_v)$ up to unitary transformation. Moreover, the
matrix elements $[S_v(\sk)]_{e_1,e_2}$, $e_1,e_2\in\cE_v$ with $\Psi_v(e_1),
\Psi_v(e_2)\in\cE$ are determined uniquely.
\end{lemma}

\begin{proof}
By Theorem \ref{verloren} the matrix elements
$[\widehat{S}_{\underline{n}}(\sk;A,B)]_{e,e^\prime}$,
$\partial(e)=\partial(e^\prime)=v$ with $\underline{n}=\underline{0}$ equals
$[S_v(\sk)]_{\Psi^{-1}_v(e),\Psi^{-1}_v(e^\prime)}$, where the map $\Psi_v$
is defined in Definition \ref{def:7.4}. Thus, the matrix elements
$[S_v(\sk)]_{e_1,e_2}$ with $\Psi_v(e_1), \Psi_v(e_2)\in\cE$ are determined
uniquely. Recall that by Theorem \ref{3:prop:4} for Haar almost all boundary
conditions the scattering matrix $S_v(\sk)$ has the form
\begin{equation}\label{rb:1}
S_v(\sk) = \frac{\sk+\ii H_v/2}{\sk-\ii H_v/2}
\end{equation}
with some Hermitian $H_v$. For any $e^\prime\in\cE_v$ the matrix element
$[S_v(\sk)]_{e^\prime,e^\prime}$ can be represented as the scalar product
$\langle \chi, S_v(\sk)\, \chi\rangle$ with $\chi$ an element of the
canonical basis in the subspace $\cL_{v}$ defined by \eqref{decomp}.

By Theorem \ref{3:prop:4} for almost all boundary conditions $(A_v, B_v)$
the vector $\chi$ is not orthogonal to any eigenvector of the matrix $H_v$.
Thus, the analytic continuation in $\sk$ of $[S_v(\sk)]_{e^\prime,e^\prime}$
with $e^\prime\in\cE_v$, $\Psi_v(e^\prime)\in\cE$ determines all poles of
$S_v(\sk)$. In turn, this determines all eigenvalues of the matrix $H_v$
and, therefore, the matrix $H_v$ up to unitary equivalence. By \eqref{rb:1}
this determines the scattering matrix $S_v(\sk)$ up to unitary equivalence.
\end{proof}

The orthogonal decomposition $\cK=\cK_{\cE}\oplus\cK_{\cI}$ with
$\cK_{\cI}=\cK_{\cI}^{(-)}\oplus\cK_{\cI}^{(+)}$ (see \eqref{K:def})
induces the orthogonal decomposition of the subspace $\cL_v$ into the
orthogonal sum $\cL_{v}^{(\cE)}\oplus \cL_{v}^{(\cI)}$, where
$\cL_{v}^{(\cE)}=\cL_v\cap\cK_{\cE}$ and
$\cL_{v}^{(\cI)}=\cL_v\cap\cK_{\cI}$. Set $p=\dim\cL_{v}^{(\cE)}$ and
$q=\dim\cL_{v}^{(\cI)}$ such that $p+q=\deg(v)$.

Denote $S_v^\prime(\sk):=S_v(\sk; A_v^\prime, B_v^\prime)$, where
$(A_v,B_v)$ corresponds to the local boundary conditions $(A^\prime,
B^\prime)$ satisfying \eqref{SS}. Lemma \ref{lem:rb:1} implies that if the
$\sk$-independent matrix
\begin{equation*}
H_v=-\ii\sk \frac{S_v(\sk)-\1}{S_v(\sk)+\1}
\end{equation*}
is represented in a block form
\begin{equation}\label{rb:H:v}
H_v= \begin{pmatrix} K_{00} & K_{01} \\ K_{10} & K_{11}  \end{pmatrix}
\end{equation}
with respect to the orthogonal decomposition $\cL_{v}=\cL_{v}^{(\cE)}\oplus
\cL_{v}^{(\cI)}$, then the matrix
\begin{equation*}
H_v^\prime := -\ii\sk \frac{S_v^\prime(\sk)-\1}{S_v^\prime(\sk)+\1}
\end{equation*}
has a block form
\begin{equation*}
H_v^\prime = \begin{pmatrix} K_{00} & K_{01} U \\ U^\dagger K_{10} &
U^\dagger K_{11} U
\end{pmatrix},
\end{equation*}
where $U$ is a unitary transformation on $\cL_{v}^{(\cI)}$. To prove
Proposition \ref{rb:propo:1} it suffices to show that $U$ is necessarily
diagonal with respect to the canonical basis in $\cL_{v}^{(\cI)}$ such that
$\1\oplus U\in {\sU_v}$ for Haar almost all local boundary conditions.

For brevity we will set
$H_{e^\prime,i}:=[H_v]_{\Psi_v(e^\prime),\Psi_v(i)}$ and
$H_{e^\prime,i}^\prime:=[H_v^\prime]_{\Psi_v(e^\prime),\Psi_v(i)}$. A
similar shorthand notation will be used for the scattering matrices.

\begin{lemma}\label{lem:rb:3}
Equation \eqref{SS} implies that for all $e\in \cE\cap\mathcal{S}(v)$ and
all $i\in \cI\cap\mathcal{S}(v)$
\begin{equation*}
H^\prime_{e,i} = H_{e,i} \e^{\ii\gamma_i}
\end{equation*}
with some real numbers $\gamma_i$ independent of $e$.
\end{lemma}

\begin{proof}
For arbitrary external edges $e,e^\prime\in\cE\cap\mathcal{S}(v)$  consider
the Fourier expansion \eqref{verloren:2} of the scattering matrix element
$[S(\sk;A,B,\underline{a})]_{e,e^\prime}$.  The coefficients associated to the
exponents $\e^{2\ii\sk a_{i}}$, $i\in \cI\cap\mathcal{S}(v)$ are equal to
\begin{equation*}
[S_{v}(\sk)]_{e,i} [S_{v_i}(\sk)]_{i,i} [S_{v}(\sk)]_{i,e^\prime},
\end{equation*}
where $v_i\in\partial(i)$, $v_i\neq v$. By Lemma \ref{lem:3.13} the leading
term of these products for $\sk\rightarrow +\infty$ equals
$H_{e,i}H_{i,e^\prime}$. This implies that
\begin{equation}\label{rb:ee:prime}
H_{e,i}H_{i,e^\prime} = H_{e,i}^\prime H_{i,e^\prime}^\prime
\end{equation}
for all $e,e^\prime\in \cE\cap\mathcal{S}(v)$ and all $i\in
\cI\cap\mathcal{S}(v)$. In particular, if $e=e^\prime$ we have $|H_{e,i}| =
|H_{e,i}^\prime|$. Thus,
\begin{equation}\label{rb:ee:prime:2}
H_{e,i}^\prime = H_{e,i} \e^{\ii\gamma_{e,i}},\qquad H_{i,e}^\prime =
H_{i,e} \e^{-\ii\gamma_{e,i}}
\end{equation}
with some real numbers $\gamma_{e,i}$. If $|\cE\cap\mathcal{S}(v)|=1$ the
claim is proven. Assume now that $|\cE\cap\mathcal{S}(v)|>1$. Plugging
\eqref{rb:ee:prime:2} into \eqref{rb:ee:prime} for $e\neq e^\prime$ we
obtain that $\gamma_{e,i}=\gamma_{e^\prime,i}$ modulo $2\pi$, which
completes the proof.
\end{proof}

To proceed further we need the following result.

\begin{lemma}\label{lem:rb:2}
Let $\{x_k\}_{k=1}^p$ and $\{y_j\}_{j=1}^q$ be orthonormal bases in
$\cL_{v}^{(\cE)}$ and $\cL_{v}^{(\cI)}$, respectively. If for some unitary
transformation $U:\,\cL_{v}^{(\cI)}\rightarrow\cL_{v}^{(\cI)}$ the relations
\begin{equation}\label{rb:gleich}
\langle x_k, K_{01} U y_j\rangle = \langle x_k, K_{01} y_j\rangle
\e^{\ii\gamma_j}
\end{equation}
hold for all $j\in\cI\cap\mathcal{S}(v)$ and all
$k\in\cE\cap\mathcal{S}(v)$ with some real numbers $\gamma_j$ independent of
$k$, then
\begin{equation}\label{eq:8:10}
U^\dagger|_{\Ran K_{10}} = \Lambda^\dagger|_{\Ran K_{10}},
\end{equation}
where $\Lambda$ is a unitary map defined by
\begin{equation*}
\langle y_k, \Lambda y_j\rangle = \delta_{jk} \e^{\ii \gamma_j}.
\end{equation*}
\end{lemma}

\begin{proof}
{}From \eqref{rb:gleich} it follows that
\begin{equation*}
\langle y_j, U^\dagger K_{10} x_k\rangle = \langle y_j, K_{10} x_k\rangle
\e^{-\ii \gamma_j}
\end{equation*}
for all $j\in\cI\cap\mathcal{S}(v)$ and all $k\in\cE\cap\mathcal{S}(v)$.
This implies that
\begin{equation*}
U^\dagger K_{10} x_k = \Lambda^\dagger K_{10} x_k.
\end{equation*}
Noting that the linear span of $\{K_{10} x_k\}_{k=1}^p$ is the whole $\Ran
K_{10}$ we obtain the claim.
\end{proof}

\begin{remark}\label{rem:rb:1}
Lemma \ref{lem:rb:2} implies that under the conditions of this lemma
\begin{equation}\label{rb:structure}
U = W \Lambda,
\end{equation}
where $W$ is a unitary transformation leaving $\Ran K_{10}$ invariant and
which satisfies $W|_{\Ran K_{10}}=\1|_{\Ran K_{10}}$. To prove this fact we
note that from \eqref{eq:8:10} it follows that
\begin{equation*}
U^\dagger = \Lambda^\dagger P_{\Ran K_{10}} + V^\dagger P_{\Ran
K_{10}}^\perp
\end{equation*}
with some linear map $V$.

Without loss of generality the map $V$ can be chosen to be unitary. Indeed,
from the unitarity of $U$ it follows that $V^\dagger P_{\Ran K_{10}}^\perp
V$ is an orthogonal projection such that on $(\Ran K_{10})^\perp$ the map
$V^\dagger$ must coincide with some unitary operator.

{}From the unitarity of $U$ we also obtain
\begin{equation*}
P_{\Ran K_{10}} \Lambda V^\dagger P_{\Ran K_{10}}^\perp =0.
\end{equation*}
Thus, $(\Ran K_{10})^\perp$ is invariant for $\Lambda V^\dagger$. Since
$\Lambda V^\dagger$ is unitary, this implies that $\Ran K_{10}$ is invariant
for $\Lambda V^\dagger$. Thus, $\Lambda V^\dagger=W_0\oplus W_1$ with
respect to the orthogonal decomposition $\cL_v^{(\cI)} = \Ran K_{10}
\oplus(\Ran K_{10})^\perp$. Therefore, $V^\dagger=\Lambda^\dagger (W_0\oplus
W_1)$. Hence, we obtain
\begin{equation*}
\begin{split}
U & = P_{\Ran K_{10}} \Lambda + P_{\Ran K_{10}}^\perp (W_0 \oplus W_1)
\Lambda \\
& =  (\1 \oplus W_1) \Lambda,
\end{split}
\end{equation*}
which proves the claim.
\end{remark}

Note that if $p\geq q$, then for Haar almost all boundary conditions $\Ran
K_{10} = \cL^{(\cI)}_{v}$. Hence, by Lemma \ref{lem:rb:2} $U = \Lambda$ such
that $(\1\oplus U)\in\mathsf{U}_v$. In this case the proof of Proposition
\ref{rb:propo:1} is complete.

Now we assume that $p < q$ which implies that $q\geq 2$.

\begin{lemma}\label{lem:rb:4}
Equation \eqref{SS} implies that for all $i_1,i_2\in
\cI\cap\mathcal{S}(v)$, $i_1\neq i_2$ either
\begin{equation*}
H^\prime_{i_1,i_2} = H_{i_1,i_2} \e^{\ii(\gamma_{i_2}-\gamma_{i_1})}
\end{equation*}
or
\begin{equation}\label{phi12:def}
H^\prime_{i_1,i_2} = \overline{H_{i_1,i_2}}
\e^{\ii(\gamma_{i_2}-\gamma_{i_1})} \e^{\ii \varphi_{i_1,i_2}},\qquad
\e^{\ii \varphi_{i_1,i_2}} = \e^{-\ii \varphi_{i_2,i_1}} :=
\frac{\overline{H_{e,i_1} H_{i_2,e}}}{H_{e,i_1} H_{i_2,e}}
\end{equation}
with the same numbers $\gamma_i$ as in Lemma \ref{lem:rb:3}.
\end{lemma}

\begin{proof}
We split the proof into several steps.

\emph{Step 1.} For an arbitrary external edge $e\in\cE\cap\mathcal{S}(v)$
consider the Fourier expansion \eqref{verloren:2} of the scattering matrix
element $[S(\sk;A,B,\underline{a})]_{e,e}$.  The quotient of the
coefficients associated to the exponents $\e^{4\ii\sk a_{i}}$ and $\e^{2\ii\sk
a_{i}}$, $i\in \cI\cap\mathcal{S}(v)$ equals
\begin{equation*}
[S_{v}(\sk)]_{ii} [S_{v_i}(\sk)]_{ii},
\end{equation*}
where $v_i\in\partial(i)$, $v_i\neq v$. Thus,
\begin{equation*}
[S_{v}(\sk)]_{ii} [S_{v_i}(\sk)]_{ii} = [S^\prime_{v}(\sk)]_{ii}
[S^\prime_{v_i}(\sk)]_{ii}
\end{equation*}
for all $i\in \cI\cap\mathcal{S}(v)$.

\emph{Step 2.} For arbitrary $i_1,i_2\in \cI\cap\mathcal{S}(v)$, $i_1\neq
i_2$ the coefficient associated to the exponent $\e^{2\ii\sk (a_{i_1} + a_{i_2})}$ of
the Fourier expansion \eqref{verloren:2} of the scattering matrix element
$[S(\sk;A,B,\underline{a})]_{e,e}$ equals
\begin{equation}\label{rb:coin}
\begin{split}
& [S_{v}(\sk)]_{e,i_2} [S_{v_2}(\sk)]_{i_2,i_2} [S_{v}(\sk)]_{i_2,i_1}
[S_{v_1}(\sk)]_{i_1,i_1}
[S_v(\sk)]_{i_1,e} \\
+ & [S_{v}(\sk)]_{e,i_1} [S_{v_1}(\sk)]_{i_1,i_1} [S_{v}(\sk)]_{i_1,i_2}
[S_{v_2}(\sk)]_{i_2,i_2} [S_v(\sk)]_{i_2,e}.
\end{split}
\end{equation}
By Lemma \ref{lem:3.13} the leading term of this expression for
$\sk\rightarrow+\infty$ equals
\begin{equation*}
-\frac{\ii}{\sk^3}(H_{e,i_2} H_{i_2,i_1} H_{i_1,e} + H_{e,i_1} H_{i_1,i_2}
H_{i_2,e}).
\end{equation*}
Thus,
\begin{equation}\label{rb:re}
\Re(H_{e,i_2} H_{i_2,i_1} H_{i_1,e}) = \Re(H_{e,i_2}^\prime
H_{i_2,i_1}^\prime H_{i_1,e}^\prime)
\end{equation}
for all $i_1,i_2\in \cI\cap\mathcal{S}(v)$, $i_1\neq i_2$.

\emph{Step 3.} For arbitrary $i_1,i_2\in \cI\cap\mathcal{S}(v)$, $i_1\neq
i_2$ the coefficient associated to the exponent $\e^{\ii\sk (4 a_{i_1} + 2 a_{i_2})}$
of the Fourier expansion \eqref{verloren:2} of the scattering matrix element
$[S(\sk;A,B,\underline{a})]_{e,e}$ equals
\begin{equation}\label{rb:third}
\begin{split}
 & [S_{v}(\sk)]_{e,i_1} [S_{v_1}(\sk)]_{i_1,i_1} [S_{v}(\sk)]_{i_1,i_2}
[S_{v_2}(\sk)]_{i_2,i_2} [S_{v}(\sk)]_{i_2,i_1} [S_{v_1}(\sk)]_{i_1,i_1}
[S_{v}(\sk)]_{i_1,e}\\
+ &  [S_{v}(\sk)]_{e,i_2} [S_{v_2}(\sk)]_{i_2,i_2} [S_{v}(\sk)]_{i_2,i_1}
[S_{v_1}(\sk)]_{i_1,i_1} [S_{v}(\sk)]_{i_1,i_1} [S_{v_1}(\sk)]_{i_1,i_1}
[S_{v}(\sk)]_{i_1,e}\\
+ & [S_{v}(\sk)]_{e,i_1} [S_{v_1}(\sk)]_{i_1,i_1} [S_{v}(\sk)]_{i_1,i_1}
[S_{v_1}(\sk)]_{i_1,i_1} [S_{v}(\sk)]_{i_1,i_2} [S_{v_2}(\sk)]_{i_2,i_2}
[S_{v}(\sk)]_{i_2,e}.
\end{split}
\end{equation}
Observe that the sum of two last terms equals
\begin{equation*}
\begin{split}
& \alpha(\sk)\Big\{ [S_{v}(\sk)]_{e,i_2} [S_{v_2}(\sk)]_{i_2,i_2}
[S_{v}(\sk)]_{i_2,i_1} [S_{v_1}(\sk)]_{i_1,i_1}
[S_{v}(\sk)]_{i_1,e}\\
+ & [S_{v}(\sk)]_{e,i_1} [S_{v_1}(\sk)]_{i_1,i_1} [S_{v}(\sk)]_{i_1,i_2}
[S_{v_2}(\sk)]_{i_2,i_2} [S_{v}(\sk)]_{i_2,e} \Big\},
\end{split}
\end{equation*}
where $\alpha(\sk):=[S_{v_1}(\sk)]_{i_1,i_1} [S_{v}(\sk)]_{i_1,i_1}$ has
been determined in Step 1. The sum in the braces coincides with
\eqref{rb:coin} and is uniquely determined in Step 2. Thus, the first term
in \eqref{rb:third} is uniquely determined by the Fourier coefficient with
the exponent $\e^{\ii\sk (4 a_{i_1} + 2 a_{i_2})}$. For $\sk\rightarrow
+\infty$ by Lemma \ref{lem:3.13} this term has the asymptotics
\begin{equation*}
\frac{1}{\sk^4} |H_{e,i_1}|^2 |H_{i_1,i_2}|^2.
\end{equation*}
Thus,
\begin{equation*}
|H_{i_1,i_2}| = |H_{i_1,i_2}^\prime|
\end{equation*}
and, in particular,
\begin{equation}\label{rb:re:2}
|H_{e,i_2} H_{i_2,i_1} H_{i_1,e}| = |H_{e,i_2}^\prime H_{i_2,i_1}^\prime
H_{i_1,e}^\prime|
\end{equation}
for all $i_1,i_2\in \cI\cap\mathcal{S}(v)$, $i_1\neq i_2$.

Comparing \eqref{rb:re} and \eqref{rb:re:2} we conclude that either
\begin{equation}\label{rb:re:3}
\Im(H_{e,i_2} H_{i_2,i_1} H_{i_1,e}) = \Im(H_{e,i_2}^\prime
H_{i_2,i_1}^\prime H_{i_1,e}^\prime)
\end{equation}
or
\begin{equation}\label{rb:re:4}
\Im(H_{e,i_2} H_{i_2,i_1} H_{i_1,e}) = -\Im(H_{e,i_2}^\prime
H_{i_2,i_1}^\prime H_{i_1,e}^\prime).
\end{equation}
{}From \eqref{rb:re} and \eqref{rb:re:4} it follows that either
\begin{equation*}
H_{e,i_2} H_{i_2,i_1} H_{i_1,e} = H_{e,i_2}^\prime H_{i_2,i_1}^\prime
H_{i_1,e}^\prime
\end{equation*}
or
\begin{equation*}
H_{e,i_2} H_{i_2,i_1} H_{i_1,e} = \overline{H_{e,i_2}^\prime
H_{i_2,i_1}^\prime H_{i_1,e}^\prime}.
\end{equation*}
Applying Lemma \ref{lem:rb:3} proves the claim.
\end{proof}

We will need the following result due to Friedland \cite{Friedland:1},
\cite{Friedland:2} (see also \cite{Alexander}) on Hermitian matrices with
prescribed off-diagonal entries.

\begin{lemma}\label{Friedland}
Let $K$ be a Hermitian $q\times q$ matrix. There are at most $q!$ different
matrices $L$ unitarily equivalent to $K$ and satisfying
\begin{equation*}
L_{jk} = K_{jk} \qquad\text{for all}\quad j\neq k.
\end{equation*}
\end{lemma}

By this result we obtain that if equality \eqref{SS} is satisfied, then for
Haar almost all boundary conditions $(A,B)$ up to trivial gauge
transformations from $\sU_v$ there is at most a finite number (depending on
$q=\dim\cL_v^{(\cI)}$ only) of unitary matrices $U\in\sU(q)$ such that
\begin{equation*}
H_v^\prime = \begin{pmatrix} \1 & 0 \\ 0 & U \end{pmatrix}^\dagger H_v
\begin{pmatrix} \1 & 0 \\ 0 & U \end{pmatrix},\qquad
H_v=\begin{pmatrix} K_{00} & K_{01} \\
K_{10} & K_{11}
\end{pmatrix},
\end{equation*}
where $H_v$ and $H_v^\prime$ are Hermitian matrices associated with the
boundary conditions $(A_v,B_v)$ and $(A_v^\prime,B_v^\prime)$ via
\begin{equation*}
\mathfrak{S}(\sk;A_v, B_v) = \frac{\sk+\ii H_v/2}{\sk-\ii H_v/2},\qquad
\mathfrak{S}(\sk;A_v^\prime, B_v^\prime) = \frac{\sk+\ii
H_v^\prime/2}{\sk-\ii H_v^\prime/2}
\end{equation*}
for an arbitrary $\sk>0$.

Lemmas \ref{lem:rb:4} and \ref{Friedland} imply that these unitary matrices
$U$ are determined only by $K_{11}$ and by the complex arguments of the
entries of $K_{10}$. On the other hand, Lemma \ref{lem:rb:2} and Remark
\ref{rem:rb:1} state that they have a special structure
\eqref{rb:structure} which is determined by the matrix $K_{10}$ and, in
particular, only by the absolute values of its entries. Intuitively, it is
clear that these two conditions on $U$ are independent and, therefore, can
be satisfied simultaneously in exceptional cases only. This observation will
allow us to prove the uniqueness of $U$.

We return to the proof of Proposition \ref{rb:propo:1}. Without loss of
generality we can choose all numbers $\gamma_i$, $i\in\cI\cap\mathcal{S}(v)$
referred to in Lemma \ref{lem:rb:3} to be zero. Note that equation
\eqref{rb:structure} implies in this case that $U$ leaves $\Ran K_{10}$
invariant and $U|_{\Ran K_{10}}=\1|_{\Ran K_{10}}$.

Let
\begin{equation*}
\begin{pmatrix} \widetilde{K}_{00} & \widetilde{K}_{01} \\
\widetilde{K}_{10} & \widetilde{K}_{11}
\end{pmatrix}
\end{equation*}
be an arbitrary $(p+q)\times(p+q)$ Hermitian matrix written in the block
form with respect to the orthogonal decomposition
$\cL_v=\cL_v^{(\cE)}\oplus\cL_v^{(\cI)}$. Denote by
$\fN(\widetilde{K}_{00},\widetilde{K}_{01},\widetilde{K}_{11})$ the set of
all Hermitian matrices $H$ of the form
\begin{equation}\label{H:repr}
H=\begin{pmatrix} \widetilde{K}_{00} & {K}_{01} \\
{K}_{10} & \widetilde{K}_{11}
\end{pmatrix}
\end{equation}
such that complex arguments of the entries of ${K}_{01}$ agree with those
of $\widetilde{K}_{01}$, that is,
\begin{equation*}
\arg\bigl([K_{01}]_{e,i}\bigr) =
\arg\bigl([\widetilde{K}_{01}]_{e,i}\bigr)\quad\text{for all}\quad
e\in\cE\cap\cS(v)\quad\text{and}\quad i\in\cI\cap\cS(v).
\end{equation*}
On this set we define the measure
\begin{equation*}
d\widetilde{\nu} = \prod_{\substack{e\in\cE\cap\cS(v) \\ i\in\cI\cap\cS(v)}}
d\,|[K_{01}]_{e,i}|.
\end{equation*}

Let $\fU(H)$ be the set of all unitary matrices $U\in\mathsf{U}(q)$ with
the property that for every $i_1,i_2\in\cI\cap\cS(v)$, $i_1\neq i_2$ either
\begin{equation*}
[U^\dagger \widetilde{K}_{11} U]_{i_1,i_2} =
[\widetilde{K}_{11}]_{i_1,i_2},\qquad [U^\dagger \widetilde{K}_{11}
U]_{i_2,i_1} = [\widetilde{K}_{11}]_{i_2,i_1}
\end{equation*}
or
\begin{equation*}
[U^\dagger \widetilde{K}_{11} U]_{i_1,i_2} = [\widetilde{K}_{11}]_{i_1,i_2}
\e^{\ii\varphi_{i_1,i_2}},\qquad [U^\dagger \widetilde{K}_{11} U]_{i_2,i_1}
= [\widetilde{K}_{11}]_{i_2,i_1} \e^{-\ii\varphi_{i_1,i_2}}
\end{equation*}
holds, where $\varphi_{i_1,i_2}$ is defined in \eqref{phi12:def}. By Lemmas
\ref{lem:rb:4} and \ref{Friedland} this set is finite and
$\fU(H_1)=\fU(H_2)$ for all $H_1, H_2
\in\fN(\widetilde{K}_{00},\widetilde{K}_{01},\widetilde{K}_{11})$.

\begin{lemma}\label{lem:rb:fin}
Assume that $p<q$. For $\widetilde{\nu}$-almost all $H\in
\fN(\widetilde{K}_{00},\widetilde{K}_{01},\widetilde{K}_{11})$ no
$U\in\fU(H)\setminus\{\1\}$ satisfies the condition
\begin{equation}\label{rb:assume}
U|_{\Ran K_{10}} = \1|_{\Ran K_{10}},
\end{equation}
where $K_{10}$ is the off-diagonal block of the matrix $H$ in the
representation \eqref{H:repr}.
\end{lemma}

\begin{proof}
Assume that condition \eqref{rb:assume} is satisfied for some unitary
matrix $U\in\fU(H)$, $U\neq\1$. Thus, $U$ has $1$ as an eigenvalue. Its
multiplicity is not less than $\dim\Ran K_{10}$. Since $U\neq\1$, there is
at least one eigenvalue $\lambda\neq 1$ of $U$. The corresponding
eigenvector $y\in\cL_v^{(\cI)}$ is orthogonal to $\Ran K_{10}$. By
assumption we have
\begin{equation*}
0=\langle y, U K_{10} x\rangle = \langle U^\dagger y, K_{10} x\rangle =
{\lambda} \langle y, K_{10} x\rangle
\end{equation*}
for any $x\in\cL_v^{(\cE)}$. Thus, $y\in\Ker K_{10}^\dagger=\Ker K_{01}$. The subspace $\Ker K_{01}$ is nontrivial, since $p<q$.

Observe that for $\widetilde{\nu}$-almost all $H\in
\fN(\widetilde{K}_{00},\widetilde{K}_{01},\widetilde{K}_{11})$ the matrix
$K_{01}$ has a rank $p\geq 1$. Recall that the complex arguments of the
entries of $K_{10}$ are fixed. Therefore, the equation $K_{01} y = 0$
represents $p$ linearly independent relations for the numbers
$|[K_{01}]_{e,i}|$, $e\in\cE\cap\cS(v)$, $i\in\cI\cap\cS(v)$. Hence, the
set of all Hermitian matrices $H\in
\fN(\widetilde{K}_{00},\widetilde{K}_{01},\widetilde{K}_{11})$ satisfying
\eqref{rb:assume} is of zero $\widetilde{\nu}$-measure.
\end{proof}

Now we are in position to complete the proof of Proposition
\ref{rb:propo:1}. Observe that the measure $\nu$ defined in \eqref{nu:def}
is a product measure containing $\widetilde{\nu}$ as a factor. Lemma
\ref{lem:rb:fin} immediately implies that the set of all Hermitian matrices
$H$ satisfying $\fU(H)\neq\{\1\}$ is of zero $\nu$-measure. Hence, by Lemma
\ref{lem:nu},
\begin{equation*}
\varkappa\bigl(\bigl\{H|\,\fU(H)\neq\{\1\}\bigr\}\bigr)=0.
\end{equation*}
By definition \eqref{kappa:def} of the measure $\varkappa$, this implies that
\begin{equation*}
\mu\bigl(\bigl\{\fS|\,\fU(H(\fS))\neq\{\1\}\bigr\}\bigr)=0,
\end{equation*}
where $\mu$ is the Haar measure on $\sU(\deg(v))$ and $H(\fS)$ is given by
\begin{equation*}
H(\fS)=-\ii \frac{\fS-\1}{\fS+\1}.
\end{equation*}
This completes the proof of Proposition \ref{rb:propo:1}.

We turn to the proof of Theorem \ref{rb:main}.

\begin{figure}[htb]
\begin{equation*}
\setlength{\unitlength}{1.2cm}
\begin{picture}(10,3)
\put(2.8,2.5){$\cG^{(k)}$} \put(7.2,2.5){$\cG^{(k+1)}$}
\put(0,1){\line(1,0){2}}
\put(0,2){\line(2,-1){2}}
\put(0,0){\line(2,1){2}}
\put(0.5,0.0){$e_{3}$}
\put(0.5,0.7){$e_{2}$}
\put(0.5,2){$e_{1}$}
\put(2,1){\circle*{0.1}}
\put(2,1){\line(1,1){0.5}}
\put(2,1){\line(1,-1){0.5}}
\put(2.1,1.5){$i_1$}
\put(2.1,0.3){$i_2$}
\put(1.8, 0.7){$v_{k}$}
\put(3,1){\oval(1,2)}
\put(3.5,1.6){\line(1,0){1}}
\put(3.5,0.4){\line(1,0){1}}
\put(6,1.5){\line(1,0){1}}
\put(6,0.5){\line(1,0){1}}
\put(7.5,1){\oval(1,2)}
\put(4.7,1.6){$e_{4}$}
\put(4.7,0.4){$e_{5}$}
\put(8,1.6){\line(1,0){1}}
\put(8,0.4){\line(1,0){1}}
\put(6.2,1.7){$e(i_1)$}
\put(6.2,0.1){$e(i_2)$}
\put(9.3,1.6){$e_{4}$}
\put(9.3,0.4){$e_{5}$}
\end{picture}
\end{equation*}
\caption{\label{fig7} The construction of the graph $\cG^{(k+1)}$ from the
graph $\cG^{(k)}$. Inside the ovals the graphs are not specified but agree
there.}
\end{figure}
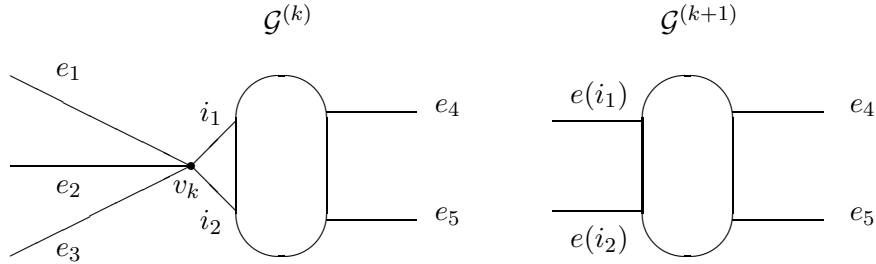

With the graph $\cG=(V,\cI,\cE,\partial)$ we associate a sequence of graphs
$\{\cG^{(k)}\}_{k=1}^{|V|}$, $\cG^{(k)}=(V_k,\cI_k,\cE_k,\partial_k)$ such
that $\cG^{(1)}=\cG$ and $\cG^{(k+1)}$ is constructed from $\cG^{(k)}$ in
the following way (see Fig.~\ref{fig7}):
\begin{itemize}
\item[(i)]{The vertex set $V_{k+1}$ of the graph $\cG^{(k+1)}$ is obtained from the
vertex set $V_k$ by removing a boundary vertex $v_k$ of $V_k$,
$v_k\in\partial V_k$;}
\item[(ii)]{The set of internal edges $\cI_{k+1}$ of the graph $\cG^{(k)}$ is obtained from
$\cI_k$ by removing all internal edges $i\in\cI_k$ incident with the vertex
$v_k$, that is,
\begin{equation*}
\cI_{k+1} = \cI_k \setminus \cS(v_k);
\end{equation*}
}
\item[(iii)]{The set of external edges $\cE_{k+1}$ of the graph $\cG^{(k)}$ is obtained from
$\cE_k$ by adding $|\cI_k|-|\cI_{k+1}|$ external edges to $\cE_k$. The set
$\cE_{k+1}\setminus\cE_{k}$ is in one-to-one correspondence with the set
$\cI_k\setminus\cI_{k+1}$, that is, for any $e\in\cE_{k+1}\setminus\cE_{k}$
there is a unique $i(e)\in\cI_k\setminus\cI_{k+1}$ and for any
$i\in\cI_k\setminus\cI_{k+1}$ there is a unique
$e(i)\in\cE_{k+1}\setminus\cE_{k}$;}
\item[(iv)]{The boundary operator $\partial_{k+1}$ coincides with $\partial_k$ on
$\cE_k\cap\cE_{k+1}$ and $\cI_k\cap\cI_{k+1}$. On $\cE_{k+1}\setminus\cE_k$
the boundary operator is defined as $\partial_{k+1}(e)=v$, where $v$ is the
vertex adjacent to $v_k$ in the graph $\cG^{(k)}$ by the internal edge
$i(e)\in\cI_k\setminus\cI_{k+1}$.}
\end{itemize}

Note that the graphs $\cG^{(k)}$ with $k>1$ are not necessarily connected
and the graph $\cG^{(|V|)}$ is a graph with no internal edges.

Given local boundary conditions
\begin{equation*}
A=\bigoplus_{v\in V} A_v,\qquad B=\bigoplus_{v\in V} B_v
\end{equation*}
on the graph $\cG$, let $S^{(k)}(\sk)$ denote the scattering matrix of the
Laplace operator on the graph $\cG_{k}$ associated with local boundary
conditions defined by
\begin{equation*}
A_k := \bigoplus_{v\in V_k} A_v,\qquad B_k := \bigoplus_{v\in V_k} B_v.
\end{equation*}
In particular, we have $S^{(1)}(\sk):=S(\sk;A,B,\underline{a})$. Let
$v_k\in V_k\setminus V_{k+1}$.

\begin{proposition}\label{propo:rb:9.1}
Given the scattering matrices $S_{v_k}(\sk)$ on the graph $\cG_{v_k}$ and
$S^{(k)}(\sk)$ on the graph $\cG^{(k)}$, respectively, the scattering
matrix $S^{(k+1)}(\sk)$ on the graph $\cG^{(k+1)}$ is defined uniquely.
\end{proposition}

\begin{proof}
Let $e,e^\prime\in\cE_{k}\cap\cE_{k+1}$. The sum of all terms of the
Fourier expansion \eqref{verloren:2} of the scattering matrix element
$[S^{(k)}(\sk)]_{e,e^\prime}$ containing none of the factors $\e^{\ii\sk
a_i}$ with $i\in\cI_k\setminus\cI_{k+1}$ gives, obviously, the scattering
matrix element $[S^{(k+1)}(\sk)]_{e,e^\prime}$.

To determine $[S^{(k+1)}(\sk)]_{e(i),e^\prime}$ for
$e^\prime\in\cE_{k}\cap\cE_{k+1}$ and $i\in\cI_k\setminus\cI_{k+1}$
consider the sum of all terms of the Fourier expansion \eqref{verloren:2}
of the scattering matrix element $[S^{(k)}(\sk)]_{e,e^\prime}$ with
arbitrary $e\in \cE_{k}\setminus\cE_{k+1}$ which contain the factor
$\e^{\ii\sk a_i}$ precisely once and none of the factors $\e^{\ii\sk
a_{i^\prime}}$, $i^\prime\in\cI_k\setminus\cI_{k+1}$, $i^\prime\neq i$. This
sum is of the form
\begin{equation*}
[S_{v_k}(\sk)]_{e,i} [S^{(k+1)}(\sk)]_{e(i),e^\prime} \e^{\ii\sk a_i}
\end{equation*}
and, thus, determines $[S^{(k+1)}(\sk)]_{e(i),e^\prime}$ uniquely. The
matrix elements $[S^{(k+1)}(\sk)]_{e,e(i)}$ can be determined similarly.

To determine $[S^{(k+1)}(\sk)]_{e(i),e(i)}$ for
$i\in\cI_k\setminus\cI_{k+1}$ we consider a scattering matrix element
$[S^{(k)}(\sk)]_{e,e}$ with $e\in\cE_k\setminus\cE_{k+1}$. Consider the sum
of all terms of the Fourier expansion \eqref{verloren:2} of this matrix
element containing the factor $\e^{2\ii\sk a_i}$ exactly once and none of
the factors $\e^{\ii\sk a_{i^\prime}}$,
$i^\prime\in\cI_k\setminus\cI_{k+1}$, $i^\prime\neq i$. This sum is of the
form
\begin{equation*}
[S_{v_k}(\sk)]_{e,i} [S^{(k+1)}(\sk)]_{e(i),e(i)} [S_{v_k}(\sk)]_{i,e}\e^{2
\ii\sk a_i}
\end{equation*}
and, thus, determines $[S^{(k+1)}(\sk)]_{e(i),e(i)}$ uniquely.

If $|\cI_k\setminus\cI_{k+1}|=1$, the proof is completed. Assume now that
$|\cI_k\setminus\cI_{k+1}|>1$. To complete the proof it suffices to show
that the matrix elements $[S^{(k+1)}(\sk)]_{e(i_1), e(i_2)}$ for all
$i_1\neq i_2$, $i_1,i_2\in\cI_k\setminus\cI_{k+1}$ are defined uniquely.

If the vertices $v^{(1)}\in\partial_k(i_1)$, $v^{(1)}\neq v_k$ and
$v^{(2)}\in\partial_k(i_2)$, $v^{(2)}\neq v_k$ belong to different
connected components of the graph $\cG^{(k+1)}$, then
$[S^{(k+1)}(\sk)]_{e(i_1), e(i_2)}$ is identically zero. Therefore, we may
assume that $v^{(1)}$ and $v^{(2)}$ belong to the same connected component
of the graph $\cG^{(k+1)}$.

The coefficient associated to the exponent $\e^{\ii\sk(a_{i_1}+a_{i_2})}$
of the Fourier expansion \eqref{verloren:2} of the scattering matrix
element $[S^{(k+1)}(\sk)]_{e,e}$ equals
\begin{equation}\label{eq:rb:1:1}
\begin{split}
&
[S_{v_k}(\sk)]_{e,e(i_2)}[S^{(k)}(\sk)]_{e(i_2),e(i_1)}[S_{v_k}(\sk)]_{e(i_1),e}\\
+ &
[S_{v_k}(\sk)]_{e,e(i_1)}[S^{(k)}(\sk)]_{e(i_1),e(i_2)}[S_{v_k}(\sk)]_{e(i_2),e}.
\end{split}
\end{equation}
The sum of all terms of the Fourier expansion \eqref{verloren:2} of the
scattering matrix element $[S^{(k+1)}(\sk)]_{e,e}$ containing the factor
$\e^{\ii\sk(3 a_{i_1}+a_{i_2})}$ is given by
\begin{equation}\label{eq:rb:1:3}
\begin{split}
& [S_{v_k}(\sk)]_{e,e(i_2)}[S^{(k)}(\sk)]_{e(i_2),e(i_2)}
[S_{v_k}(\sk)]_{e(i_2),e(i_2)} [S^{(k)}(\sk)]_{e(i_2),e(i_1)}
[S_{v_k}(\sk)]_{e(i_2),e}\\
+ & [S_{v_k}(\sk)]_{e,e(i_2)}[S^{(k)}(\sk)]_{e(i_2),e(i_1)}
[S_{v_k}(\sk)]_{e(i_1),e(i_2)} [S^{(k)}(\sk)]_{e(i_2),e(i_2)}
[S_{v_k}(\sk)]_{e(i_2),e}\\
+ & [S_{v_k}(\sk)]_{e,e(i_1)}[S^{(k)}(\sk)]_{e(i_1),e(i_2)}
[S_{v_k}(\sk)]_{e(i_2),e(i_2)} [S^{(k)}(\sk)]_{e(i_2),e(i_2)}
[S_{v_k}(\sk)]_{e(i_2),e}.
\end{split}
\end{equation}
Observe that the only unknown factors in \eqref{eq:rb:1:1} and
\eqref{eq:rb:1:3} are
\begin{equation*}
[S^{(k)}(\sk)]_{e(i_1),e(i_2)}\qquad \text{and} \qquad
[S^{(k)}(\sk)]_{e(i_2),e(i_1)}.
\end{equation*}

It is easy to see that \eqref{eq:rb:1:1} and \eqref{eq:rb:1:3} considered as
linear combinations of the variables $[S^{(k)}(\sk)]_{e(i_1),e(i_2)}$ and
$[S^{(k)}(\sk)]_{e(i_2),e(i_1)}$, are linearly independent. Thus,
\eqref{eq:rb:1:1} and \eqref{eq:rb:1:3} determine
$[S^{(k)}(\sk)]_{e(i_1),e(i_2)}$ and $[S^{(k)}(\sk)]_{e(i_2),e(i_1)}$
uniquely.
\end{proof}

Combining Propositions \ref{rb:propo:1} and \ref{propo:rb:9.1} concludes the
proof of Theorem \ref{rb:main}.

\section{The Traveling Salesman Problem}\label{sec:5}

In this section we will present applications of the theory of Laplace
operators on non-compact graphs developed in the present work to some
well-known combinatorial problems. In particular, we provide a new approach
to solving the Traveling Salesman Problem (TSP) as well as some other
combinatorial problems. The graph $\cG=\cG(V,\cI,\cE,\partial)$ with
$\cI\neq\emptyset$ and $\cE\neq\emptyset$ is now assumed to satisfy
Assumption \ref{con:graph}.

We start with a formulation of these problems adapted to the context of
walks on noncompact graphs.

\medskip

\textbf{The K\"{o}nigsberger Br\"{u}cken Problem (KBP).} \textit{For given
$e,e^{\prime}\in\cE$ determine whether there is a walk $\bw$ (called Euler
path) from $e^{\prime}$ to $e$ which traverses each internal edge $i\in\cI$
exactly once, i.e.\ $n_{i}({\bf w})=1$ for all $i\in\cI$. There is no
limitation on the number of times a vertex $v\in V$ may be visited.}

\medskip

Recall that the solution of KBP for $e=e^\prime$ is given by the celebrated
Euler theorem (see, e.g., \cite{Diestel}): An Euler path $\bw\in\cW_{e,e}$
exists if and only if $\deg_{\cG_{\mathrm{int}}}(v)$ is even for all $v\in
V$. Here $\deg_{\cG_{\mathrm{int}}}(v)$ is the degree of the vertex $v\in
V$ in the interior $\cG_{\mathrm{int}}$ of the graph $\cG$. If the graph
$\cG_{\mathrm{int}}$ has more than two vertices of odd degree, then there
is no Euler path for all $e,e^\prime\in\cE$. If the graph
$\cG_{\mathrm{int}}$ has exactly two vertices of odd degree, then the Euler
path exists if and only if $\partial(e)\neq \partial(e^\prime)$ and both
vertices $\partial(e)$ and $\partial(e^\prime)$ in the interior
$\cG_{\mathrm{int}}$ of the graph $\cG$ have an odd degree.

\medskip

\textbf{The Hamiltonian Path Problem (HPP).} \textit{For given
$e,e^{\prime}\in\cE$ determine whether there is a walk $\bw$ from
$e^\prime$ to $e$ such that
\begin{itemize}
\item[(a)]{it visits each vertex $v\in V$ exactly once if $\partial(e)\neq\partial(e^\prime)$,}
\item[(b)]{it visits each vertex $v\neq\partial(e)$ exactly once and the vertex
$v=\partial(e)$ exactly twice if $\partial(e)=\partial(e^\prime)$.}
\end{itemize}
}

\medskip

There are no known simple necessary and sufficient conditions for the
existence of Hamiltonian paths. Some sufficient conditions can be found,
e.g., in \cite{Diestel}. The HPP is NP-complete.

In contrast to KBP and HPP, the TSP involves the metric structure on the
graph $\cG$. This problem is also NP-complete. There are several slightly
different versions considered in the literature. Here we give three
formulations suited for the present context.

Assume that a metric $\underline{a}_0\in(\R_+)^{|\cI|}$ on the graph $\cG$
is given.

\medskip

\textbf{The Traveling Salesman Problem (TSP I).} \textit{For given
$e,e^{\prime}\in\cE$ find a walk $\bw$ from $e^{\prime}$ to $e$ with
minimal metric length $|\bw|$ which visits each vertex $v\in V$ of the
graph $\cG$ at least once.}

\medskip

The reader is referred to \cite{Applegate}, \cite{LLKS} and the references
quoted therein for known combinatorial solutions of TSP I. The second
version of TSP is HPP extended by a metric length condition.

\medskip

\textbf{The Traveling Salesman Problem (TSP II).} \textit{For given
$e,e^{\prime}\in\cE$ determine whether there is a walk $\bw$ from
$e^{\prime}$ to $e$ satisfying the same conditions as in HPP and which, in
addition, is of shortest metric length.}

\medskip

The third version is the same as TSP I but formulated as a decision problem involving an additional
length condition.

\medskip

\textbf{The Traveling Salesman Problem (TSP III).} \textit{For given
$e,e^{\prime}\in\cE$ and a given number $L>0$ decide whether there is a walk
$\bw$ from $e^\prime$ to $e$ which visits each vertex $v\in V$ of the graph
$\cG$ at least once and $|\bw|\leq L$.}

\medskip

Recall that the definition of walks on the graph $\cG$ is independent of
the particular choice of the orientation of the edges (see Remark
\ref{richtung}). Therefore, we deal with \emph{symmetric} versions of the
combinatorial problems listed above, where the length of an edge is
independent of the direction in which this edge is traversed.

\subsection{Solution of KBP}
Choose any of the boundary conditions $(A,B)$ referred to in Theorem
\ref{4:theo:2}. For an arbitrary metric structure
$\underline{a}\in(\R_+)^{|\cI|}$ they define a self-adjoint Laplace
operator $\Delta(A,B;\underline{a})$. The associated scattering matrix
$S(\sk; A,B,\underline{a})$ we will denote for brevity by
$S(\sk;\underline{a})$. Set
\begin{equation*}
\underline{1} := \{1\}_{i\in\cI}\in(\N_0)^{|\cI|}.
\end{equation*}
By Theorem \ref{4:theo:2} to solve KBP, we only have to see whether the
matrix element of the Fourier coefficient
$[\widehat{S}_{\underline{1}}(\sk)]_{e,e^{\prime}}$ defined in
\eqref{fourier:coef} of the scattering matrix $S(\sk;\underline{a})$
vanishes identically or not. In particular, evaluating integrals in
\eqref{fourier:coef}, one can give a purely analytic proof of the Euler
theorem.

\subsection{Solution of HPP} First, assume that the graph $\cG$ has a vertex
$v$ with $\deg(v)=1$. Let $i\in\cI$ be the unique internal edge incident
with this vertex. Then, HPP has a solution if and only if $|\cI|=1$,
$|V|=2$, and $\partial(e)\in\partial(i)$ for all $e\in\cE$. Indeed, assume
that for some graph $\cG$ with minimum degree $1$ there exists a walk $\bw$
satisfying the conditions of HPP. Let $\deg(v)=1$, $v\in\partial(i)$.
Denote by $v^\prime$ the vertex adjacent to $v$ by the edge $i$. Thus, the
walk $\bw$ visits $v^\prime$ twice. If $v^\prime$ is not a boundary vertex,
this is a contradiction. If not, there are external edges
$e,e^\prime\in\cE$ (possibly $e=e^\prime$) such that
$\bw=\{e^\prime,i,i,e\}$. Thus, if $|V|>2$ the walk $\bw$ does not visit
any other vertex different from $v$ and $v^\prime$.

By this observation, we may assume that $\deg(v)>1$ for all $v\in V$.
Obviously, any solution $\bw$ of HPP traverses any internal edge at most
once. We call a score $\underline{n}=\{n_i\}_{i\in\cI}$ \emph{simple} if
$n_{i}\in\{0,1\}$ for all $i\in\cI$. Let
$\cN_{e,e^{\prime}}^{\mathrm{simple}}$ be the (possibly empty) set of
simple scores in the set $\cN_{e,e^{\prime}}$ defined in \eqref{enn}.
Obviously, $|\cN_{e,e^{\prime}}^{\mathrm{simple}}| \leq 2^{|\cI|}$.

\begin{lemma}\label{lem:9:1:neu}
If a walk $\bw$ solves HPP, then its score $\underline{n}(\bw)$ satisfies
\begin{equation}\label{nn}
|\underline{n}(\bw)| =\begin{cases}
|V|\quad&\text{if}\quad \partial(e)=\partial(e^{\prime})\\
|V|-1\quad&\text{if}\quad \partial(e)\neq \partial(e^{\prime}).
\end{cases}
\end{equation}
\end{lemma}

\begin{proof}
Assume $\bw\in\cW_{e,e^{\prime}}(\underline{n})$ is a walk solving HPP. Then
the score $\underline{n}\equiv\{n_i\}_{i\in\cI} := \underline{n}(\bw)$ is
simple. Let $\cG^\prime=(V^\prime,\cI^\prime,\cE^\prime,\partial^\prime)$ be
the (noncompact) graph obtained from $\cG=(V,\cI,\cE,\partial)$ by removing
the internal edges $i\in\cI$ with $n_i=0$. Obviously, $V^\prime = V$ and
$\cE^\prime = \cE$. If $\partial(e)\neq
\partial(e^{\prime})$, then the degree $\deg_{\cG^\prime_{\mathrm{int}}}(v)$ of any vertex
$v\notin\{\partial(e),\partial(e^\prime)\}$ in the interior
$\cG^\prime_{\mathrm{int}}$ of the graph $\cG^\prime$ equals $2$. The
degree of the vertices $\partial(e)$ and $\partial(e^\prime)$ in
$\cG^\prime_{\mathrm{int}}$ equals $1$. Thus, applying the First Theorem of
Graph Theory to $\cG^\prime_{\mathrm{int}}$ we obtain
\begin{equation*}
|\underline{n}| = |\cI^\prime| = \frac{1}{2} \sum_{v\in
V}\deg_{\cG^\prime_{\mathrm{int}}}(v) = |V|-1.
\end{equation*}
If $\partial(e) = \partial(e^{\prime})$, then
$\deg_{\cG^\prime_{\mathrm{int}}}(v)=2$ for all $v\in V$. Therefore, we get
\begin{equation*}
|\underline{n}| = |\cI^\prime| = \frac{1}{2} \sum_{v\in
V}\deg_{\cG^\prime_{\mathrm{int}}}(v) = |V|.
\end{equation*}
This proves \eqref{nn}.
\end{proof}

Let
\begin{equation*}
\cN_{e,e^\prime}^{\mathrm{(HPP)}}=\{\underline{n}\in\cN_{e,e^\prime}^{\mathrm{simple}}
\mid |\underline{n}| \quad\text{satisfies \eqref{nn}}\}.
\end{equation*}
Obviously,
\begin{equation*}
|\cN_{e,e^\prime}^{\mathrm{(HPP)}}| \leq \begin{cases}\frac{|\cI|!}{|V|!(|\cI|-|V|)!}\quad & \text{if}\quad
\partial(e)=\partial(e^{\prime}),\\[1ex] \frac{|\cI|!}{(|V|-1)!(|\cI|-|V|+1)!}\quad & \text{if}\quad
\partial(e)\neq\partial(e^{\prime}).\end{cases}
\end{equation*}
Note that to determine the number of elements in the set
$\cN_{e,e^\prime}^{\mathrm{(HPP)}}$ there are simple procedures based on
the calculation of the generating function. Some counting problems of this
type are considered in \cite{KS6}.

We emphasize that $\underline{n}\in\cN_{e,e^\prime}^{\mathrm{(HPP)}}$ in
general does not imply that a walk $\bw$ from $e^\prime$ to $e$ with the
score $\underline{n}$ solves HPP.

Let $\bw$ be an arbitrary walk from $e^\prime\in\cE$ to $e\in\cE$,
$\underline{n}=\underline{n}(\bw)$ its score. For any vertex $v\in V$ define
\begin{equation}\label{Nv1}
N_{e,e^{\prime}}(v;\underline{n})=\frac{1}{2}\sum_{i\in\cI:
v\in\partial(i)}n_{i} + \frac{n_{e,e^{\prime}}(v)}{2},\qquad \underline{n} =
\{n_i\}_{i\in\cI},
\end{equation}
where
\begin{equation}\label{Nv2}
n_{e,e^{\prime}}(v)=\begin{cases}
2\quad\mbox{if}\quad v=\partial(e)=\partial(e^{\prime}),\\
1\quad\mbox{if}\quad \partial(e)\neq
\partial(e^{\prime})\quad\mbox{and}\quad
v=\partial(e)\quad\mbox{or}\quad v=\partial(e^{\prime}),\\
0\quad\mbox{otherwise}.
\end{cases}
\end{equation}
It is easy to check that $N_{e,e^{\prime}}(v;\underline{n})$ is the number
of times a walk $\bw\in\cW_{e,e^{\prime}}$ visits the vertex $v\in V$.

\begin{theorem}\label{5:theo:1}
Assume that the minimum degree of the graph $\cG$ is bigger than $1$. The
HPP has a solution if and only if there is $\underline{n}\in
\cN_{e,e^{\prime}}^{\mathrm{(HPP)}}$ such that
$N_{e,e^{\prime}}(v;\underline{n})\neq 0$ for all $v\in V$.
\end{theorem}

\begin{proof}
It remains to prove the ``if'' part of the statement. Assume that there is
$\underline{n}\equiv\{n_i\}_{i\in\cI}\in
\cN_{e,e^{\prime}}^{\mathrm{(HPP)}}$. Therefore, there is a walk
$\bw\in\cW_{e,e^\prime}(\underline{n})$ from $e^\prime$ to $e$ traversing
any internal edge of the graph $\cG$ at most once. By the assumption
$N_{e,e^{\prime}}(v;\underline{n})\neq 0$ this walk visits any vertex of
the graph at least once. Again, let $\cG^\prime$ be the graph obtained from
$\cG$ by removing the internal edges $i\in\cI$ with $n_i=0$.

If $\partial(e)=\partial(e^\prime)$, then the minimum degree of the graph
$\cG^\prime_{\mathrm{int}}$ is not less than $2$. Indeed,
$\deg_{\cG^\prime_{\mathrm{int}}}(v)\geq 2$ for any vertex
$v\neq\partial(e)$ by assumption. On the other hand for $v=\partial(e)$,
the walk $\bw$ visits $\partial(e)$ at least twice and, hence, there must
be at least two internal edges $i\in\cI$ incident with $\partial(e)$.
Condition \eqref{nn} and the First Theorem of Graph Theory imply that
\begin{equation*}
\sum_{v\in V} \deg_{\cG^\prime_{\mathrm{int}}}(v) = 2 |\cI^\prime| = 2|V|.
\end{equation*}
Therefore, $\deg_{\cG^\prime_{\mathrm{int}}}(v)=2$ for all $v\in V$. This
implies that the walk $\bw$ visits every vertex $v\neq\partial(e)$ exactly
once and the vertex $\partial(e)$ twice.

If $\partial(e)\neq\partial(e^\prime)$, then
\begin{equation*}
\deg_{\cG^\prime_{\mathrm{int}}}(v) \geq 2\quad\text{for all}\quad v\in
V\setminus\{\partial(e),\partial(e^\prime)\},
\end{equation*}
and
\begin{equation*}
\deg_{\cG^\prime_{\mathrm{int}}}(v) \geq 1\quad\text{if}\quad v\in
\{\partial(e),\partial(e^\prime)\}.
\end{equation*}
Again by \eqref{nn} and the First Theorem of Graph Theory we have that
\begin{equation*}
\sum_{v\in V} \deg_{\cG^\prime_{\mathrm{int}}}(v)= 2 |\cI^\prime| = 2|V|-2.
\end{equation*}
Therefore, $\deg_{\cG^\prime_{\mathrm{int}}}(v)=2$ for all $v\in
V\setminus\{\partial(e),\partial(e^\prime)\}$ and
$\deg_{\cG^\prime_{\mathrm{int}}}(v)=1$ for
$v\in\{\partial(e),\partial(e^\prime)\}$. This implies that the walk $\bw$
visits every vertex $v\in V$ exactly once.
\end{proof}


Note that the condition of Theorem \ref{5:theo:1} can be verified by
computing the Fourier coefficients of the scattering matrix
$S(\sk;A,B,\underline{a})$ for the Laplace operator associated with any of
the boundary conditions $(A,B)$ referred to in Theorem \ref{4:theo:2}.
However, one is still faced with the combinatorial problem for scores
$\underline{n}\in\cN_{e,e^\prime}^{\mathrm{HPP}}$. Our next aim is to solve
this problem by purely analytic means.

\begin{figure}
\begin{equation*}
\setlength{\unitlength}{1.5cm}
\begin{picture}(2,2)
\put(0,1){\line(1,0){1}} \put(1,1){\circle*{0.1}} \put(0.7, 0.7){$v$}
\put(2.6,1){$b_{v}$} \put(1,1){\line(-1,1){1}} \put(1,1){\line(0,-1){1}}

\curve(1.0,1.0,2.0,1.5,2.5,1.0,2.0,0.5,1.0,1.0)
\end{picture}
\end{equation*}
\caption{\label{figtad:neu} A tadpole (penalty lap) attached to a vertex $v$
which was originally incident with three edges.}
\end{figure}
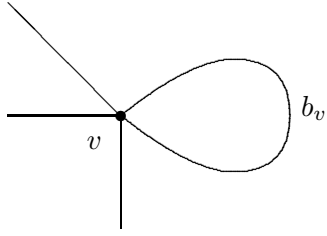

At each vertex $v\in V\setminus\{\partial(e),\partial(e^\prime)\}$ of the
graph $\cG$ we attach a tadpole of metric length $b_{v} > 0$ (see
Fig.~\ref{figtad:neu}). Having in mind some similarity with biathlon
competition, we will call these tadpoles \emph{penalty laps}. The resulting
graph $\widetilde{\cG}=(V,\widetilde{\cI},\cE,\widetilde{\partial})$ has
the same vertex set $V$ and the same set of the external edges $\cE$. The
degree of any vertex $v\notin\{\partial(e),\partial(e^\prime)\}$ in the
graph $\widetilde{\cG}$ is given by
$\deg_{\widetilde{\cG}}(v)=\deg_{\cG}(v)+2$. The set of the internal edges
$\widetilde{\cI}$ is the union $\cI\cup\cJ$ of the set $\cI$ of the
internal edges of the graph $\cG$ and the set $\cJ$ of the penalty laps. By
construction, there is a bijective mapping $\phi:\
V\setminus\{\partial(e),\partial(e^\prime)\}\rightarrow \cJ$. The boundary
operator $\widetilde{\partial}$ is defined as follows
\begin{equation*}
\widetilde{\partial}(i) = \begin{cases}
\partial(e) & \text{if}\quad e\in\cE,\\
\partial(i) & \text{if}\quad i\in\cI,\\
\{\phi^{-1}(i),\phi^{-1}(i)\} & \text{if}\quad i\in\cJ.
                          \end{cases}
\end{equation*}
The metric structure on the graph $\widetilde{\cG}$ is given by
$(\underline{a},\underline{b})$, where $\underline{a}=\{a_i\}_{i\in\cI}$,
$\underline{b}=\{b_j\}_{j\in \cJ}$.

Choose any boundary conditions on the graph $\widetilde{\cG}$ of the type
referred to in Theorem \ref{4:theo:2}. These boundary conditions define a
self-adjoint Laplace operator. For brevity we denote by
$\widetilde{S}(\sk;\underline{a},\underline{b})$ the associated scattering
matrix. Let $\widetilde{\cN}_{e,e^\prime}$ be the set of scores of all
walks from $e^\prime$ to $e$ defined similarly to \eqref{enn},
\begin{equation*}
\widetilde{\cN}_{e,e^\prime} = \{(\underline{n},\underline{m})\mid
\text{there is walk}\,
\bw\in\widetilde{\cW}_{e,e^\prime}(\underline{n},\underline{m})\},
\end{equation*}
where $\widetilde{\cW}_{e,e^\prime}(\underline{n},\underline{m})$ is a
subset of all walks on the graph $\widetilde{\cG}$ from $e^\prime$ to $e$
having the score $(\underline{n},\underline{m})$. Recall that
$\underline{n}=\{n_i\}_{i\in\cI}$ and $\underline{m}=\{m_j\}_{j\in \cJ}$,
where $n_i\in\N_0$ is the number of times a walk $\bw$ traverses the edge
$i\in\cI$ and $m_j\in\N_0$ is the number of times a walk $\bw$ traverses the
penalty lap $j\in\cJ$.

Theorems \ref{thm:main:harmony} and \ref{verloren} immediately imply that
the scattering matrix $\widetilde{S}(\sk;\underline{a},\underline{b})$ is
given by the absolutely converging Fourier series
\begin{equation}\label{tadasym:neu}
[\widetilde{S}(\sk;\underline{a},\underline{b})]_{e,e^{\prime}}=
\sum_{(\underline{n},\underline{m})\in\widetilde{\cN}_{e,e^{\prime}}}
[\widehat{\widetilde{S}}_{\underline{n},\underline{m}}(\sk)]_{e,e^{\prime}}
\;\e^{\ii\sk\langle\underline{n},\underline{a}\rangle}
\e^{\ii\sk\langle\underline{m},\underline{b}\rangle},
\end{equation}
where
\begin{equation*}
[\widehat{\widetilde{S}}_{\underline{n},\underline{m}}(\sk)]_{e,e^{\prime}}
= \left(\frac{\sk}{2\pi}\right)^{|\cI|+|\cJ|} \int_{[0,2\pi/\sk]^{|\cI|}}
d\underline{a} \int_{[0,2\pi/\sk]^{|\cJ|}} d\underline{b}\,\,
[\widetilde{S}(\sk;\underline{a},\underline{b})]_{e,e^{\prime}}
\e^{-\ii\sk\langle\underline{n},\underline{a}\rangle}
\e^{-\ii\sk\langle\underline{m},\underline{b}\rangle}.
\end{equation*}
By Theorem \ref{4:theo:2} none of the Fourier coefficients
$\widehat{\widetilde{S}}_{\underline{n},\underline{m}}(\sk)_{e,e^{\prime}}$
vanishes identically whenever $(\underline{n},\underline{m})$
$\in\widetilde{\cN}_{e,e^{\prime}}$.

We are interested in the terms with
$\underline{m}=\underline{1}\equiv\{1\}_{j\in \cJ}$, i.e., walks traversing
each penalty lap exactly once. Such walks visit each vertex of the graph at
least once.


\begin{theorem}\label{thm:9:3:aha}
Assume that the minimum degree of the graph $\cG$ is bigger than $1$. The
HPP has a solution if and only if the Fourier coefficient
$[\widehat{\widetilde{S}}_{\underline{n}^{(0)},\underline{1}}(\sk)]_{e,e^\prime}$
does not vanish identically for some $\underline{n}^{(0)}\in(\N_0)^{|\cI|}$
with
\begin{equation}\label{nn:neu}
|\underline{n}^{(0)}|= \begin{cases}
|V|\quad&\text{if}\quad \partial(e)=\partial(e^{\prime}),\\
|V|-1\quad&\text{if}\quad \partial(e)\neq \partial(e^{\prime}).
\end{cases}
\end{equation}
In this case the Fourier coefficients
$[\widehat{\widetilde{S}}_{\underline{n},\underline{1}}(\sk)]_{e,e^\prime}$
vanish identically for all $\underline{n}\in(\N_0)^{|\cI|}$ with
$|\underline{n}| < |\underline{n}^{(0)}|$.
\end{theorem}

\begin{remark}
By Remark \ref{Taylor} the natural number $|\underline{n}_0|$ is the order
of the first nontrivial coefficient in the Taylor expansion of a function
holomorphic in a polydisc. Once the existence of a solution of HPP is
established, one can uniquely reconstruct the walk from its score
$\underline{n}^{(0)}$.
\end{remark}

\begin{proof}
The ``only if'' part follows from Lemma \ref{lem:9:1:neu} and Theorem
\ref{5:theo:1}. We turn to the proof of the ``if'' part. Assume there is
$\underline{n}^{(0)}\equiv\{n^{(0)}_i\}_{i\in\cI}\in(\N_0)^{|\cI|}$
satisfying \eqref{nn:neu} such that
$\widehat{S}_{\underline{n}^{(0)},\underline{1}}(\sk)$ does not vanish
identically. Then by Theorem \ref{4:theo:2} there is walk $\widetilde{\bw}$
on the graph $\widetilde{\cG}$ with the score
$(\underline{n}^{(0)},\underline{1})\in(\N_0)^{|\cI|+|\cJ|}$ visiting every
vertex $v\in V$ at least once. Erasing the ``penalty laps'' in this walk we
obtain a walk $\bw$ from $e^\prime$ to $e$ on the graph $\cG$ with a score
$\underline{n}^{(0)}$ visiting every vertex $v\in V$ at least once. Thus,
$N_{e,e^\prime}(v,\underline{n}^{(0)})\neq 0$ for all $v\in V$.

Let $\cG^\prime$ be the graph obtained from $\cG$ by removing the internal
edges $i\in\cI$ with $n_i=0$.

\textit{Claim 1:} $\deg_{\cG^\prime_{\mathrm{int}}}(v)\geq 2$ holds for all
$v\in V$ if $\partial(e)=\partial(e^\prime)$ and for all $v\in
V\setminus\{\partial(e),\partial(e^\prime)\}$ if
$\partial(e)\neq\partial(e^\prime)$. Consider the case
$\partial(e)=\partial(e^\prime)$. Assume to the contrary that there is at
least one vertex with $\deg_{\cG^\prime_{\mathrm{int}}}(v)=1$. Let
$i^\prime\in\cI^\prime$ be the edge incident with $v$. Obviously, the walk
$\bw$ traverses $i^\prime$ at least twice, that is, $n^{(0)}_{i^\prime}\geq
2$. Therefore,
\begin{equation*}
\begin{split}
|\underline{n}^{(0)}| & \geq |\cI^\prime| + |\{v\in V| \deg_{\cG^\prime_{\mathrm{int}}}(v)=1\}|\\
& = \frac{1}{2} \sum_{v\in V} \deg_{\cG^\prime_{\mathrm{int}}}(v) +
|\{v\in V| \deg_{\cG^\prime_{\mathrm{int}}}(v)=1\}|\\
& = \frac{1}{2} \sum_{v\in V} \max\{\deg_{\cG^\prime_{\mathrm{int}}}(v),2\}
+ \frac{1}{2} |\{v\in V| \deg_{\cG^\prime_{\mathrm{int}}}(v)=1\}|
 > |V|,
\end{split}
\end{equation*}
which contradicts the assumption $|\underline{n}^{(0)}|=|V|$. Similarly, if
$\partial(e)\neq\partial(e^\prime)$ we obtain the inequality
$|\underline{n}^{(0)}| > |V| - 1$ which contradicts the assumption
$|\underline{n}^{(0)}| = |V| - 1$. The claim is proven.

\textit{Claim 2:} $\underline{n}^{(0)}$ is simple. Assume not, that is,
there is $i^\prime\in\cI^\prime$ with $n_{i^\prime}^{(0)}\geq 2$. Using
Claim 1 we obtain the inequality
\begin{equation*}
|\underline{n}^{(0)}| > |\cI^\prime| = \frac{1}{2}\sum_{v\in V}
\deg_{\cG^\prime_{\mathrm{int}}}(v) \geq \begin{cases}
|V| & \text{if}\quad \partial(e)=\partial(e^\prime),\\
|V|-1 & \text{if}\quad \partial(e)\neq\partial(e^\prime),
\end{cases}
\end{equation*}
which contradicts the assumption of the theorem.

Applying Theorem \ref{5:theo:1} we obtain the first assertion of the
theorem. Assume now that there is $\underline{n}\in(\N_0)^{|\cI|}$ with
$|\underline{n}| < |\underline{n}^{(0)}|$ such that
$[\widehat{\widetilde{S}}_{\underline{n},\underline{1}}(\sk)]_{e,e^\prime}$
does not vanish identically. Then there is a walk from $e^\prime\in\cE$ to
$e\in\cE$ on the graph $\cG$ with the score
$\underline{n}=\{n_i\}_{i\in\cI}$ visiting every vertex of the graph at
least once.

Again by $\cG^\prime$ we denote the graph obtained from $\cG$ by removing
the internal edges $i\in\cI$ with $n_i=0$. Repeating the arguments used to
prove Claim 1 we obtain that $\deg_{\cG^\prime_{\mathrm{int}}}(v)\geq 2$
holds for all $v\in V$ if $\partial(e)=\partial(e^\prime)$ and for all $v\in
V\setminus\{\partial(e),\partial(e^\prime)\}$ if
$\partial(e)\neq\partial(e^\prime)$. But this immediately leads to the
inequality
\begin{equation*}
|\underline{n}| \geq|\cI^\prime| = \frac{1}{2}\sum_{v\in V}
\deg_{\cG^\prime_{\mathrm{int}}}(v) \geq  |\underline{n}^{(0)}|,
\end{equation*}
which is a contradiction.
\end{proof}

\subsection{Solution of TSP} Recall that the metric
$\underline{a}_0\in(\R_+)^{|\cI|}$ on the graph $\cG$ is given. Choose
arbitrary boundary conditions $(A,B)$ referred to in Theorem
\ref{4:theo:2}. For an arbitrary metric structure
$\underline{a}\in(\R_+)^{|\cI|}$ they define a self-adjoint Laplace
operator $\Delta(A,B;\underline{a})$. The associated scattering matrix
$S(\sk; A,B,\underline{a})$ we will denote for brevity by
$S(\sk;\underline{a})$.

For arbitrary $e,e^\prime\in\cE$ let $\cN_{e,e^\prime}^{\mathrm{(I)}}$ be
the set of all vectors
$\underline{n}\equiv\{n_i\}_{i\in\cI}\in(\N_0)^{|\cI|}$ satisfying the
following conditions:
\begin{itemize}
\item[(i)]{the matrix element $\widehat{S}_{\underline{n}}(\sk)_{e,e^\prime}$
of the Fourier coefficient \eqref{fourier:coef} of the scattering matrix
$S(\sk;\underline{a})$ does not vanish identically for all $\sk>0$,}
\item[(ii)]{$n_i\in\{0,1,2\}$ for all $i\in\cI$,}
\item[(iii)]{$N_{e,e^\prime}(v;\underline{n})\neq 0$ for all $v\in V$.}
\end{itemize}

Note that by Theorem \ref{4:theo:2} condition (i) implies that
$\underline{n}$ is the score of some walk from $e^\prime$ to $e$. Therefore, the quantity
$N_{e,e^\prime}(v;\underline{n})$ in condition (iii) is well-defined. Obviously,
\begin{equation*}
\cN_{e,e^{\prime}}^{\mathrm{(I)}}\neq
\emptyset\qquad\text{and}\qquad|\cN_{e,e^{\prime}}^{\mathrm{(I)}}| \leq 3^{|\cI|}.
\end{equation*}

\begin{theorem}\label{5:theo:2}
A vector $\underline{n}\in\cN_{e,e^\prime}^{\mathrm{(I)}}$ satisfying
\begin{equation}\label{nI}
\langle\underline{n},\underline{a}_0\rangle \leq
\langle\underline{n}^\prime,\underline{a}_0\rangle
\end{equation}
for all $\underline{n}^\prime\in\cN_{e,e^\prime}^{\mathrm{(I)}}$ is the
score of a walk from $e^\prime$ to $e$ solving TSP I. Conversely, if a walk
$\bw$ from $e^\prime$ to $e$ solves TSP I, its score $\underline{n}(\bw)$
belongs to $\cN_{e,e^\prime}^{\mathrm{(I)}}$ and satisfies \eqref{nI}.
\end{theorem}

The solution of TSP I is unique if the lengths of internal edges are
rationally independent.

We expect the following result to be known.

\begin{lemma}\label{lem:12:4}
The score $\underline{n}(\bw)$ of any solution $\bw$ of TSP I is in
$\cN_{e,e^{\prime}}^{\mathrm{(I)}}$.
\end{lemma}

\begin{proof}
It suffices to prove that the score
$\underline{n}(\bw)\equiv\{n_i\}_{i\in\cI}$ of any solution $\bw$ of TSP I
satisfies $n_i\leq 2$ for all $i\in\cI$. To show this we will assume on the
contrary that there is at least one internal edge of the graph $\cG$ which
is traversed by the walk $\bw$ at least three times. Obviously, this edge is
traversed at least twice in the same direction. We will construct another
walk $\bw^\prime$ from $e^\prime$ to $e$ with metric length
$|\bw^\prime|<|\bw|$ which visits every vertex at least once.

By assumption the walk $\bw$ is of the form
\begin{equation*}
\bw = \{e^\prime,i_1,\ldots, i_p,\ldots, i_q,\ldots, i_N,e \},
\end{equation*}
where $i_p=i_q$ for some $p < q$ and either
\begin{equation*}
\partial^-(i_p)\in\partial(i_{p-1}),\quad
\partial^+(i_p)\in\partial(i_{p+1}),\quad
\partial^-(i_q)\in\partial(i_{q-1}),\quad \partial^+(i_q)\in\partial(i_{q+1})
\end{equation*}
or
\begin{equation*}
\partial^-(i_p)\in\partial(i_{p+1}),\quad
\partial^+(i_p)\in\partial(i_{p-1}),\quad
\partial^-(i_q)\in\partial(i_{q+1}),\quad
\partial^+(i_q)\in\partial(i_{q-1}).
\end{equation*}

Consider the walk from $e^\prime$ to $e$
\begin{equation*}
\bw^\prime =
\{e^\prime,i_1,\ldots,i_{p-1},i_{q-1},\ldots,i_{p+1},i_{q+1},\ldots,i_N\},
\end{equation*}
where the sequence $i_{q-1},\ldots,i_{p+1}$ is the part
$i_{p+1},\ldots,i_{q-1}$ of the walk $\bw$ taken in the reversed order.
Obviously, the metric length of the walk $\bw^\prime$ is given by
\begin{equation*}
|\bw^\prime| = |\bw| - 2 a_{i_p} < |\bw|.
\end{equation*}
This completes the proof.
\end{proof}

\begin{proof}[Proof of Theorem \ref{5:theo:2}]
Let $\underline{n}\in\cN_{e,e^\prime}^{\mathrm{(I)}}$ satisfy \eqref{nI}.
Then there is a walk $\bw$ from $e^\prime$ to $e$ with the score
$\underline{n}$ having the shortest length among all walks from $e^\prime$
to $e$ with scores in $\cN_{e,e^\prime}^{\mathrm{(I)}}$. By Lemma
\ref{lem:12:4} the walk $\bw$ is a solution of TSP I.
\end{proof}

The \textit{solutions of TSP II and III} are similar. We formulate the
final results.

\begin{theorem}\label{5:theo:3}
A walk $\bw$ from $e^{\prime}$ to $e$ solving TSP II exists if and only if
there is a solution of HPP. Among the solutions of HPP it is one with
shortest metric length. The corresponding score $\underline{n}(\bw)$ is
unique whenever the metric lengths $\underline{a}_0$ of internal edges are
rationally independent.
\end{theorem}

\begin{theorem}\label{5:theo:4}
A walk $\bw$ from $e^{\prime}$ to $e$ solving TSP III exists if and only if
there is a score $\underline{n}\in \cN_{e,e^{\prime}}^{\mathrm{(I)}}$ such
that $\langle\underline{n},\underline{a}_0\rangle\leq L$.
\end{theorem}

As in the case of HPP one can solve the optimization problems referred to
in Theorems \ref{5:theo:2}, \ref{5:theo:3}, and \ref{5:theo:4} in an
analytic way. In what follows we assume that the metric lengths
$\underline{a}_0$ of internal edges satisfy Assumption \ref{2cond}.
Obviously, given $\underline{a}\in(\R_+)^{|\cI|}$, this assumption can
always be satisfied by changing the lengths by arbitrarily small amount.
Consider the graph $\widetilde{\cG}$ obtained from $\cG$ by adding penalty
laps at each vertex $v\in V\setminus\{\partial(e),\partial(e^\prime)\}$ and
the scattering matrix $\widetilde{S}(\sk;\underline{a},\underline{b})$
associated with some boundary conditions referred to in Theorem
\ref{4:theo:2} (see the previous subsection).

We consider the partial Fourier transform of the scattering matrix
\eqref{tadasym:neu} with respect to the variables $\underline{b}$
\begin{equation*}
\widetilde{S}_{\sharp}(\sk;\underline{a})
=\left(\frac{\sk}{2\pi}\right)^{|\cJ|}\int_{[0,2\pi/\sk]^{|\cJ|}}
d\underline{b}\: \widetilde{S}(\sk;\underline{a},\underline{b})\:
\e^{-\ii\sk\langle\underline{1},\underline{b}\rangle}
\end{equation*}
such that
\begin{equation*}
\widehat{\widetilde{S}}_{\underline{n}, \underline{1}}(\sk)  =
\left(\frac{\sk}{2\pi}\right)^{|\cI|}
\int_{[0,2\pi/\sk]^{|\cI|}}d\underline{a}\:
\widetilde{S}_{\sharp}(\sk;\underline{a})\:
\e^{-\ii\sk\langle\underline{n},\underline{a}\rangle}\\
\end{equation*}
and the series
\begin{equation*}
[\widetilde{S}_{\sharp}(\sk;\underline{a})]_{e,e^{\prime}} =
\sum_{\underline{n}\in\cN_{e,e^\prime}}
[\widehat{\widetilde{S}}_{\underline{n},
\underline{1}}(\sk)]_{e,e^{\prime}}\;
\e^{\ii\sk\langle\underline{n},\underline{a}\rangle}
\end{equation*}
is absolutely convergent for all $\sk>0$. By Theorem \ref{4:theo:2}
$\widehat{\widetilde{S}}_{\underline{n},
\underline{1}}(\sk)_{e,e^{\prime}}$ does not vanish identically if and only
if there is a walk on the graph $\cG$ from $e^\prime$ to $e$ visiting every
vertex of the graph at least once.

By Theorem \ref{eindeutigkeit} and Remark \ref{rem} the function
$\sk\mapsto [\widetilde{S}_{\sharp}(\sk;\underline{a}_0)]_{e,e^{\prime}}$
determines the lengths of all walks from $e^{\prime}$ to $e$ on the graph
$\cG$ visiting every vertex of the graph at least once. Let $w$ be the
smallest number in this set. In particular, we have
\begin{equation*}
w = \lim_{\Im\, \sk\rightarrow
  \infty}\frac{1}{\ii}\frac{\partial}{\partial \sk}
\log [\widetilde{S}_{\sharp}(\sk;\underline{a}_0)]_{e,e^{\prime}}.
\end{equation*}
By Assumption \ref{2cond} there is a unique $\underline{n}\in(\N_0)^{|\cI|}$
such that $w = \langle \underline{n}, \underline{a}_0\rangle$. Obviously,
$\underline{n}$ is the score of a walk from $e^{\prime}$ to $e$ on the
graph $\cG$ solving TSP I. If HPP has a solution, then, by Theorem
\ref{thm:9:3:aha}, the solution of TSP II is a walk with the score
satisfying \ref{nn:neu} and having the shortest metric length. TSP III has
a solution if and only if $w\leq L$.


\end{document}